\documentclass[12pt]{article}

\usepackage{epsfig,multicol,multirow,cite}
\usepackage{amsmath,subfigure,latexsym,amssymb}

\def\lsi{\raise0.3ex\hbox{$<$\kern-0.75em\raise-1.1ex\hbox{$\sim$}}}
\def\gsi{\raise0.3ex\hbox{$>$\kern-0.75em\raise-1.1ex\hbox{$\sim$}}}
\def\backder{\raise1.4ex\hbox{$\leftarrow$\kern-0.75em\raise-1.4ex\hbox{$\partial$}}}
\def\forder{\raise1.4ex\hbox{$\rightarrow$\kern-0.75em\raise-1.4ex\hbox{$\partial$}}}

\newcommand{\lsim}{\mathop{\lsi}}
\newcommand{\gsim}{\mathop{\gsi}}

\newcommand{\be}{\begin{equation}}
\newcommand{\ee}{\end{equation}}
\newcommand{\nn}{\nonumber}
\newcommand{\bea}{\begin{eqnarray}}
\newcommand{\eea}{\end{eqnarray}}

\newcommand{\R}{{\kern+.25em\sf{R}\kern-.78em\sf{I} \kern+.78em\kern-.25em}}
\newcommand{\RR}{{\kern+.25em\sf{R}\kern-.6em\sf{I} \kern+.6em\kern-.25em}}
\newcommand{\N}{{\kern+.25em\sf{N}\kern-.78em\sf{I} \kern+.78em\kern-.25em}}
\newcommand{\C}{{\kern+.25em\sf{C}\kern-.50em\sf{I} \kern+.50em\kern-.25em}}

\newcommand{\ri}{{\rm i}}

\makeatletter
\@addtoreset{equation}{section}
\makeatother

\begin{document}

\begin{flushright} 
BI-TP 2011/13
\end{flushright}

\vspace*{3mm}

\begin{center}

{\Large\bf A Numerical Study of the 2-Flavour} \\
\vspace*{4mm}
{\Large\bf Schwinger Model with Dynamical} \\
\vspace*{4mm}
{\Large\bf Overlap Hypercube Fermions}

\vspace*{10mm}

Wolfgang Bietenholz$^{\rm \, a}$, Ivan Hip$^{\rm \, b}$, \\
Stanislav Shcheredin$^{\rm \, c}$ and Jan Volkholz$^{\rm \, d}$

\vspace*{7mm}

{\small 
$^{\rm \, a}$  Instituto de Ciencias Nucleares \\
Universidad Nacional Aut\'{o}noma de M\'{e}xico \\
A.P. 70-543, C.P. 04510 Distrito Federal, Mexico \\

\vspace*{2mm}

$^{\rm \, b}$ Faculty of Geotechnical Engineering, University of Zagreb \\
Hallerova aleja 7, 42000 Vara\v{z}din, Croatia \\

\vspace*{2mm}

$^{\rm \, c}$ 
Fakult\"{a}t f\"{u}r Physik, Universit\"{a}t Bielefeld \\
D-33615 Bielefeld, Germany \\

\vspace*{2mm}
$^{\rm \, d}$ Potsdam Institute for Climate Impact Research \\
Telegrafenberg A62, D-14412 Potsdam, Germany

\vspace*{5mm}

}

\end{center}

\vspace*{3mm}

\noindent

\noindent
We present numerical results for the 2-flavour Schwinger model 
with dynamical chiral lattice fermions. We insert an approximately
chiral hypercube Dirac operator into the overlap formula to construct
the overlap hypercube operator. This is an exact solution to
the Ginsparg-Wilson relation, with an excellent level of locality
and scaling. Due to its similarity with the hypercubic
kernel, a low polynomial in this kernel provides a numerically 
efficient Hybrid Monte Carlo force.
We measure the microscopic Dirac spectrum 
and discuss the corresponding scale-invariant parameter,
which takes a surprising form.
This is an interesting case, since Random 
Matrix Theory is unexplored for this setting, where the chiral
condensate $\Sigma$ vanishes in the chiral limit.
We also measure $\Sigma$ and the ``pion'' mass,
in distinct topological sectors. 
In this context we discuss and probe the topological summation of 
observables by various methods, as well as the evaluation of the 
topological susceptibility. The feasibility of
this summation is essential for the prospects of 
dynamical overlap fermions in QCD.\\

%\noindent
%{\footnotesize{Keywords: Lattice gauge theory; chiral fermions;
%Hybrid Monte Carlo simulation; \\
%\hspace*{1.63cm} Dirac spectrum; topological summation}}

\newpage

\tableofcontents

\section{The Schwinger model}

The Schwinger model describes QED in two dimensions, {\it i.e.}\
2-component Dirac fermions interacting through a $U(1)$ gauge field
\cite{schwingmod}. In a Euclidean plane, the Lagrangian reads
\be
{\cal L} (\bar \psi , \psi , A_{\mu}) =
\bar \psi (x) \Big[ \gamma_{\mu} ( {\rm i} \partial_{\mu} + g A_{\mu}(x) ) 
+ m \Big] \psi (x) + \frac{1}{2} F_{\mu \nu}(x) F_{\mu \nu}(x) \ .
\ee
It is a popular toy model for QCD; in particular 
it shares the property of confinement \cite{schwingmod,confine}.
On the other hand there are fundamental differences, such as 
the super-renormalisability of the Schwinger model, and a 
non-running gauge coupling $g$.

A further qualitative difference --- which is of particular interest 
in this work --- is the spontaneous chiral symmetry breaking
in QCD with massless quarks. In $d=2$ this effect can be
mimicked to some extent for instance by the Gross-Neveu model,
where a discrete chiral symmetry breaks spontaneously.
However, in the Schwinger model with fermion mass $m=0$ the
chiral symmetry is continuous, and therefore it cannot
undergo spontaneous symmetry breaking (SSB) due to the
Mermin-Wagner Theorem \cite{MW}. Nevertheless the chiral condensate
$\Sigma = - \langle \bar \psi \psi \rangle \, $,
which acts as the order parameter for chiral symmetry breaking,
takes a non-vanishing value in the 1-flavour case,
because of the explicit symmetry breaking due to
the axial anomaly \cite{schwingmod}. This leads to  
$\Sigma (m =0) = (e^{\gamma}/ 2 \pi^{3/2}) \, g \simeq 0.16 \, g\, $
(where $\gamma$ is Euler's constant).

For $N_{f} \geq 2$, however, the massless limit has an {\em unbroken} 
chiral symmetry. For $N_{f}$ degenerate fermion flavours
of mass $m$ the chiral condensate behaves as \cite{Smilga92}
\be  \label{deltatheo}
\Sigma (m) % \equiv - \langle \bar \psi \, \psi \rangle 
\propto \Big( \frac{m^{N_{f}-1}}{\beta} \Big)^{1/(N_{f}+1)} 
\quad \Rightarrow \quad \delta = \frac{N_{f}+1}{N_{f}-1} \ ,
%\qquad ( \beta = 1 /g^{2})
\ee
where $\beta = 1 /g^{2}$, and $\delta$ is the critical exponent. 

In our study we consider $N_{f}=2$. 
Here there are analytical evaluations
of the proportionality constant for the case of light
fermions, $m \ll g$, based on bosonisation and low energy 
assumptions,
\be  \label{Sigmam}
\Sigma (m) = {\rm const.} \ \Big( \frac{m}{\beta} \Big) ^{1/3} \ , \quad
{\rm const.} = \left\{ \begin{array}{lcr}
%0.372 \dots && \hspace*{3mm} {\rm Ref.~\protect{\cite{HHI}}} \\
%0.388 \dots && \hspace*{3mm} {\rm Ref.~\protect{\cite{Smilga}}} 
0.372 \dots && \hspace*{3mm} {\rm Ref.~[5]} \\
0.388 \dots && \hspace*{3mm} {\rm Ref.~[6]} 
\end{array} \right. \ .
\ee
Under the same assumptions the mass of the iso-triplet (``pion'')
\cite{Smilga} and of the iso-singlet (``$\eta$ particle'')
\cite{Ivan} are predicted as
\be  \label{mpieta}
M_{\pi} = 2.008 \dots (m^{2} g)^{1/3} \ , \quad
M_{\eta} = \sqrt{M_{\pi}^{2} + \frac{2 g^{2}}{\pi} }  \ .
\ee
The relation $M_{\pi} \propto m^{2/3}$ replaces
the Gell-Mann--Oakes--Renner relation ($M_{\pi} \propto \sqrt{m}$)
of QCD \cite{GMOR}.

For models with a finite condensate $\Sigma (m=0)$, its value can 
be determined from the Banks-Casher plateau of the Dirac eigenvalue
density at zero \cite{BC} (or near zero in a finite volume). 
Moreover, chiral Random Matrix Theory (RMT) has elaborated 
subtle techniques to predict a wiggle structure on this plateau,
which allows for a refined determination of
$\Sigma$ from the densities of the
low-lying Dirac eigenvalues \cite{RMT1,RMT2}. This method has been
tested successfully in the $\epsilon$-regime of QCD with
quenched \cite{BJS,RMTqQCD,BS06}\footnote{Strictly speaking $\Sigma$
diverges logarithmically for increasing volume in the quenched
approximation \cite{logsig}. Still quenched QCD in boxes of length 
$\gsim 1.2 ~{\rm fm}$ yields sensible results for the chiral condensate.} 
and with dynamical \cite{RMTdQCD1,RMTdQCD2}
quarks, and also in the 1-flavour Schwinger model \cite{DuHo,Hell}.
The latter studies were based on configurations,
which were generated quenched and later
re-weighted with the fermion determinant;
we denote this method as ``quenched re-weighted''.\footnote{That 
method was pioneered in the Schwinger model in Ref.\ \cite{Pany}.
It worked successfully in some cases, but it runs into trouble
when the fermion determinant fluctuates strongly, as it happens
for very light fermions. Hence the study presented here requires
the simulation of truly dynamical fermions, as it was first
attempted in Ref.\ \cite{Dubna}.\label{rewei}}

However, the established RMT techniques are not applicable
in the chiral limit of the 2-flavour Schwinger model;
RMT for situations with
$\Sigma ( m = 0) = 0$ awaits to be worked out.
Nevertheless we confirm the usual RMT prediction for the unfolded
level spacing distribution in a unitary ensemble. On the other hand, 
the microscopic spectrum does not exhibit a Banks-Casher
plateau. Instead we observe to high precision the scale-invariance
of the product $\lambda V^{5/8}$, where $\lambda$ is a low-lying
Dirac eigenvalue in the volume $V$.
%which corresponds to a microscopic
%spectral density $\propto \lambda_{1}^{3/5}$. 
This result remains to be understood from the RMT perspective, 
since it cannot be explained simply with the critical exponent 
$\delta =3$ given in eq.\ (\ref{deltatheo}).
%$\Sigma (m) \propto m^{1/\delta}$, $\delta = (N_{f}+1)/(N_{f}-1) 
%= 3$. 
We also discuss the densities of the Dirac eigenvalues in the bulk
and their scaling behaviour.

Next we confront the measurement of $\Sigma (m)$, 
based on the full Dirac spectrum, with eq.\ (\ref{Sigmam}). 
%Finally we present and test a conjecture about
%the volume dependence of $\Sigma$.
This requires an %appropriate
(approximate) summation
of the values in all topological sectors, guided by the measurements
in a few sectors.
%--- that issue is generic in simulations with dynamical chiral fermions. 
We probe several methods for this purpose and apply them to
$\Sigma$ and to the ``pion'' mass given in eq.\ (\ref{mpieta}).
These approaches also involve a determination of the topological 
susceptibility. The requirement of a topological summation
is generic for simulations with dynamical
chiral fermions, because the Monte Carlo histories tend to
perform only very few topological transitions.
Also other lattice fermion formulations, such as the
Wilson fermion and variants thereof, will run into the same problem
when they represent light fermions on very fine lattices 
\cite{MLtopo} (say $a \lsim 0.05~{\rm fm}$ in QCD).
Therefore the applicability  of these techniques is relevant, 
particularly in view of QCD simulations with light quarks
close to the continuum limit.

Section 2 describes our lattice formulation of the Schwinger
model and discusses some of its properties, in particular
locality and scaling.
Section 3 presents our version of the Hybrid Monte Carlo
algorithm which we used in this study. We discuss its
properties regarding conceptual conditions, and practical
aspects of its performance. Section 4 deals with the
Dirac spectrum, the construction of a scale-invariant
variable and the link to RMT. Section 5 discusses 
%three methods for 
the summation of $\Sigma$ and $M_{\pi}$ over the topological sectors,
along with the evaluation of the topological susceptibility.
%The results in Section 4 and 5
%are confronted with eqs.\ (\ref{Sigmam}), (\ref{mpieta}),
%and with a theoretical conjecture about the finite size effects.
%%which is also presented in Section 5. 
Section 6 is devoted to our conclusions.
%along with an outlook. 
Progress reports of this project have appeared in several proceeding
contributions \cite{procs}.

\section{Lattice formulation with overlap Hypercube Fermions}

We consider the lattice formulation of the Schwinger model
with compact link variables
$U_{x, \mu}\in U(1)$, and with the plaquette gauge action. 
Remarkably, for the pure gauge theory this is indeed a
perfect lattice action \cite{WBUJW}.
For the fermions we employ an overlap hypercube fermion 
(overlap-HF) Dirac operator, which is an exact solution to the
Ginsparg-Wilson Relation (GWR).

The GWR is a criterion for a lattice modified, exact chiral symmetry 
\cite{ML}, which first emerged from the study of perfect actions 
for lattice fermions \cite{GW,perf1,WBUJW,Has}.
Independently, chiral lattice fermions were constructed in the
Domain Wall Fermion formulation \cite{DWF}, which separates
the zero modes of opposite chirality in an extra 
``dimension''.\footnote{An extra direction is introduced, which
appears as a dimension for the free fermion, but which does
not carry gauge fields.} Integrating out this extra direction
leads to the overlap formula \cite{Kiku}, which provides yet 
another way to represent a chiral vector theory on the lattice \cite{Neu}. 
The lattice Dirac operator for Domain Wall Fermions (in the limit
of an infinite wall separation) and for overlap fermions turned out 
to be solutions to the GWR as well \cite{Neu}.

Its importance as a general chirality criterion was first pointed 
out in Refs.\ \cite{Has}, which showed in addition that classically
perfect fermion actions obey this criterion as well.
Since those formulations involve couplings over an infinite range
(in $d >1$), a truncation is needed, which marginally
distorts the perfect symmetry and scaling properties. % somewhat.
For the free, optimally local, perfect fermion \cite{WBUJW} the 
truncation to a unit hypercube by means of periodic boundary conditions 
over three lattice spacings 
preserves excellent scaling \cite{BBCW} and chirality \cite{EPJC}. 
It leads to a lattice Dirac operator of the form 
\be  \label{DHF}
D_{{\rm HF}}(x, x+r) = \rho_{\mu}(r) \gamma_{\mu} + \lambda (r) \ ,
\ee
{\it i.e.}\ a vector term plus a scalar term, as in the case of
the Wilson fermion, but with an extended structure
($x$ and $x+r$ are lattice sites).

In $d=2$ these terms include only couplings to nearest neighbour
lattice sites and across the plaquette diagonals. We are using 
here the version denoted as CO-HF (Chirally Optimised 
Hypercube Fermion) in Ref.\ \cite{WBIH}, which is
optimal for our algorithm to be described in Section 3.
For convenience we display in Table \ref{tabcop} the
couplings in the notation of eq.\ (\ref{DHF}).

We gauge $D_{\rm HF}$ by multiplying the compact link variables along 
the shortest lattice paths connecting $x$ and $y = x + r\,$; for 
the diagonal the two shortest paths are averaged \cite{WBIH}. 
Thus we arrive at the operator $D_{{\rm HF}, xy}(U)$, which 
characterises the interacting Hypercube Fermion (HF).

\begin{table}
\begin{center}
\begin{tabular}{|c||c|c|}
\hline
$r$ & $\rho_{1}(r)$ & $\lambda (r)$ \\
\hline
\hline
$(0,0)$ & 0 & $~~~1.49090692$ \\
\hline
$(1,0)$ & $0.30583220$ & $-0.24771369$ \\
\hline
$(1,1)$ & $0.09708390$ & $-0.12501304$ \\
\hline
\end{tabular}
\end{center}
\caption{\it{The coupling constants of the Chirally Optimised
Hypercube Fermion (CO-HF) \cite{WBIH}. Note that $\rho_{1}(r)$
is even in $r_{1}$ and odd in $r_{2}$ (and vice versa for 
$\rho_{2}(r)$), while $\lambda (r)$ is even in both 
components of $r$.}}
\label{tabcop}
\end{table}

\begin{figure}[h!]
%\vspace*{-3mm}
%\begin{center}
\hspace*{-5mm}
\includegraphics[angle=270,width=.55\linewidth]{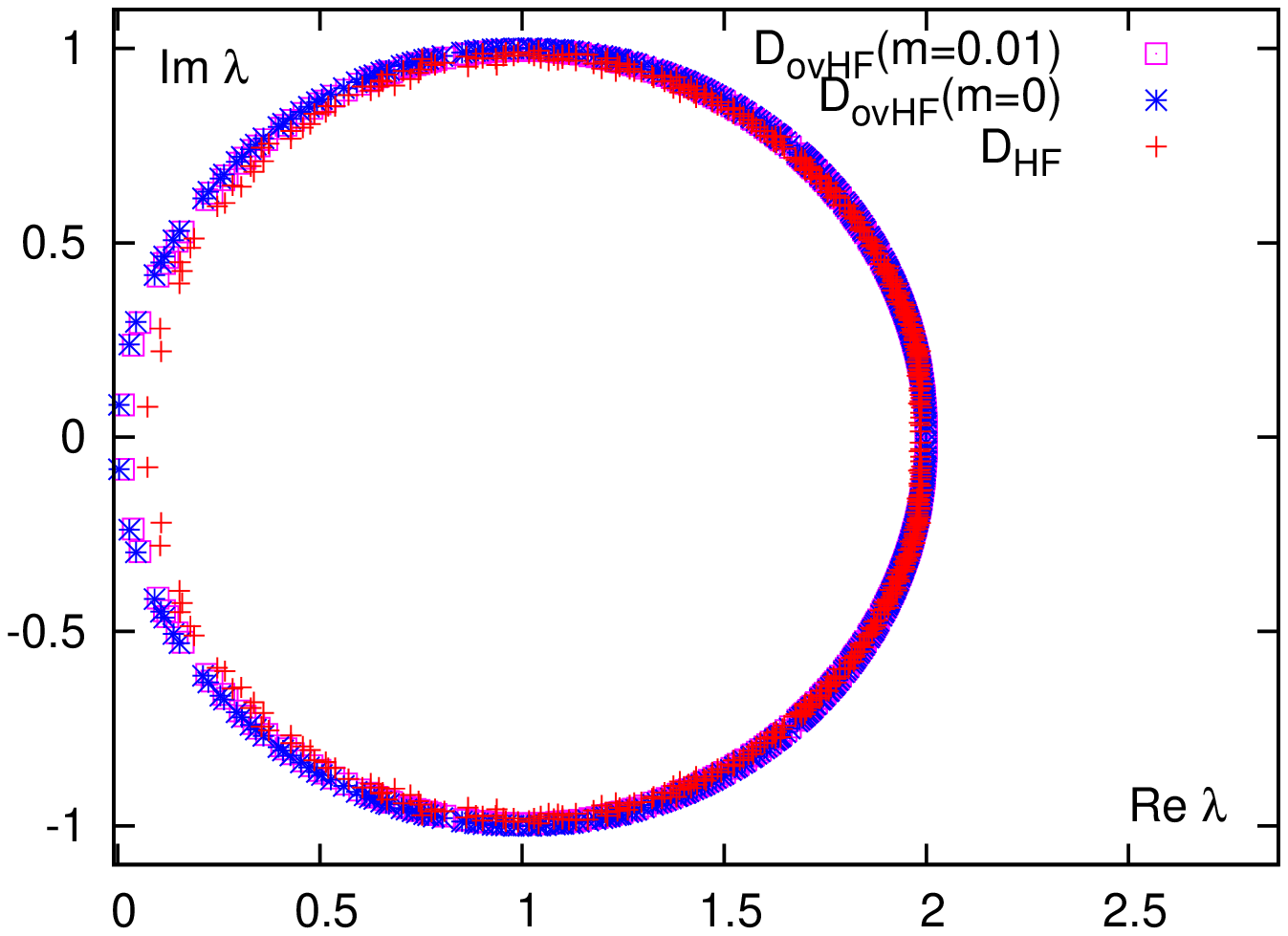}
\hspace*{-6mm}
\includegraphics[angle=270,width=.55\linewidth]{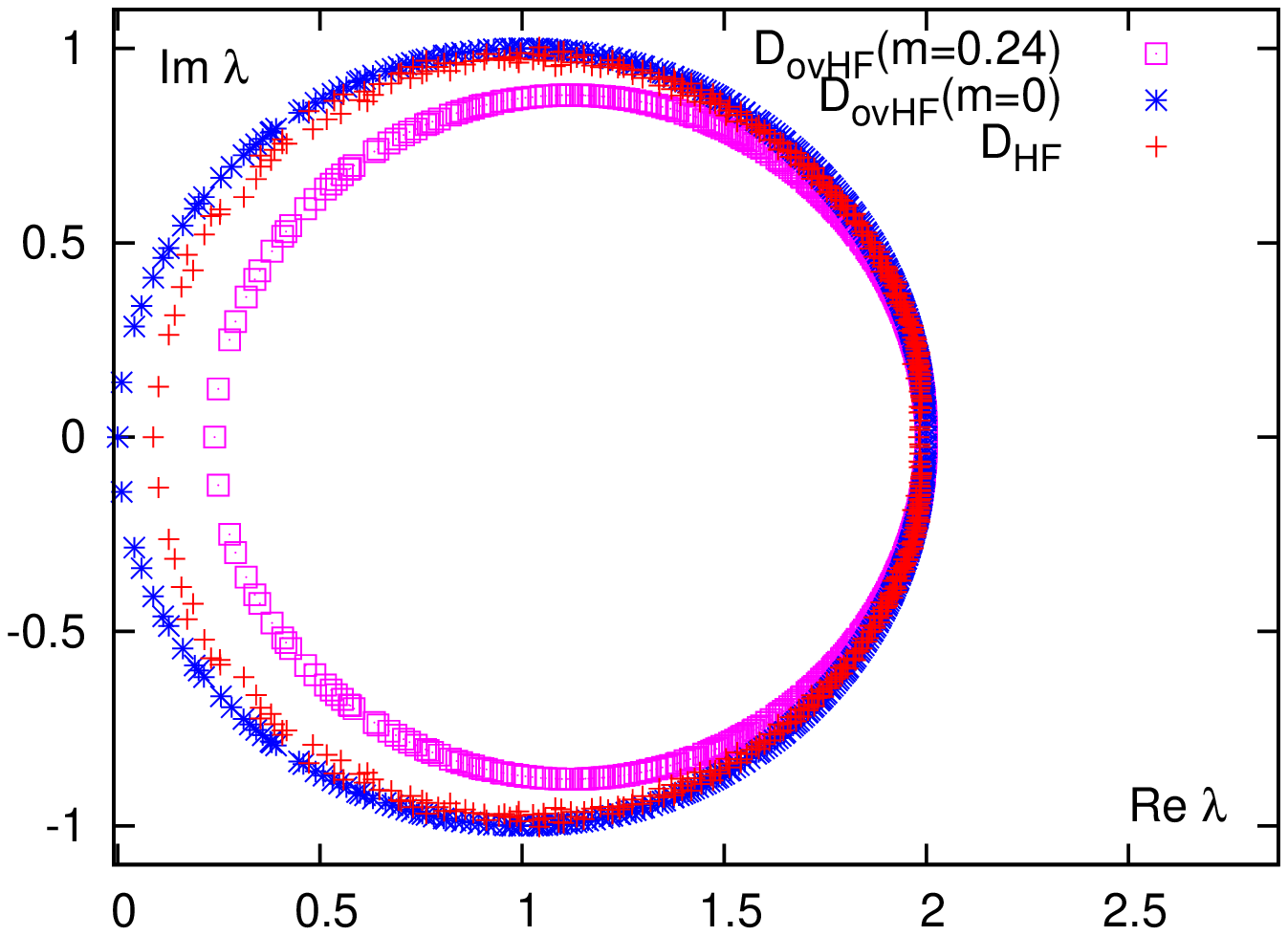}
%\end{center}
 \caption{{\it The spectra of $D_{\rm HF}$ and of $D_{\rm ovHF}$
(with and without mass)
in the complex plane, for a typical configuration generated 
at $\beta =5$ on a $16 \times 16$ lattice with
$m=0.01$ (on the left) and at $m=0.24$ (on the right).
The similarity of $D_{\rm HF}$ to $D_{\rm ovHF}^{(0)}$
and to $D_{\rm ovHF}(m)$ shows that the HF is approximately
chiral, and useful for an efficient computation of the Hybrid Monte
Carlo force (see Section 3).}}
\label{specfig}
\end{figure}

Since the operator $D_{\rm HF}$ is ``$\gamma_{3}$-Hermitian'', 
$D_{\rm HF}^{\dagger} = \gamma_{3} D_{\rm HF} \gamma_{3}$,
the exact chirality (which got lost in the truncation) can be
restored by inserting $D_{\rm HF}$ into the overlap formula \cite{Neu}.
This yields the overlap-HF operator \cite{EPJC,WBIH}
\bea
D_{\rm ovHF}(m) &=& \Big( 1 - \frac{m}{2} \Big) D_{\rm ovHF}^{(0)} + m \ ,
\nn \\
\quad D_{\rm ovHF}^{(0)} &=& 1 + \gamma_{3} \frac{H_{\rm HF}}
{\sqrt{H_{\rm HF}^{2}}} \ , \quad H_{\rm HF} = \gamma_{3} (D_{\rm HF}-1) \ .
\label{overlap} 
\eea
$H_{\rm HF}$ is Hermitian\footnote{The constant $1$ in the formulation
of $H_{\rm HF}$ is fine since we deal with smooth gauge
configurations. For stronger gauge couplings one would prefer
to increase this constant.} and $D_{\rm ovHF}^{(0)}$ fulfils the
GWR in its simplest form, 
\be
\{ D_{\rm ovHF}^{(0)} , \gamma_{3} \} = 
D_{\rm ovHF}^{(0)} \gamma_{3} D_{\rm ovHF}^{(0)} \ .
\ee
In practice we evaluate this overlap operator by means of rational
Zolotarev polynomials, as suggested in Ref.\ \cite{Zolo},
after projecting out the lowest two modes of $D_{\rm HF}^{\dagger} 
D_{\rm HF}$, which are treated separately. 
%The polynomial approximation
%was driven to some absolute accuracy of $\varepsilon$ (see below), so 
%we deal with $D_{{\rm ovHF}, \varepsilon}$.

Compared to H.\ Neuberger's standard overlap operator $D_{\rm N}$ 
\cite{Neu}, we replace the Wilson kernel $D_{\rm W}$ 
by $D_{\rm HF}$ \cite{EPJC}. 
Since the latter is an approximate solution to the GWR already, the
transition $D_{\rm HF} \to D_{\rm ovHF}$ is only a modest
{\em chiral correction,}
\be  \label{approxi}
D_{\rm ovHF} \approx D_{\rm HF} \ ,
\ee
in contrast to the drastic transition $D_{\rm W} \to D_{\rm N}$. 
This property is illustrated in Figure \ref{specfig},
which compares the spectra of $D_{\rm HF}$ and $D_{\rm ovHF}$ for
typical gauge configurations generated at $m=0.01$ and at $m=0.24$, both
at $\beta =5$ on a $16\times 16$ lattice. The similarity of
$D_{\rm HF}$ to $D_{\rm ovHF}^{(0)}$ is useful for the computation
of the overlap operator and for its favourable properties in addition
to chirality (see below), while the similarity to $D_{\rm ovHF}(m)$
is helpful for our algorithm to be discussed in Section 3. 

Due to its perfect action background, $D_{\rm HF}$ is also endowed
with a good scaling behaviour and approximate rotation symmetry, which
are inherited by $D_{\rm ovHF}$ thanks to relation
(\ref{approxi}). That relation further provides a high level of 
locality for $D_{\rm ovHF}^{(0)}\,$, since it deviates only little
from the ultralocal operator $D_{\rm HF}\,$. All these properties have
been tested and confirmed extensively in the quenched re-weighted study 
of Ref.\ \cite{WBIH}.

\begin{figure}
\begin{center}
%\hspace*{-2mm}
  \includegraphics[angle=270,width=.62\linewidth]{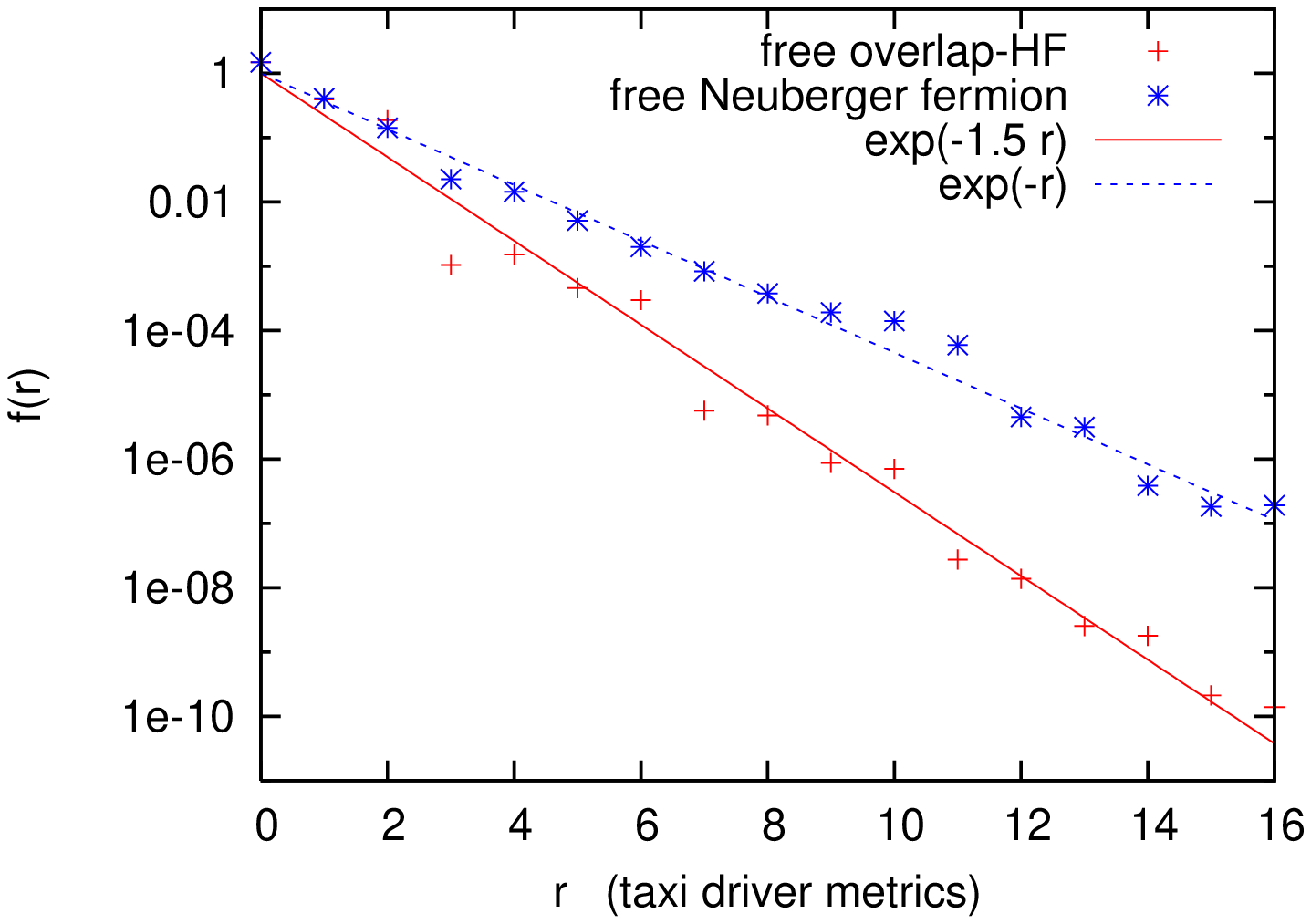} \\
 \hspace*{-4mm}
  \includegraphics[angle=270,width=.62\linewidth]{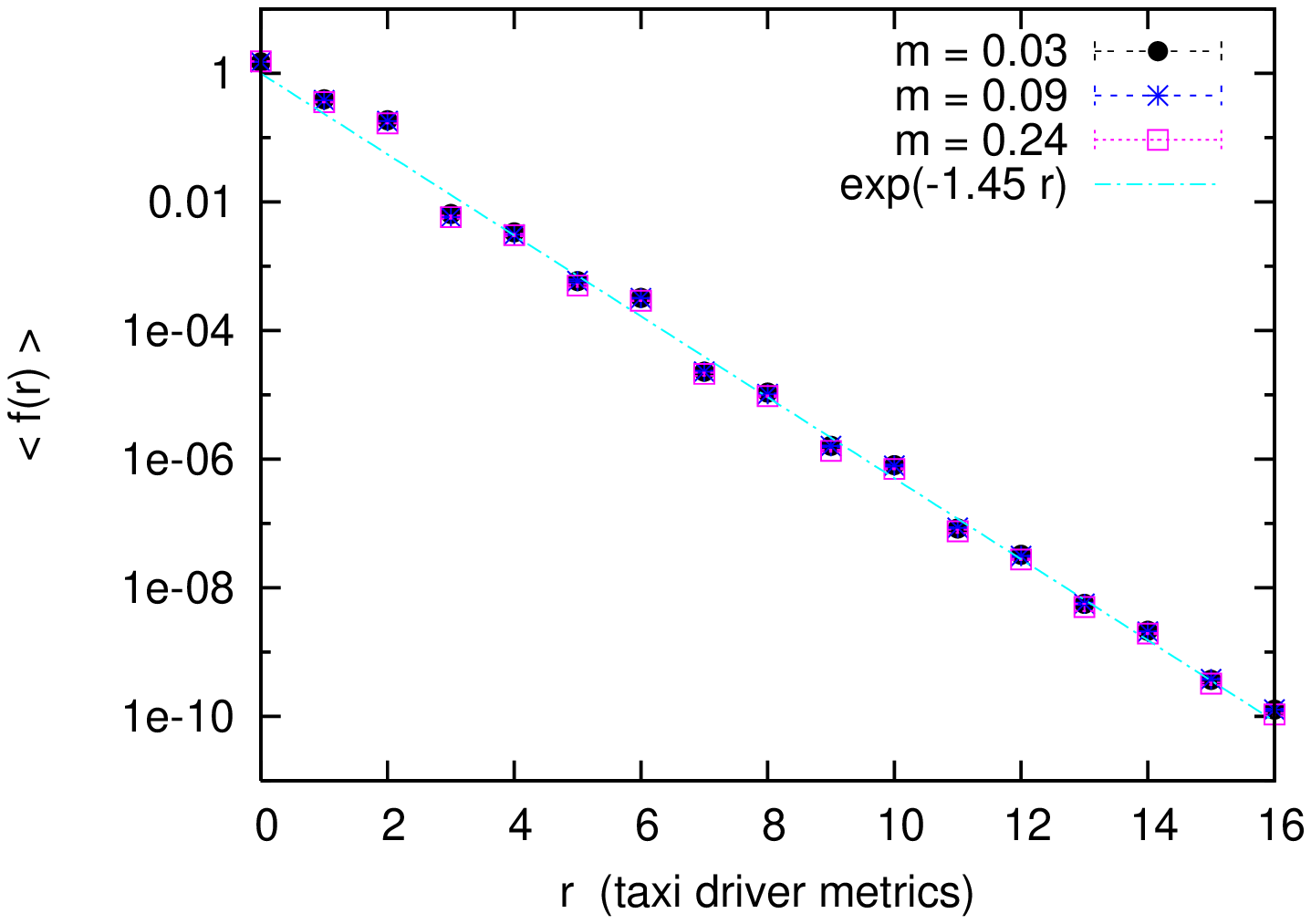}
\hspace*{-5mm}
\end{center}
\caption{\it{The locality of the overlap Dirac operators,
tested by the decay of the function (\ref{floc}),
against the taxi driver distance ${\rm r} = |r_{1}| + |r_{2}|\,$.
On top we compare our overlap-HF operator $D_{\rm ovHF}^{(0)}$ to 
Neuberger's standard overlap operator $D_{\rm N}$
(with a Wilson kernel) in the free case. At $\, {\rm r}=3$
(${\rm r} = 7$), $f({\rm r})$ is already
suppressed by more than one (two) order(s) of magnitude for
$D_{\rm ovHF}^{(0)}$. The plot below shows the 
exponential decay of $\langle f({\rm r}) \rangle$ based on our
overlap-HF simulations %with various fermion masses 
at $\beta =5$.
The gauge interaction reduces the decay rate just marginally, with
hardly any dependence on the fermion mass $m$.
%{\tt ev.\ $m=0.01$ instead of $0.03$}
}}
\label{locfig}
\end{figure}

Let us reconsider here the level of {\em locality,} which is
a key criterion in the comparison of different chiral lattice fermion
formulations. We test it in the usual way \cite{HJL}, 
by applying  $D_{{\rm ovHF}}^{(0)}$ on a unit source $\eta$
%\left( \begin{array}{c} 1 \\ 0 \end{array} \right)$ 
and measuring the decay of the function
\be  \label{floc}
f({\rm r}) = \ ^{\rm max}_{~ \, x} \ \Big\{ 
D^{(0)}_{{\rm ovHF},xy}(U) \, \eta_{y} \ \Big| \ 
%\Vert x - y \Vert_{1} 
\sum_{\mu =1}^{2} | x_{\mu} - y_{\mu} |
= {\rm r} \Big\} \ , \ \ \eta_{y} = 
\delta_{y0}\left( \begin{array}{c} 1 \\ 0 \end{array} \right) \ . \
\ee
We first consider the free fermion and demonstrate that this decay 
is clearly faster for the overlap-HF operator $D_{\rm ovHF}$
than for the Neuberger operator $D_{\rm N}$, 
see Figure \ref{locfig} on top. The plot below shows 
that the decay is still exponential for the configurations
that we generated with dynamical fermions at $\beta = 5$, 
which confirms the locality of our Dirac operator. This assures 
that our lattice fermion formulation is conceptually 
on safe grounds. In the range that we studied, the mass has practically
no influence on this decay rate.
%A previous quenched re-weighted study revealed that the overlap-HF
%operator has a much higher degree of locality than the standard
%overlap operator $D_{\rm N}$ \cite{WBIH}. 
We observe that $D_{\rm ovHF}$ has a higher degree of 
locality than $D_{\rm N}$, since $D_{\rm ovHF}$ at 
$\beta =5$ is still clearly more local than even the free $D_{\rm N}\,$: 
the decay for the free $D_{\rm ovHF}$,
$f ({\rm r}) \propto \exp(-1.5 \, {\rm r})$, is reduced just slightly 
to $\exp(-1.45 \, {\rm r})$ by the gauge interaction,
whereas $D_{\rm N}$ only decays as $\exp(- {\rm r})$ even in the
absence of gauge fields. 

This virtue also holds for the overlap-HF in quenched QCD 
\cite{QCD,BS06}: at $\beta =6$ the exponent is increased by almost
a factor of $2$ compared to $D_{\rm N}$,
and the locality of overlap-HF is manifest down to
$\beta = 5.6$. This enables the formulation 
of chiral fermions on coarser lattices than the use of $D_{\rm N}$,
which is of importance in view of QCD simulations
at finite temperature. In that case, simulations are
performed with a very small number $N_{t}$ of lattice sites in the
Euclidean time direction. Its extension is extremely expensive; the 
computational effort grows $\propto N_{t}^{12}$ \cite{Fodor}.
The application of the $D_{\rm HF}$ --- and in future also
of $D_{\rm ovHF}$ --- is therefore most %particularly 
promising in finite temperature QCD \cite{StaniEdwin}.

The scaling behaviour was found to be excellent for both, the
HF and the overlap-HF, by considering dispersion relations in
the free case and in the 2-flavour Schwinger model with 
quenched re-weighted configurations, which were generated
at $\beta =6$ \cite{WBIH}.
The HF and the overlap-HF have an even better 
scaling behaviour than the (truncated) classically perfect action, 
which was constructed and tested for the Schwinger model in 
Ref.\ \cite{Pany} (although that concept was actually designed
for asymptotically free theories \cite{HasNieO3}).
Another quenched re-weighted scaling test was added 
in Ref.\ \cite{Nils}.

\begin{figure}
\begin{center}
  \includegraphics[angle=270,width=.7\linewidth]{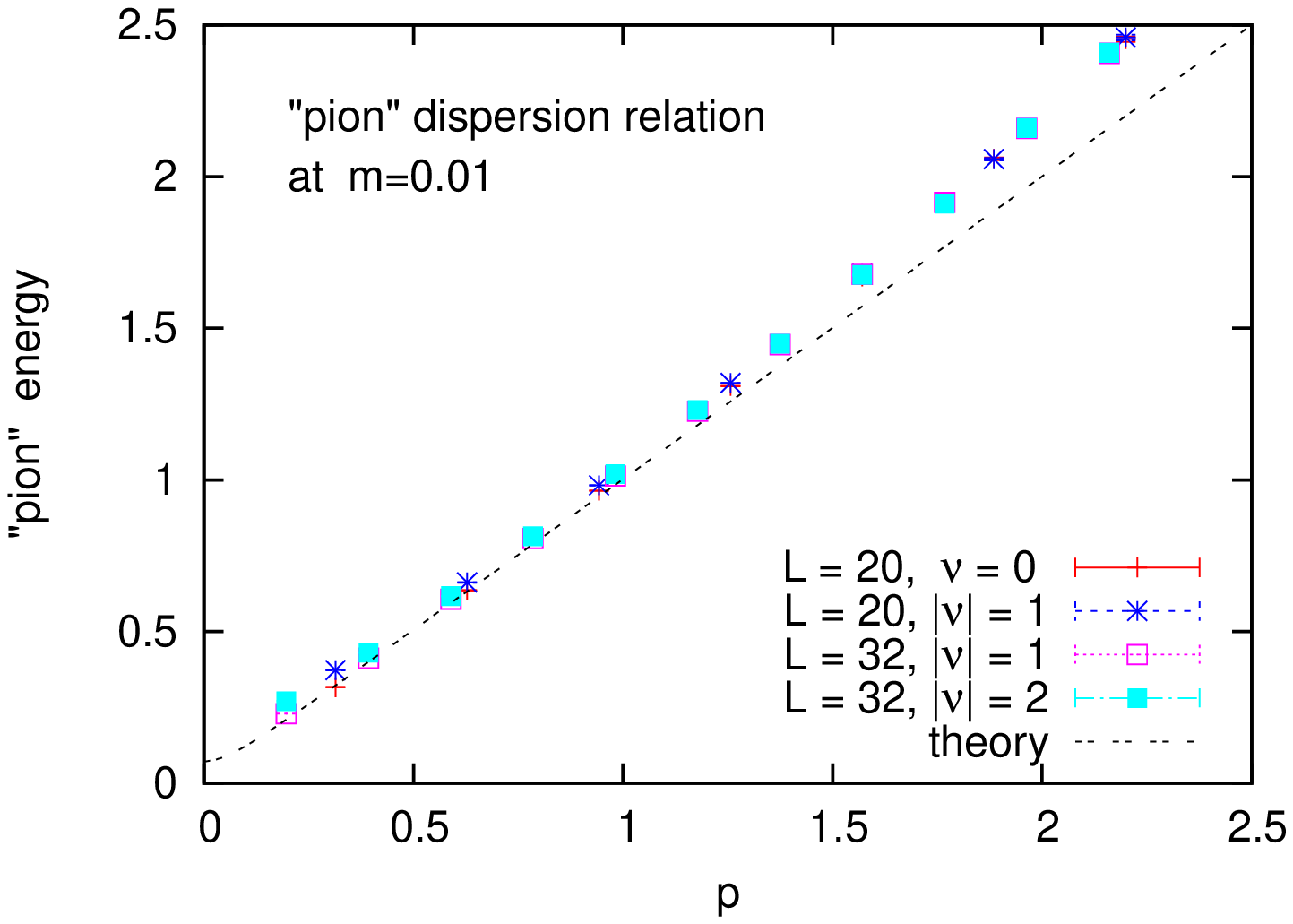}
  \includegraphics[angle=270,width=.7\linewidth]{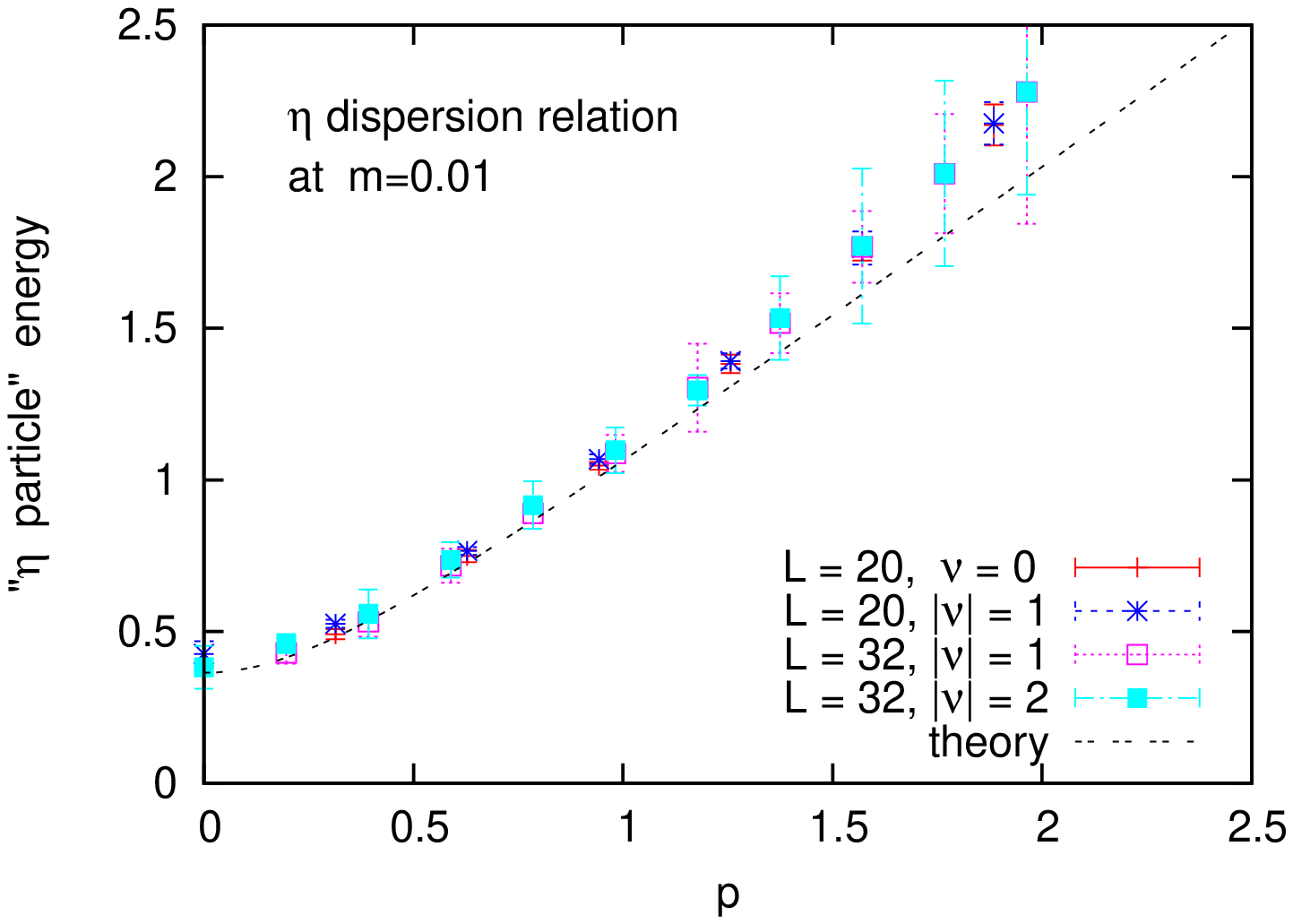}
\end{center}
\caption{\it{The dispersion relations for the ``pion'' and the 
``$\eta$-particle'' (isospin triplet and singlet), 
measured at fermion mass $m=0.01$ at $L=20$
and $L=32$, in different topological sectors. Irrespective of
the small differences, they follow in all cases very closely
the theoretical curve in the continuum, given by 
eq.\ (\ref{mpieta}). Up to momentum $p \approx \pi /2$
lattice artifacts are tiny, which confirms an excellent
scaling behaviour. This scaling quality is similar to the overlap-HF 
in our previous quenched re-weighted study at $\beta =6$, but in that
case the Wilson fermion and the Neuberger fermion show sizable
scaling artifacts setting in at $p \approx 0.9$ \cite{WBIH}.}}
\label{mesodisp}
\end{figure}

Throughout this work we fix $\beta =5$ and study the effects of
varying the lattice size and the fermion mass. So we do
not investigate explicitly the continuum extrapolation,
since the scaling artifacts due to the finite lattice spacing
turned out to be very small. This is illustrated by the dispersion 
relations of the ``meson'' masses shown in Figure \ref{mesodisp}: 
they follow the continuum behaviour up to quite large momenta,
much further than the Wilson fermion or the Neuberger %standard overlap
fermion \cite{WBIH}.
Moreover scaling artifacts are expected to be negligible
also based on the large plaquette values
near $0.9$, see Table \ref{algotab}. Hence the configurations 
are smooth, which corresponds to a fine lattice.

\begin{figure}
\begin{center}
  \includegraphics[angle=270,width=.6\linewidth]{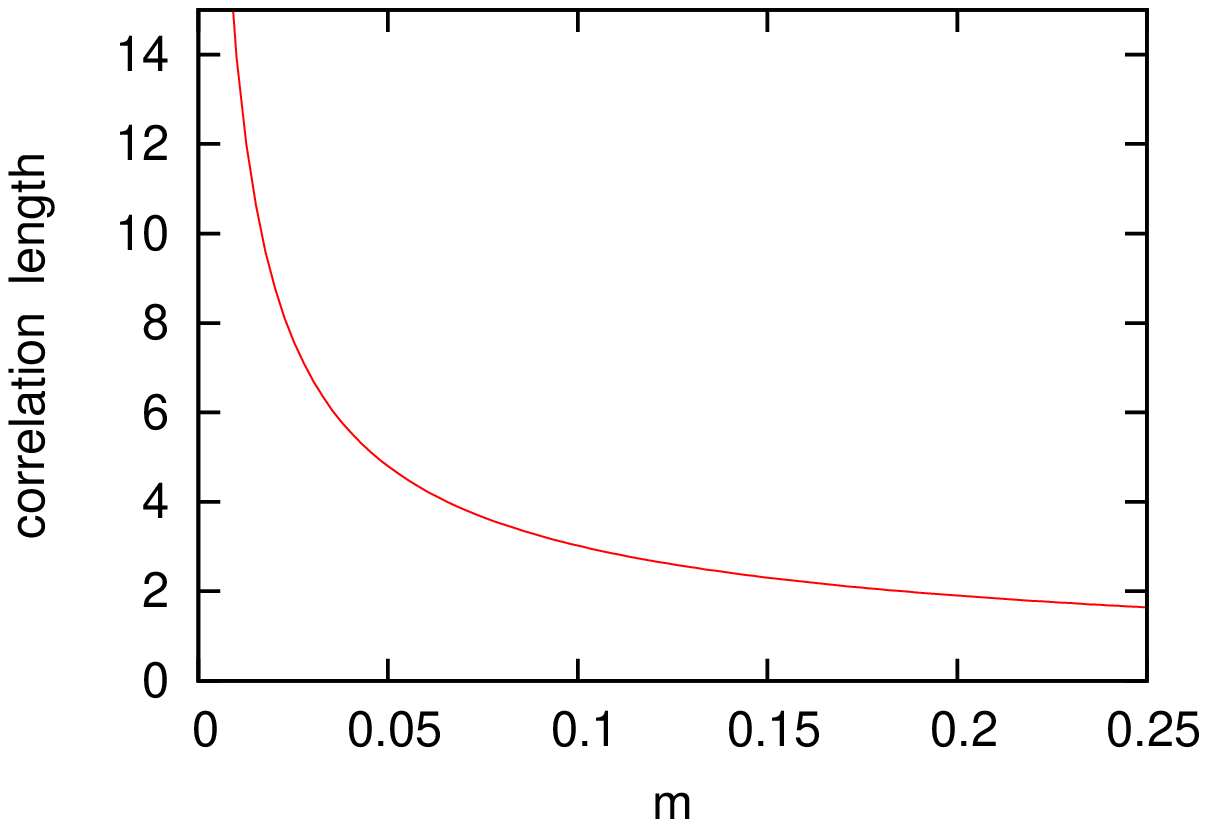}
\end{center}
\caption{\it{The correlation length $\xi = 1/ M_{\pi}$
(in infinite volume) as a function of the degenerate fermion mass
$m$, according to 
eq.\ (\ref{mpieta}). For instance for our lightest fermion mass, 
$m = 0.01$, it amounts to
%a correlation length 
$\xi \simeq 14$.}}
\label{correfig}
\end{figure}

On the other hand, the issue of finite size effects is relevant
here, and we will address it extensively in Sections 4 and 5.
Figure \ref{correfig} shows the correlation length 
$\xi = 1/M_{\pi}$ as a function of the fermion mass, as expected 
in infinite volume according to eq.\ (\ref{mpieta}).
It reveals that significant finite size effects may occur
for our smallest fermion masses and volumes.
These effects can be very useful to investigate the distinction 
between topological sectors. In QCD they have been used to
determine some of the Low Energy Constants by means of simulations 
in --- or close to --- the $\epsilon$-regime 
\cite{BJS,RMTqQCD,BS06,RMTdQCD1,RMTdQCD2,GHLW,AApap,zeromodes}
and the $\delta$-regime \cite{delta}.
In our study the finite size effects are useful since they
provide a suitable laboratory to probe methods of summing up 
observables measured separately in a few topological sectors. 
Section 5 presents pilot studies of such procedures, which might 
become relevant in lattice QCD.

\section{Hybrid Monte Carlo with a preconditioned force}

\subsection{Algorithm}

In order to simulate overlap-HFs dynamically, the standard 
Hybrid Monte Carlo (HMC) algorithm\footnote{The original work 
on the HMC algorithm is Ref.\ \cite{DKPR}; pedagogical descriptions 
can be found for instance in Refs.\ \cite{MM}.} 
would use the fermionic force term
\be
\hspace*{-1cm} \bar \psi \, Q^{-1}_{\rm ovHF} \Big( Q^{-1}_{\rm ovHF} 
\, \frac{\partial Q_{\rm ovHF}}{\partial A_{x,\mu}} + 
\frac{\partial Q_{\rm ovHF}}{\partial A_{x,\mu}} \, Q^{-1}_{\rm ovHF} 
\Big) Q^{-1}_{\rm ovHF} \, \psi \ , \label{HMCforce}
\ee
where $Q_{\rm ovHF} = \gamma_{5} D_{\rm ovHF}$ is the Hermitian overlap-HF 
operator, and $A_{x,\mu}$ are the non-compact gauge link variables.
However, this force term is computationally expensive. 
In particular in view of prospects for QCD we are going to explore a 
hopefully efficient alternative. In addition, the force (\ref{HMCforce})
is conceptually problematic due to the discontinuous sign 
function $H_{\rm HF} / \sqrt{ H_{\rm HF}^{2} } \, $ 
in $Q_{\rm ovHF}$, see eq.\ (\ref{overlap}). Under a continuous
deformation of the configurations, the denominator vanishes
at the transition between topological sectors. Here we refer to
the proper definition of the topological charge by means of the
fermionic index \cite{Has}
\be
\nu = n_{-} - n_{+} \ ,
\ee
where $n_{+}$ ($n_{-}$) is the number of zero modes of
$D_{\rm ovHF}^{(0)}$ with positive (negative) chirality.\footnote{We
fix the sign of the index such that it matches 
the continuum gauge formulation of the topological charge, 
$\int d^{2}x \, \epsilon_{12} F_{12}$. However,
in all considerations of this work only $| \nu |$ matters.}
Regarding the spectrum of $H_{\rm HF}$, a topological
transition means that an eigenvalue crosses $0$, 
so its map by the overlap formula (\ref{overlap})
flips between $m$ and $2$, {\it i.e.}\ it either appears as a zero 
mode of $D_{\rm ovHF}^{(0)}\,$, or it is sent to the cutoff scale.
Indeed the QCD simulations that have been performed with this 
HMC force hardly ever achieve topological
transitions, see in particular Refs.\ \cite{RMTdQCD2,Wupp};
this issue is discussed in detail in Ref. \cite{Cundytunnel}.
We repeat that the same problem occurs also with other lattice Dirac
operators for light dynamical fermions when the lattice becomes very 
fine \cite{MLtopo}: the HMC history tends to get 
trapped in one topological sector. % (usually the trivial one).

Here we report on HMC simulations, which  are again facilitated
by the powerful property (\ref{approxi}).
Our algorithmic concept follows the HMC version, which was applied 
to improved staggered fermions of the HF-type in Ref.\ \cite{Dilg}.
It used a sophisticated Dirac operator in the accept/reject step,
but a simplified formulation for the force, which is quick to
evaluate. In order to simulate the dynamical overlap-HF
we render the force term continuous and computationally cheap
by inserting only approximate overlap operators in the term 
(\ref{HMCforce}). To be explicit, we apply an overlap-HF to a low 
precision $\varepsilon '$ in the external factors $Q^{-1}_{\rm ovHF}$, 
and we use $H_{\rm HF}$ instead of $Q_{\rm ovHF}$ in the 
derivatives (although this could be extended to a polynomial 
as well),\footnote{The simplified force that we are using
is not only computationally cheaper, but it is also expected to be
helpful to achieve at least a few topological transitions.
Hence it might not even be favourable to extend $H_{\rm HF}$ in eq.\
(\ref{HFforce}) to a polynomial, which would move the force
formulation closer to the dangerous sign function.}
\be
\bar \psi \, Q^{-1}_{{\rm ovHF}, \varepsilon '} 
\Big( Q^{-1}_{{\rm ovHF}, \varepsilon '} \,
\frac{\partial H_{\rm HF}}{\partial A_{x,\mu}} +
\frac{\partial H_{\rm HF}}{\partial A_{x,\mu}} \, 
Q^{-1}_{{\rm ovHF}, \varepsilon '}  
\Big) Q^{-1}_{{\rm ovHF}, \varepsilon '} \, \psi \ . 
\label{HFforce}
\ee
For the integration we applied the Sexton-Weingarten scheme 
\cite{SexandWein} with a partial $(\delta \tau )^{3}$ correction
%error cancellation 
(where $\delta \tau$ is the step size). The time scales for the 
fermionic vs.\ gauge force had the ratio of $1\!:\!5$, but 
we did not observe much sensitivity to this ratio.

The Metropolis accept/reject step is performed with the high precision
overlap operator $D_{{\rm ovHF}, \varepsilon}$. Hence the deviations
in the force are corrected, and the point to worry about is just 
the acceptance rate. The complete simplification, which reduces
$Q_{{\rm ovHF}, \varepsilon '}$ to $\gamma_{5} D_{\rm HF}$,
was considered in Ref.\ \cite{Nils2}, which found a decreasing
acceptance rate for increasing volume; that study was based on the
Scaling Optimised Hypercube Fermion (SO-HF) of Ref.\ \cite{WBIH}.
However, in this respect it turns out to be highly profitable
--- and still cheap --- to correct the external factors
to a modest precision. We chose the algorithmic parameters for the 
(absolute) precisions as
\be  \label{epsprimeps}
\varepsilon ' = 0.005 \quad {\rm (force~term)} \ , \quad
\varepsilon = 10^{-16} \quad {\rm (Metropolis~step)} \ ,
\ee
which increases the acceptance rate by an order of magnitude
compared to the simple use of $H_{\rm HF}$ throughout the force 
term. The force we obtain in this way is not based on
Hamiltonian dynamics, but the way it deviates from it (by 
proceeding from $\gamma_{5} D_{\rm HF}$ to $Q_{{\rm ovHF}, 
\varepsilon '}$) does maintain the property
of area conservation in phase space.

\subsection{Statistics}

With this algorithm, we performed production runs at $\beta =5$
on $L \times L$ lattices of the sizes $L=16,\ 20,\ 24,\ 28$ and $32$.
On the $16 \times 16$ lattice we collected statistics at 
seven fermion masses in the range $m=0.01 \dots 0.24$
in distinct topological sectors,
as displayed in Table \ref{stat-tabL16}.
At our lightest mass, $m=0.01$, we performed additional
simulations on larger lattices of size $L=20 \dots 32$,
plus runs at $m=0.06$, $L=32$, see Table \ref{stat-tabL20-32}.

\begin{table}
\centering
\begin{tabular}{|c||c|c|r||c||c|}
\hline
$m$ & \multicolumn{4}{|c||}{number of configurations} & topological \\
    & $\nu =0$   & $|\nu |= 1$ & $|\nu |= 2$ & total  & transitions \\
\hline
\hline
0.01 & 2428  &  307 &     & 2735 &  7 \\  
\hline
0.03 & 1070  &  508 &     & 1578 &  2 \\
\hline
0.06 &  741  &  660 &     & 1401 &  7 \\ 
\hline
0.09 &  919  &  587 &   1 & 1507 &  7 \\
\hline
0.12 &  664  &  501 & 248 & 1413 &  8 \\
\hline
0.18 &  791  &  563 &  50 & 1404 & 15 \\
\hline
0.24 &  576  &  978 &  56 & 1637 & 17 \\
\hline
\end{tabular}
\caption{\it Our statistics of %thermalised
configurations at $L=16$ and seven fermion masses $m$. 
The HMC trajectory lengths $\ell$
%was in general $\ell = 1/8$ for $m=0.03 \dots 0.18$, and $\ell = 1/16$
%for $m=0.24$ (cf.\ Table \ref{algotab}), in all cases
are given in Table \ref{algotab}. In all cases they consist of
20 integration steps ($\delta \tau = \ell /20$).
The measurements are separated by at least 200 trajectories. 
For $m=0.01$ this separation was enlarged to 600 trajectories
for a better decorrelation.
As a generic property of dynamical overlap fermion simulations,
topological transitions are rare, as we see in particular for
our lightest fermion masses.}
\label{stat-tabL16}
\end{table}

\begin{table}
\centering
\begin{tabular}{|c|c||r|r|r|r||r||}
\hline
$L$ & $m$ & \multicolumn{5}{|c||}{number of configurations} \\
    &     & $\nu =0$   & $|\nu |= 1$ &
$|\nu |= 2$ & $|\nu |= 3$ & total \\
\hline
\hline
20 & 0.01 & 435 & 304 &     &     & 739 \\
\hline
24 & 0.01 &     &     & 278 & 273 & 551  \\
\hline
28 & 0.01 &     & 240 &     & 180 & 420  \\
\hline
32 & 0.01 & 138 &  98 &  82 &     & 318 \\
\hline
\hline
32 & 0.06 &  91 & 293 &     &     & 384 \\
\hline
\end{tabular}
\caption{\it Our statistics for the lattice sizes $L=20 \dots 32$
at fermion masses $m=0.01$ and $0.06$.
The trajectory length $\ell$ was reduced
for increasing $L$, see Table \ref{algotab},
while the integration step was always fixed as 
$\delta \tau = \ell /20$. After thermalisation, the
configurations are separated by at least 200 trajectories.
We show our statistics in the topological sectors
$|\nu | = 0 \dots 3$. In particular at $m=0.01$
topological transitions were very rare, so we captured 
various sectors by a multitude of ``hot starts''.}
\label{stat-tabL20-32}
\end{table}

Since the force term (\ref{HFforce}) tends to push the trajectory a bit 
off the hyper-surface of constant energy, we kept the trajectory length 
$\ell$ (between the Metropolis steps) short. At $L=16$, $m \leq 0.18$,
we chose $\ell = 1/8$, which is divided into 20 integration 
steps ($\delta \tau = 0.00625$). This was a compromise in view 
of precision and dynamics between the trajectory 
end-points. On the larger lattices the trajectory length $\ell$ 
was further reduced, see Table \ref{algotab}.\footnote{At 
a few points in the HMC histories, 
where the algorithm run into convergence problems,
we temporarily reduced $\ell$ below the trajectory length given 
in Table \ref{algotab}, always maintaining the dissection into
$\delta \tau = \ell /20$.}

The configurations used for the measurements 
were separated by at least 200 trajectories.
Still the autocorrelation with respect to the observables in 
Sections 4 and 5 is not always negligible, see Table \ref{autotab}.
In particular some problems show up for $L>20$ and higher
topological charges, which suggests that an application of this
algorithmic approach to QCD might require further refinements.
Here autocorrelations were taken into account by a jackknife 
analysis of the measured data. That method was used for the
error calculations throughout this work; we probed a variety of
bin sizes and took the maximal error in each case.

\subsection{Reversibility, acceptance rate and computational effort}

{\em Reversibility} --- to a good precision ---
is a requirement of the HMC algorithm. 
The possible danger for this crucial property could be an instability
in the molecular dynamics trajectory due to directions with
positive Lyapunov exponents, such that certain deviations from the
exact trajectory are amplified exponentially \cite{Lyapu}.

\begin{figure}[h!]
%\vspace*{-3mm}
\begin{center}
\includegraphics[angle=0,width=.6\linewidth]{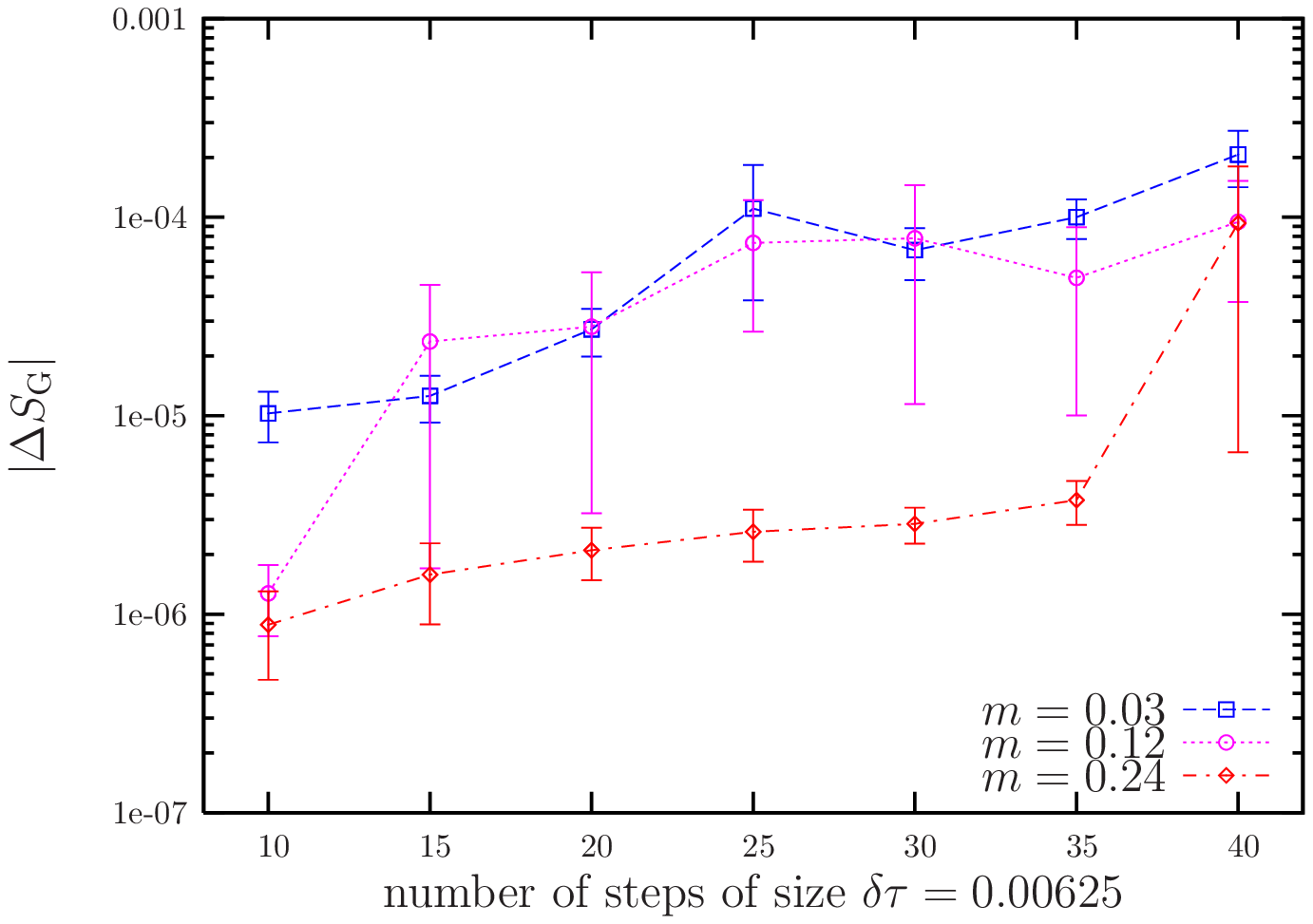}
\hspace*{-2mm} 
\includegraphics[angle=0,width=.6\linewidth]{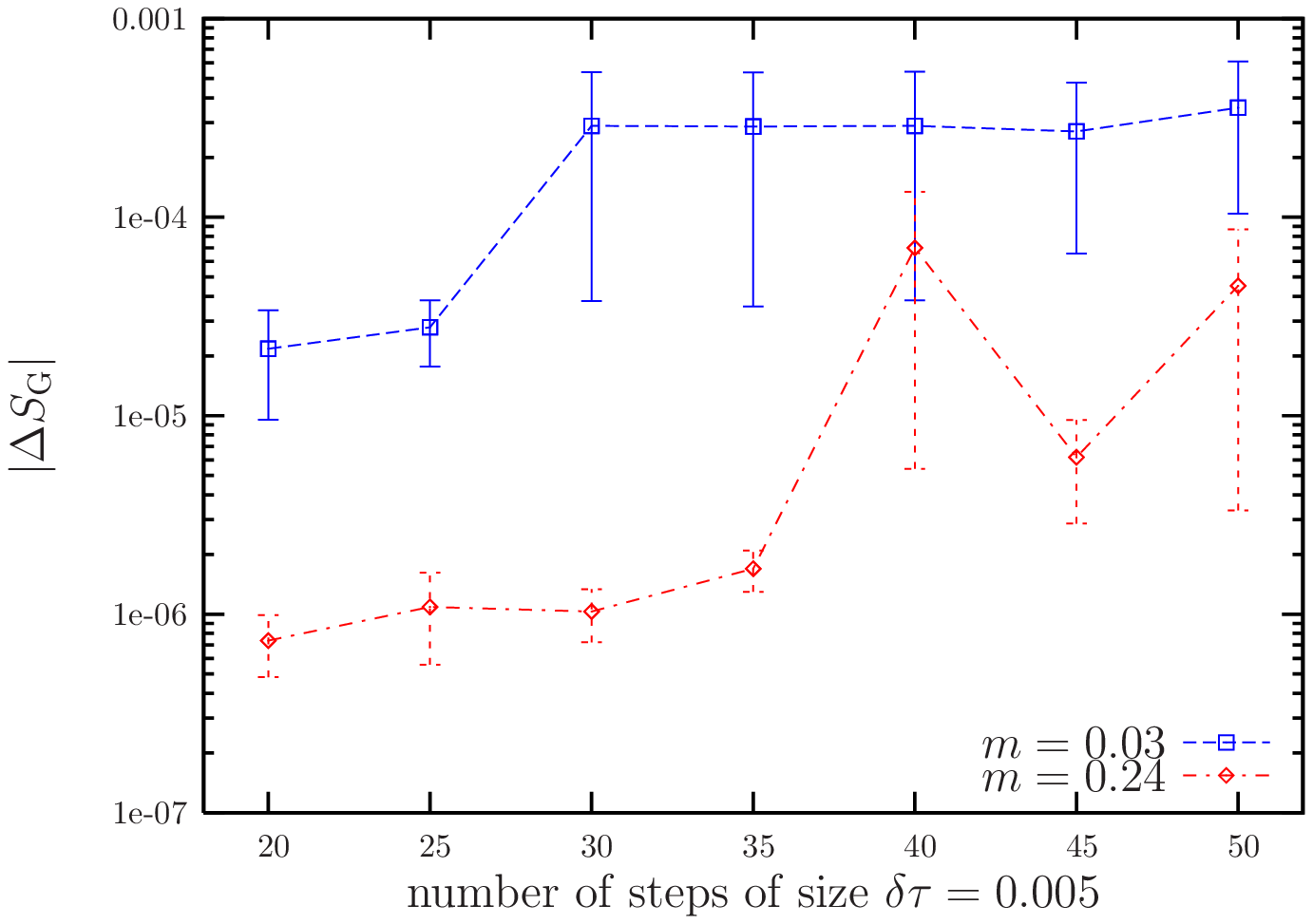}
\end{center}
\caption{\it{The reversibility precision with respect to the gauge 
action $S_{G}$ for a variable number of steps of length 
$\delta \tau =0.00625$ (on top) and $\delta \tau =0.005$ (below), 
both on a $16 \times 16$ lattice at $\beta =5$. 
We show results for very different masses.
There is no indication for any positive 
Lyapunov exponent. The precision improves as we
increase $m$, but it is satisfactory in all cases.}}
\label{revfig}
\end{figure}

To test if we are confronted with this problem,
we moved forth and back with a variable number of steps, 
and measured the (absolute) modification of the gauge action, 
$| \Delta S_{G} |$. As examples,
Figure \ref{revfig} (on top) shows our results 
for the reversibility precision at $L=16$,
$\delta  \tau = 0.00625$, for the masses $m=0.03$, $0.12$ and $0.24$.
The precision is satisfactory in all cases.
It still improves significantly for increasing mass, %$m = 0.24$,
as we also observe for $\delta \tau = 0.005$, see Figure 
\ref{revfig} (below). Our results do not indicate the 
presence of any positive Lyapunov exponent.

\begin{figure}[h!]
\hspace*{-2mm} 
\includegraphics[angle=270,width=.53\linewidth]{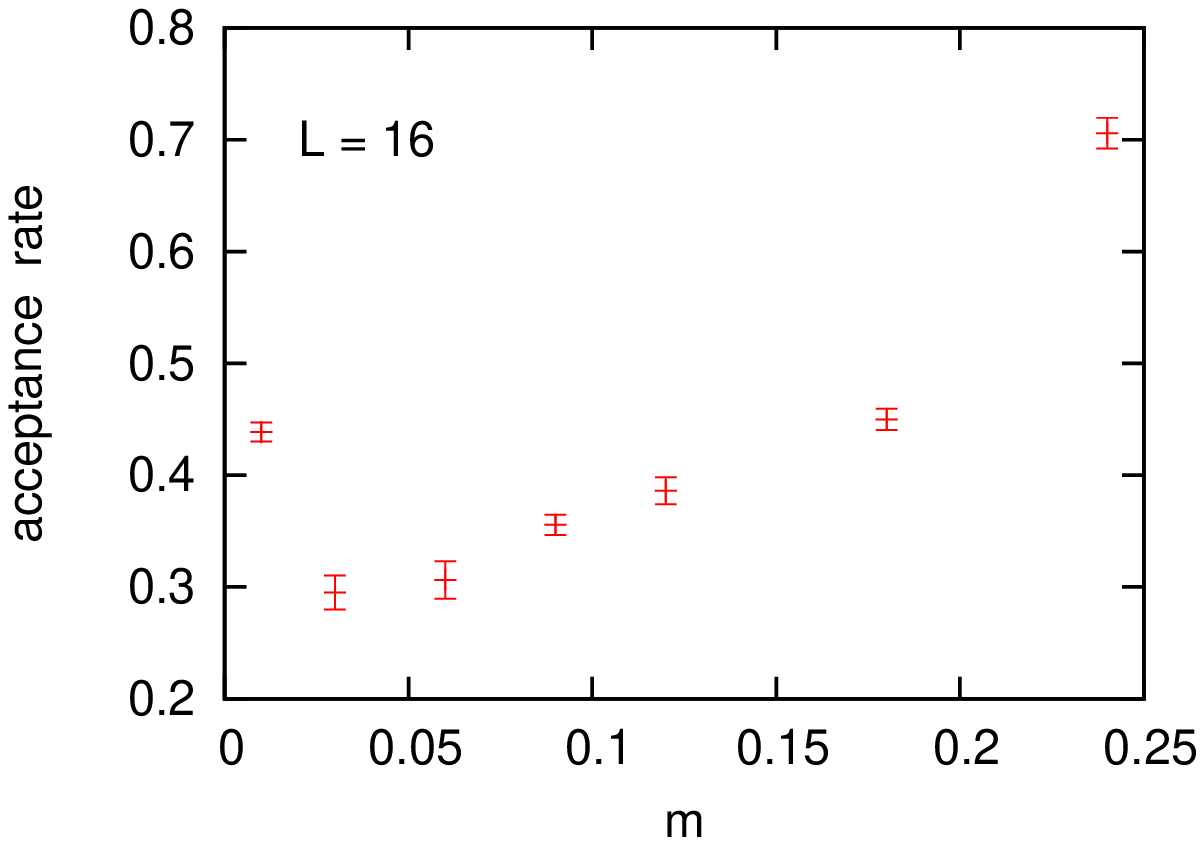}
\hspace*{-4mm}  
\includegraphics[angle=270,width=.53\linewidth]{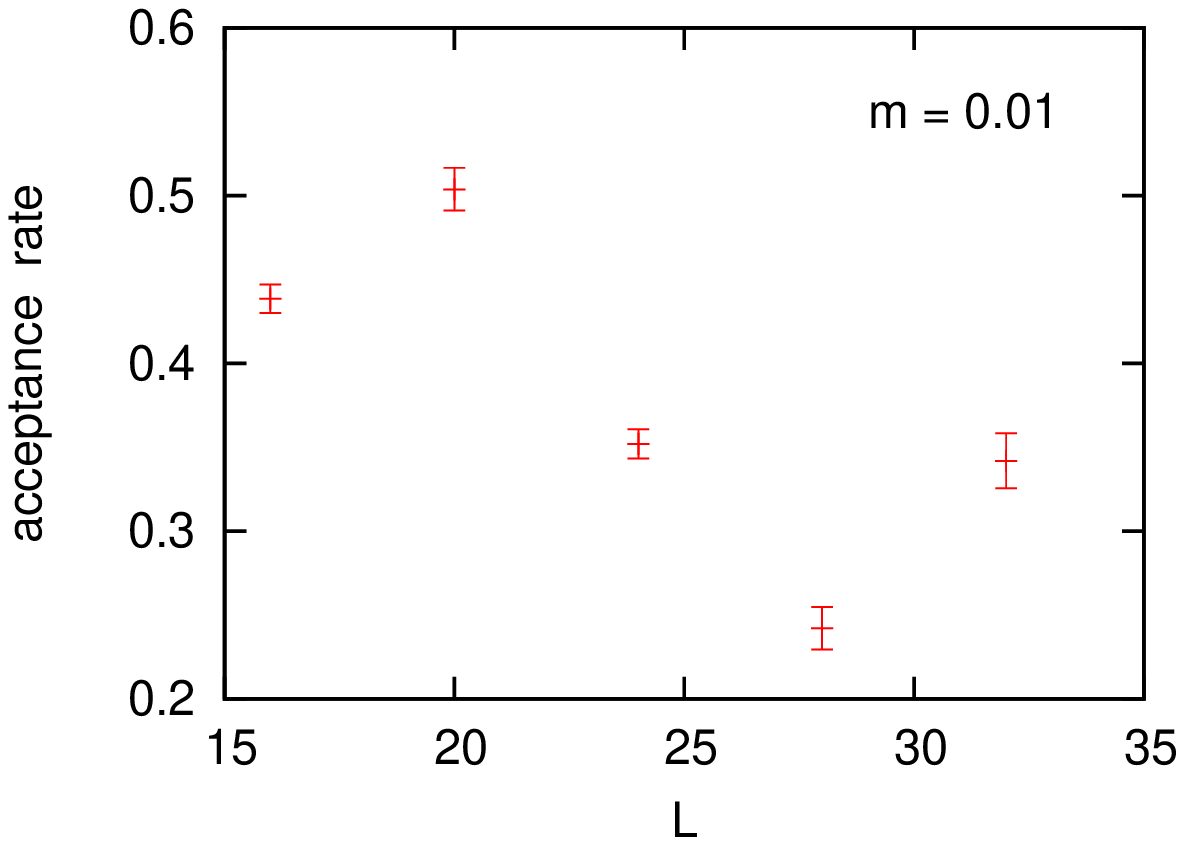}
 \caption{\it{The Metropolis acceptance rate on the $L=16$ lattice 
at seven different masses (on the left), and at $m=0.01$ on five
lattice sizes (on the right). Note that the trajectory length
varies, as specified in Table \ref{algotab}.}}
\label{algofig}
\end{figure}

Being confident that our algorithm is safe, we now
proceed to the question of its efficiency.
The plots in Figure \ref{algofig} show the {\em acceptance
rates} in the Metropolis step.\footnote{We evaluated the 
acceptance rate by considering
the acceptance probability in each Metropolis step, regardless of the 
actual accept/reject decision. This is statistically
more conclusive than just counting the acceptance ratio.}
In some sectors they were somewhat modest for the parameters 
chosen here, which is related to the aforementioned cases of 
rather long autocorrelation times.
%For the trajectory lengths chosen, we found these rates to
%be useful in all simulations.

\begin{figure}[h!]
\hspace*{-2mm} 
\includegraphics[angle=270,width=.53\linewidth]{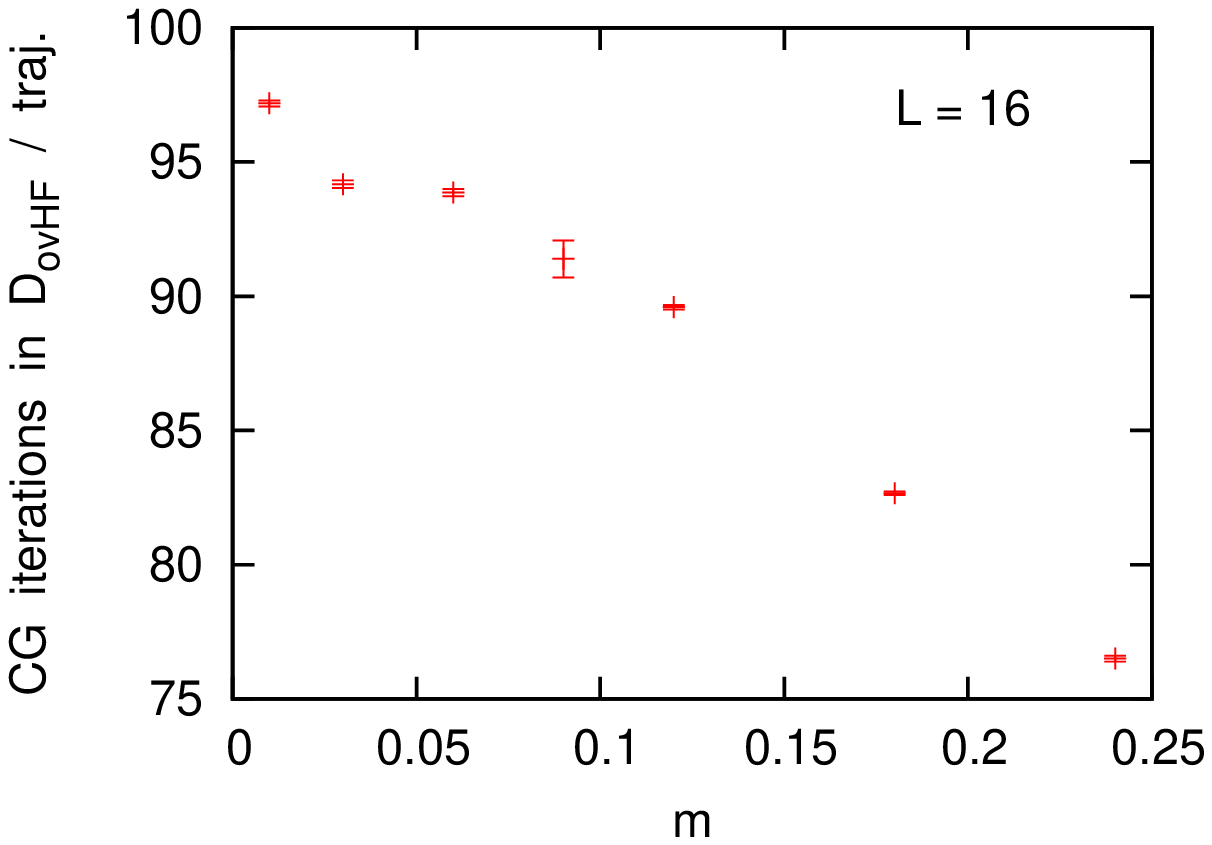}
\hspace*{-4mm}  
\includegraphics[angle=270,width=.53\linewidth]{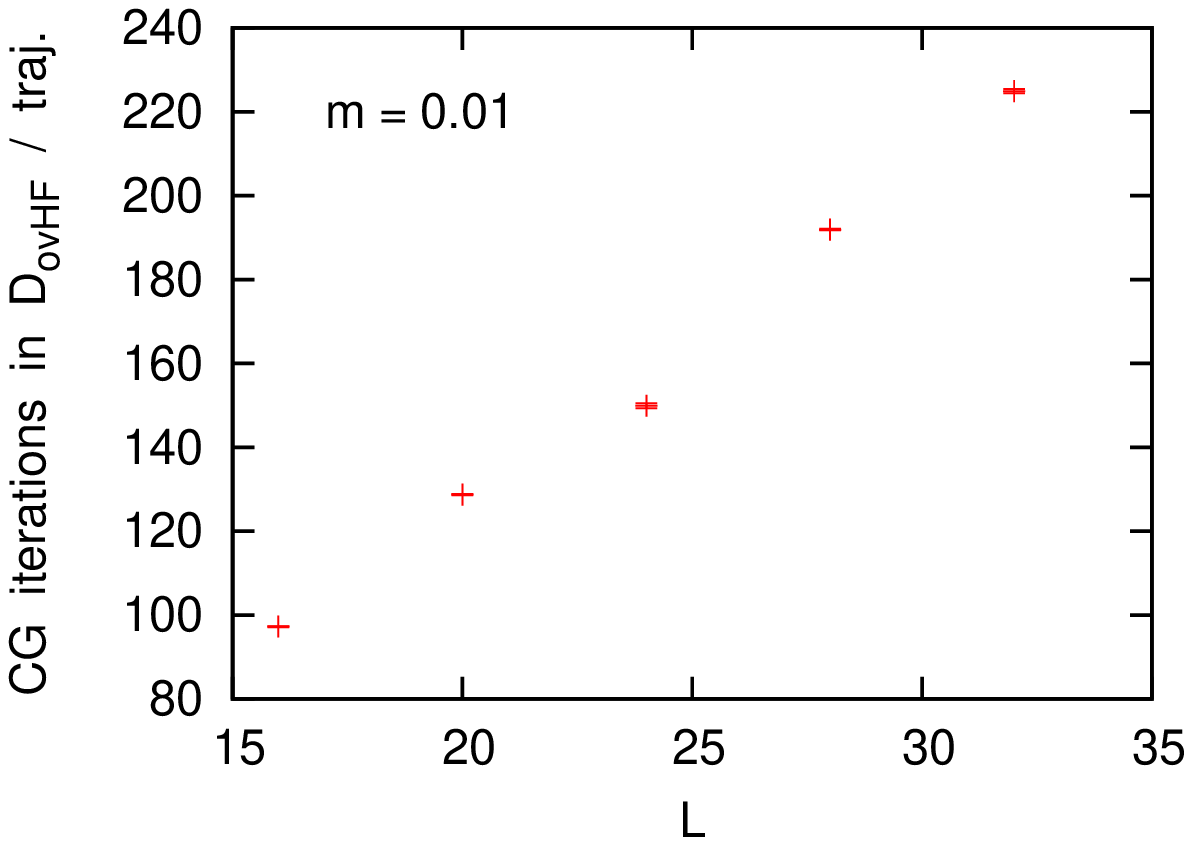} \\
\hspace*{-2mm}
\includegraphics[angle=270,width=.54\linewidth]{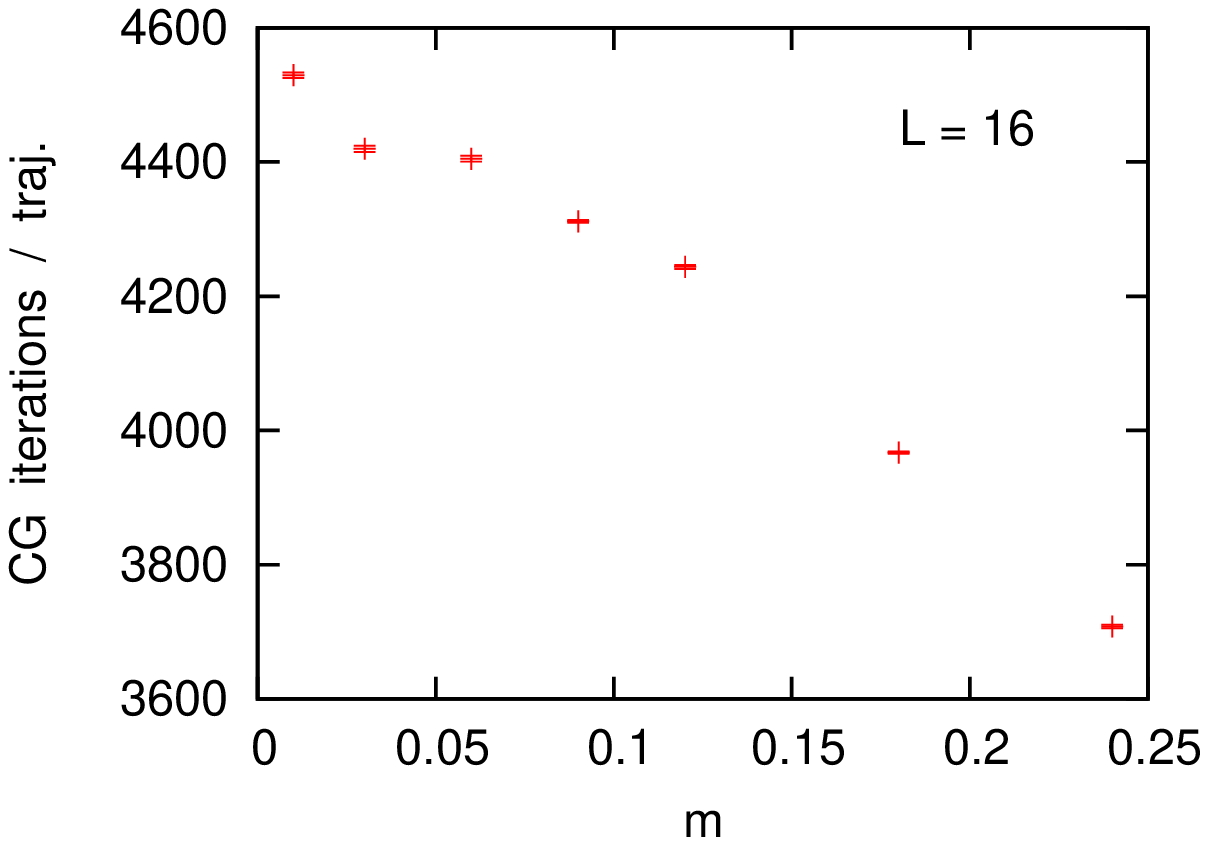}
\hspace*{-4mm}  
\includegraphics[angle=270,width=.54\linewidth]{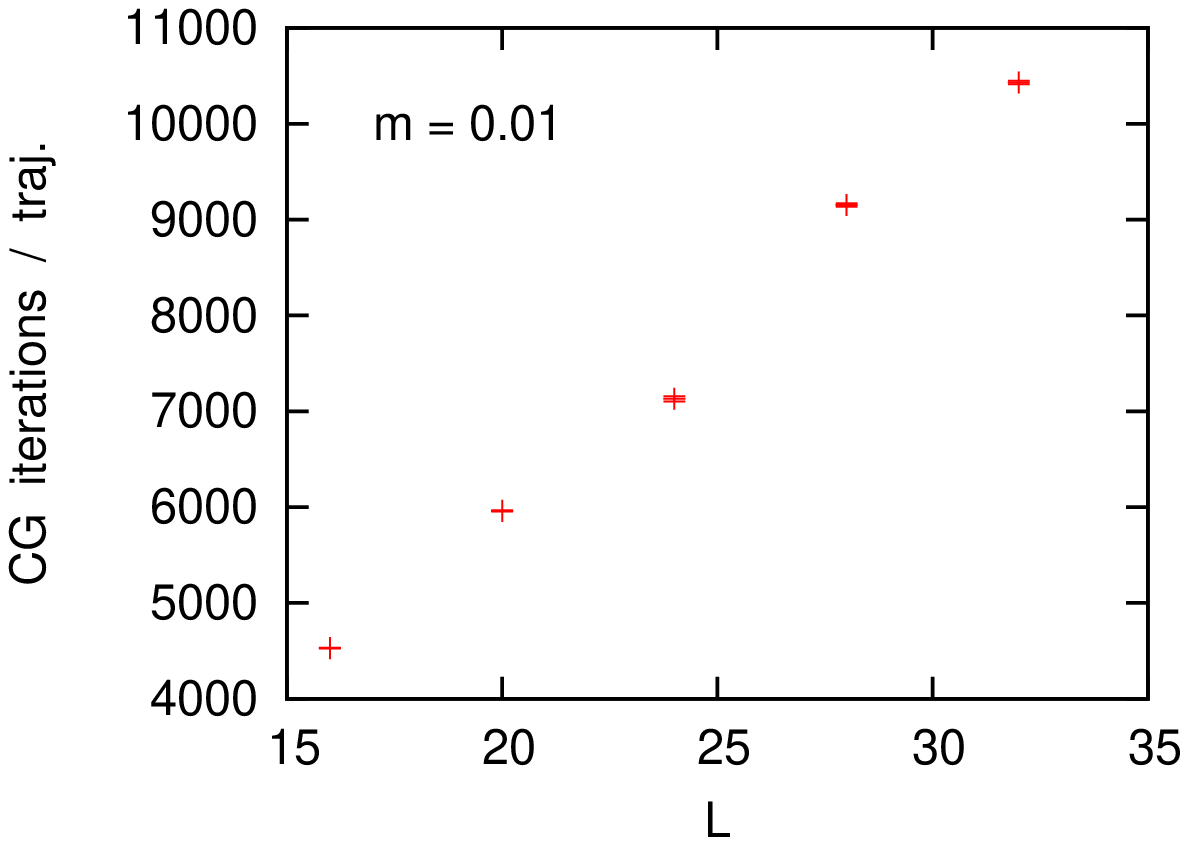}
%\end{center}
\caption{\it{The number of Conjugate Gradient iterations
per trajectory in $D_{\rm ovHF}$ (upper plots), and
including all operations (lower plots). We show results at
$L=16$ and various masses (plots on the left), 
and at $m=0.01$ on various lattice sizes (plots on the right).}}
\label{CGfig}
\end{figure}

Figure \ref{CGfig} displays the number of {\em Conjugate Gradient 
iterations} that was required per trajectory,
specifically in the evaluation of $D_{\rm ovHF}$ (upper plots)
and in total (lower plots).
Table \ref{algotab} summarises the acceptance rates, as well
as the number of Conjugate Gradient steps in the evaluation
of the overlap operator and in total. We add the plaquette
value to characterise the smoothness of the configurations;
this serves as a point of orientation for comparison with
other models in lattice gauge theory.

\begin{table}
\centering
\begin{tabular}{|c|c|l||l||r|r||c|}
\hline
$L$ & $m $ & ~~~$\ell $ 
& acceptance & \multicolumn{2}{|c||}{number of CG iterations}
& plaquette \\
    &      & & rate      &  in $D_{\rm ovHF}$ & total & value \\
\hline
\hline
16 & 0.01 & 0.125 & 0.439(8) &  97.2(1) & 4529(4) & 0.8971(2) \\
\hline
16 & 0.03 & 0.125 & 0.295(15) & 94.2(1) & 4419(5) & 0.8965(4) \\
\hline
16 & 0.06 & 0.125 & 0.306(17) & 93.9(1) & 4405(4) & 0.8974(4) \\
\hline
16 & 0.09 & 0.125 & 0.355(9) &  91.4(1) & 4312(2) & 0.8963(2) \\
\hline
16 & 0.12 & 0.125 & 0.386(12) & 89.6(1) & 4244(3) & 0.8961(3) \\
\hline
16 & 0.18 & 0.125 & 0.450(9) & 82.7(1) & 3966(2) & 0.8965(2)  \\
\hline
16 & 0.24 & 0.0625 & 0.706(13) & 76.5(1) & 3708(3) & 0.8947(4)  \\
\hline
\hline
20 & 0.01 & 0.0625 & 0.504(13) & 128.7(2) & 5961(6) & 0.8972(3)  \\
\hline
24 & 0.01 & 0.05 & 0.352(9)  & 149.9(6) & 7129(27) & 0.8969(5) \\
\hline
28 & 0.01 & 0.04 & 0.242(13) & 191.9(2) & 9152(17) & 0.8972(5) \\ 
\hline
32 & 0.01 & 0.03 & 0.342(16) & 224.9(5) & 10432(18) & 0.8968(5) \\
\hline
\hline
32 & 0.06 & 0.03 & 0.619(10) & 199.0(2) & 9546(8)   & 0.8971(3) \\
\hline
\end{tabular}
\caption{\it The characteristic indicators for the
performance of the preconditioned HMC algorithm:
first we give the acceptance rate in the Metropolis step
at the end of each trajectory of length $\ell$. The accept/reject
step uses $D_{\rm ovHF}$ to machine precision 
(see eq.\ (\ref{epsprimeps})). We add the number of Conjugate 
Gradient iterations needed to evaluate the operator $D_{\rm ovHF}$ 
and in total. Finally we quantify the smoothness of the
configurations by the mean plaquette value.}
\label{algotab}
\end{table}

\begin{table}
\centering
%\begin{tabular}{|c|c||c|c|c|c|c|c|}
%\hline
%$L$ & $m $ & \multicolumn{4}{c|}{$\langle ~{\rm plaquette}~ \rangle$} 
%& %Dirac eigenvalue 
%$\lambda_{1}$ & %chiral condensate 
%$\Sigma$ \\
%\hline
%\multicolumn{2}{|c||}{$|\nu |$} & 0 & 1 & 2 & 3 & & \\
%\hline
%\hline
%16 & 0.01 & $0.6$ & $0.7$ & & & & \\
%\hline
%16 & 0.03 & $0.6$ & $0.5$ & & & & \\
%\hline
%16 & 0.06 & $0.6$ & $0.7$ & & & & \\
%\hline
%16 & 0.09 & $<0.5$ & $<0.5$ & & & & \\
%\hline
%16 & 0.12 & $<0.5$ & $0.5$ & & & & \\
%\hline
%16 & 0.18 & $<0.5$ & $0.5$ & $0.5$ & & & \\
%\hline
%16 & 0.24 & $<0.5$ & $<0.5$ &  $<0.5$ & & & \\
%\hline
%\hline
%20 & 0.01 & & & & & & \\
%\hline
%24 & 0.01 & & & & & & \\
%\hline
%28 & 0.01 & & & & & & \\
%\hline
%32 & 0.01 & & & & & & \\
%\hline
%\hline
%32 & 0.06 & & & & & & \\
%\hline
\begin{tabular}{|c|c|c||c|c|c|}
\hline
$L$ & $m $ & $|\nu |$ & plaquette 
& Dirac eigenvalue $\lambda_{1}$ & chiral condensate $\Sigma$ \\
\hline
\hline
16 & 0.01 & 0 & $0.6$ & $0.7$ & $1.1$ \\
\hline
16 & 0.01 & 1 & $0.7$ & $0.5$ & $0.8$ \\
\hline
16 & 0.03 & 0 & $0.6$ & $1.2$ & $1.2$ \\
\hline
16 & 0.03 & 1 & $0.5$ & $1.0$ & $1.1$ \\
\hline
16 & 0.06 & 0 & $0.6$ & $1.2$ & $1.2$ \\
\hline
16 & 0.06 & 1 & $0.7$ & $0.9$ & $0.9$ \\
\hline
16 & 0.09 & 0 & $<0.5$ & $1.2$ & $1.2$ \\
\hline
16 & 0.09 & 1 & $<0.5$ & $0.9$ & $1.0$ \\
\hline
16 & 0.12 & 0 & $<0.5$ & $1.5$ & $1.3$ \\
\hline
16 & 0.12 & 1 & $0.5$ & $0.9$ & $0.9$ \\
\hline
16 & 0.12 & 2 & $0.5$ & $0.8$ & $0.9$ \\
\hline
16 & 0.18 & 0 & $<0.5$ & $1.1$ & $0.9$ \\
\hline
16 & 0.18 & 1 & $0.6$ & $0.7$ & $0.7$ \\
\hline
16 & 0.18 & 2 & $<0.5$ & $0.6$ & $0.6$ \\
\hline
16 & 0.24 & 0 & $<0.5$ & $1.1$ & $1.1$ \\
\hline
16 & 0.24 & 1 & $<0.5$ & $0.9$ & $0.7$ \\
\hline
16 & 0.24 & 2 & $<0.5$ & $0.5$ & $<0.5$ \\
\hline
\hline
20 & 0.01 & 0 & $<0.5$ & $1.3$ & $0.9$ \\
\hline
20 & 0.01 & 1 & $<0.5$ & $1.2$ & $1.1$ \\
\hline
24 & 0.01 & 2 & $<0.5$ & $2.5$ & $2.5$ \\
\hline
24 & 0.01 & 3 & $0.6$ & $2.8$ & $2.4$ \\
\hline
28 & 0.01 & 1 & $0.5$ & $2.7$ & $1.7$ \\
\hline
28 & 0.01 & 3 & $0.5$ & $4.5$ & $5.0$ \\
\hline
32 & 0.01 & 0 & $<0.5$ & $1.3$ & $1.2$ \\
\hline
32 & 0.01 & 1 & $<0.5$ & $1.8$ & $1.6$ \\
\hline
32 & 0.01 & 2 & $1.3$ &  $1.7$ & $1.9$ \\
\hline
\hline
32 & 0.06 & 0 & $<0.5$ & $0.9$ & $0.8$ \\
\hline
32 & 0.06 & 1 & $<0.5$ & $1.9$ & $2.7$ \\
\hline
\end{tabular}
\caption{\it The integrated autocorrelation times 
$\tau_{\rm int}= \frac{1}{2} + \sum_{\tau \geq 1} C(\tau )$
(where $C(\tau )$ is the autocorrelation function) over
a total trajectory length 25, for the mean plaquette value,
the leading non-zero Dirac eigenvalue $\lambda_{1}$
(relevant in Section 4), and the chiral condensate 
$\Sigma$ (relevant in Section 5).}
\label{autotab}
\end{table}

As usual, the simulation becomes faster when we proceed from
light to moderate fermion mass. However, the number of Conjugate 
Gradient iterations rises only mildly as we approach the chiral 
limit, even down to very light masses.
%and even down to very light masses we obtained useful
%acceptance rates (on the $L=16$ lattice it even rises again
%as we move down to $m=0.01$).
The reason is that finite size effects
prevent the non-zero eigenvalues from becoming really tiny
(this virtue is obviously reduced if the volume increases).
The low-lying non-zero eigenvalues will be 
discussed in detail in the next section.

\section{The Dirac spectrum}

In this section we discuss our results for the eigenvalues
of the Dirac operator $D_{\rm ovHF}^{(0)}$ in eq.\ (\ref{overlap}), 
after stereographic projection (a M\"{o}bius transform) onto the 
half-line $\R_{+}$,
\be  \label{stereo}
\lambda_{i} \to \left| \frac{\lambda_{i}}{1 - \lambda_{i}/2} \right| \ .
\ee

\subsection{Unfolded level spacing distribution}

We first consider the full spectra and the resulting
unfolded level spacing distribution. This distribution is obtained
as follows: one first numerates the eigenvalues of single
configurations in ascending order, $\lambda_{i}$ (here we omit the 
eigenvalues with negative imaginary parts before the mapping
(\ref{stereo})).
%on $\R_{+}$). 
Next we put all eigenvalues 
in a set of configurations together and numerate them again.
Now we consider pairs of eigenvalues $\lambda_{i}, \ \lambda_{i+1}$
(of the same configuration), and count by how many ordinal numbers they
differ in the overall order. This difference --- divided by the
number of configurations involved --- is the {\em unfolded level
spacing.}

Random Matrix Theory (RMT) predicts three shapes
for the statistical distribution of these level spacings.
They refer to the three possible patterns of spontaneous
chiral flavour symmetry breaking (for $N_{f}$ flavours),
\bea
SU(2N_{f}) \to SO(2N_{f}) && {\rm orthogonal} \nn \\
SU(N_{f}) \otimes SU(N_{f}) \to SU(N_{f}) && {\rm unitary} \nn \\
SU(2N_{f}) \to Sp(2N_{f}) && {\rm symplectic} \ .
\eea 
At least in 4d Yang-Mills theory with chiral fermions,
this set of patterns is complete \cite{chiSB}.\footnote{An
overview of the conceivable types of chiral symmetry breaking
with an isomorphic representation by non-unitary groups is
given in Ref.\ \cite{chiSBnonu}.}
They correspond to the real, complex and pseudo-real fermion
representation of the gauge group. (In the real and pseudo-real
case, fermion and anti-fermion representations are equivalent,
hence the unbroken symmetry is enlarged to $SU(2N_{f})$.)
Ref.\ \cite{HalVer} elaborated the corresponding formulae for the
unfolded level spacing distributions. For lattice QCD 
(with chiral quarks) the prediction of the unitary ensemble has 
been confirmed \cite{EHKN,BJS}, but the case of the
2-flavour Schwinger model is theoretically less clear, because
there is no spontaneous breaking of the chiral flavour symmetry.
%it does not match the usual RMT assumption $\Sigma (m=0) \neq 0$.
\begin{figure}
\begin{center}
\includegraphics[angle=270,width=1.\linewidth]{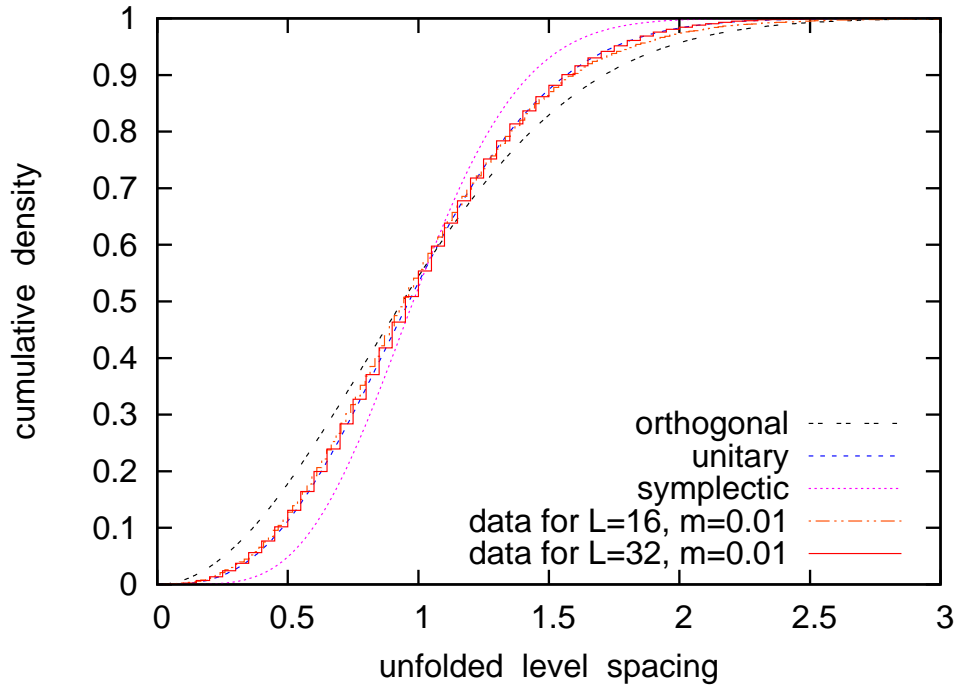}
\end{center}
\caption{\it{The cumulative density of the unfolded level spacing 
distribution. We show the curves corresponding to the RMT prediction
for the orthogonal, the unitary and the symplectic ensemble, and 
confront them with our simulation data (for our lightest
fermion mass, $m=0.01$). We clearly observe agreement with the
RMT formula for the unitary ensemble. For $L=16$ there is still
a slight deviation for level spacings $\gsim 1.5$. 
As we increase the size to $L=32$, even this deviation practically
disappears, {\it i.e.}\ the agreement becomes very precise.}}
\label{unfold}
\end{figure}

Our results (for $m=0.01$, as an example) are shown in Figure 
\ref{unfold}. We see very clear agreement with the RMT formula for 
the unitary ensemble. The tiny deviation that we observe for $L=16$
is a finite size effect, which is manifestly suppressed
as we enlarge the lattice size to $L=32$.

\subsection{The microscopic Dirac spectrum}

We now focus on the leading non-zero eigenvalue $\lambda_{1}$,
based on the statistics presented in Tables \ref{stat-tabL16}
and \ref{stat-tabL20-32}.
%of the Dirac operator 
%$D_{\rm ovHF}^{(0)}$ in eq.\ (\ref{overlap}), 
%after stereographic projection onto a line,
%\be  \label{stereo}
%\lambda_{1} \to \left| \frac{\lambda_{1}}{1 - \lambda_{1}/2} \right| \ .
%\ee
The results for $\langle \lambda_{1}\rangle$ are given in Table 
\ref{lamstat}.
We mentioned before that chiral RMT has been worked out for 
the case of a finite condensate $\Sigma (m \to 0)$ \cite{RMT1,RMT2},
with successful applications in the $\epsilon$-regime of QCD.
This is not the situation we are dealing with; here $\Sigma$ 
vanishes in the chiral limit, as eq.\ (\ref{Sigmam}) shows.
The situation is qualitatively similar for fermions interacting
through Yang-Mills gauge fields above the critical temperature 
of the chiral phase transition. Also there the understanding
of the Dirac spectra is controversial; for numerical studies
and conjectures we refer to Refs.\ \cite{airyref,YMhighT}.
\begin{table}
\centering
\begin{tabular}{|c|c||c|c|c|c|}
\hline
$L$ & $m$ & $\langle \lambda_{1, \, \nu =0} \rangle$ &  
$\langle \lambda_{1, \, |\nu | =1} \rangle$ &
$\langle \lambda_{1, \, |\nu | =2} \rangle$ & 
$\langle \lambda_{1, \, |\nu | =3} \rangle$ \\
\hline
\hline
16 & 0.01 & 0.1328(6)  & 0.175(2) &  &  \\  
\hline
16 & 0.03 & 0.130(2)~ & 0.173(2) &  &  \\
\hline
16 & 0.06 & 0.125(2)~ & 0.173(1) &  &  \\ 
\hline
16 & 0.09 & 0.115(2)~ & 0.172(2) &  &  \\
\hline
16 & 0.12 & 0.108(2)~ & 0.166(2) & 0.216(3) &  \\
\hline
16 & 0.18 & 0.108(2)~ & 0.166(2) & 0.221(4) &  \\
\hline
16 & 0.24 & 0.109(3)~ & 0.163(2) & 0.215(4) &  \\
\hline
\hline
20 & 0.01 & 0.102(2) & 0.127(2) &  &  \\ 
\hline
24 & 0.01 &  &  & 0.125(4) & 0.148(6) \\ 
\hline
28 & 0.01 &  & 0.082(3) &  & 0.120(5) \\ 
\hline
32 & 0.01 & 0.057(3) & 0.076(3) & 0.084(3) &  \\ 
\hline
\hline
32 & 0.06 & 0.030(3) & 0.054(3) &  &  \\ 
\hline
\end{tabular}
\caption{\it The lowest non-zero eigenvalue of $D_{\rm ovHF}^{(0)}$ 
(after the stereographic projection (\ref{stereo})) 
for different masses and lattice sizes, in distinct topological sectors.}
\label{lamstat}
\end{table}

In infinite volume, $V \to \infty$, 
the chiral condensate is given by the Dirac spectrum as
\be  \label{sigspec}
\Sigma = \int d \lambda \ \frac{\rho (\lambda )}{\lambda +m}
\qquad (\rho ~ : ~ {\rm eigenvalue~density}) \ .
\ee
Along with the prediction quoted in Section 1, 
$\Sigma \propto m^{1/3}$, this suggests \cite{Hell}
\be  \label{rholam3}
\rho ( \lambda \gsim 0 ) \propto \lambda^{1/3} \ , 
\ee
in contrast to the Banks-Casher plateau \cite{BC} that one obtains in the
standard setting (with $\Sigma (m \to 0) \neq 0$). In that case,
the density for the re-scaled small eigenvalues $\lambda_{i}\Sigma V$
is scale-invariant (at fixed $m \Sigma V$) \cite{LeuSmi}.
In our case, the generic relation 
$\langle \lambda_{i} \rangle \propto [ V \rho ( \lambda \gsim 0 )]^{-1}$
suggests that the parameter
\be
\zeta_{i} = \lambda_{i} V^{3/4} W_{\zeta} 
\qquad ({\rm for~small~} \lambda_{i}\ ; \ V = L^2)
\ee
should adopt this r\^{o}le, at fixed 
$\mu_{\zeta} = m V^{3/4}  W_{\zeta}$ --- or simply at small $m$.
$W_{\zeta}$ is a constant of dimension [mass]$^{1/2}$,
which is (in this context) analogous to $\Sigma$ in the standard 
setting.

\begin{figure}[h!]
\hspace*{-6mm} 
\includegraphics[angle=270,width=.54\linewidth]{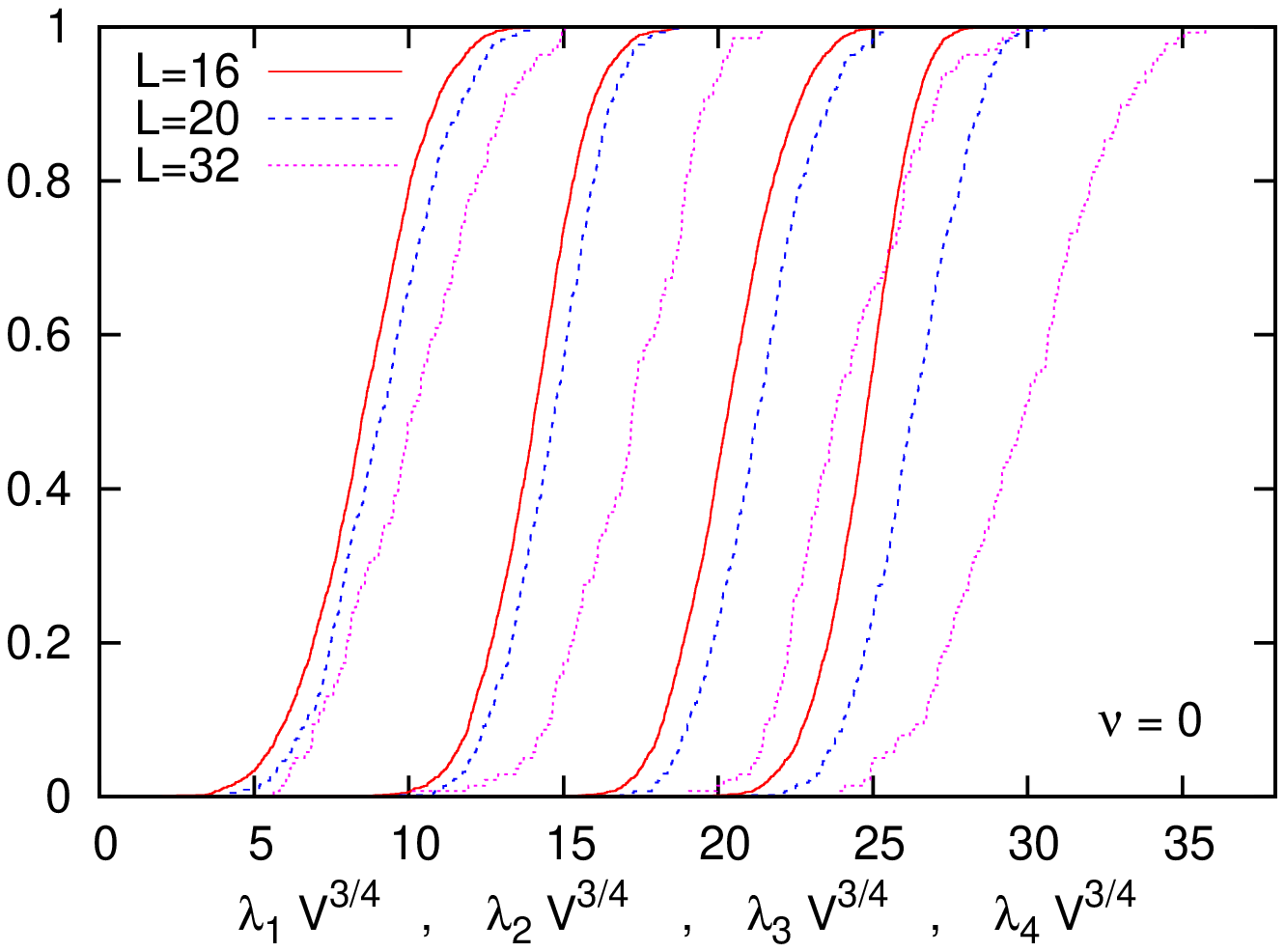}
\hspace*{-5mm}
\includegraphics[angle=270,width=.54\linewidth]{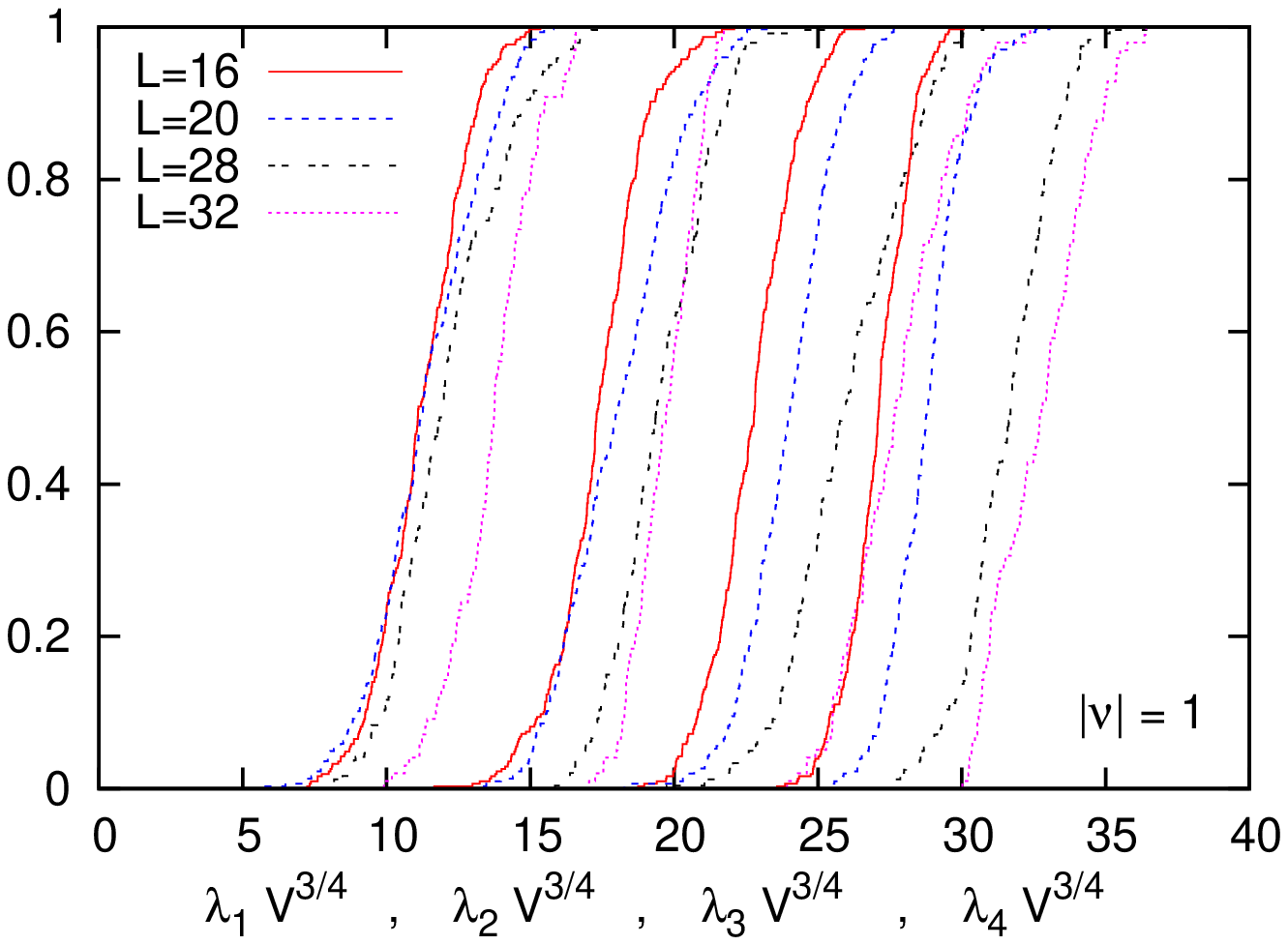}
\caption{\it{Cumulative densities of $\lambda_{i}
V^{3/4} \propto \zeta_{i}$, for $i=1 \dots 4$, at mass $m=0.01$ and 
topological charge $\nu =0$ (on the left) resp.\ $|\nu | = 1$ (on the 
right). We see that $\zeta_{i}$ deviates from scale-invariance.}}
\label{lam1rs.75}
\end{figure}
Hence we probed the corresponding finite-size scaling, but it
is {\em not} confirmed. This is illustrated in Figure 
\ref{lam1rs.75} for our lightest fermion mass, $m=0.01$, 
in the sectors of topological charge $\nu =0$ and $|\nu |= 1$.
As a quantitative measure, the integrated variance is given
--- and compared to better approaches --- in Table \ref{intvar}.
Note, however, that the derivation of relation (\ref{rholam3})
may be invalidated by inserting an explicitly mass-dependent
spectral density, $\rho (\lambda, m)$, in eq.\ (\ref{sigspec}).\\

Next we consider another scenario, with a reduced 
exponent of $V$ in the re-scaling factor. 
Now we assume the scale-invariant variable
to be $Z_{i} = \lambda_{i} V^{2/3} W_{Z}$ ($W_{Z}$ of dimension
[mass]$^{1/3}$). This scenario is motivated by the fact that
it belongs to a theoretically well
explored universality class: it corresponds to 
$\rho ( \lambda \gsim 0) \propto \lambda^{1/2}$, which is the spectral
density obtained by RMT in the {\em Gaussian approximation}. There
is a precise prediction for the corresponding spectral density in 
terms of Airy functions \cite{airyref},
\be \label{airyfun}
\rho_{\rm Airy} (Z) \propto Z [ {\rm Ai}(-Z)]^{2} + [ {\rm Ai}'(-Z)]^{2} 
\quad \ (\sim \sqrt{Z} /\pi \ {\rm at~}Z \gg 1) \ .
\ee
%The behaviour of $Z_{1}$ is shown in Figure \ref{lam1Z} (plots above); 
%as in Figure \ref{lam1rs.75} we fix $m=0.01$ and consider 
%$|\nu | = 0$ and $1$. The finite size scaling quality is 
%better than the one of $\zeta_{1}$, but still not really convincing.
%The finite size scaling quality of $Z_{1}$ is also captured in 
%Table \ref{intvar}. It is significantly better
%than the one of $\zeta_{1}$, but still not fully satisfactory.
\begin{figure}[h!]
%\hspace*{-6mm}
%\includegraphics[angle=270,width=.54\linewidth]{cumdense_m.01nu0rescal.67.eps}
%\hspace*{-5mm}
%\includegraphics[angle=270,width=.54\linewidth]{cumdense_m.01nu1rescal.67.eps}
%\vspace*{-11mm} \\
\begin{center} 
\includegraphics[angle=270,width=.6\linewidth]{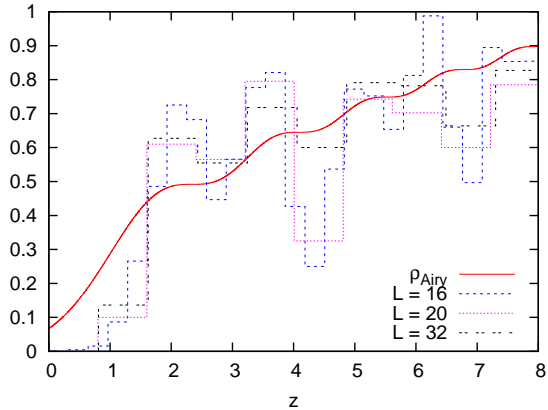}
\vspace*{-4mm}
\end{center}
\caption{\it{%Finite size scaling for $Z_{1}
%\propto \lambda_{1} V^{2/3}$, at $m=0.01$ and
%$\nu = 0$ (left), or $|\nu |= 1$ (right). Regarding scale-invariance,
%$Z_{1}$ performs better than $\zeta_{1}$, but still not fully
%satisfactory.}}
%\newline 
%Below: 
Eigenvalue histograms for $Z_{1} \propto \lambda_{1} V^{2/3}$,
at $m=0.01$, $\nu =0$ compared to the
spectral density $\rho_{\rm Airy}$ in eq.\ (\ref{airyfun}),
which RMT predicts in the Gaussian approximation.
There is no convincing support for this scenario.}}
%None of these three plots supports this scenario convincingly.}}
\label{lam1Z}
\vspace*{-4mm}
\end{figure}

\begin{table}
\vspace*{1mm}
\centering
\begin{tabular}{|c||c|c|c||c|c|c|}
\hline
index $|\nu |$ & \multicolumn{3}{|c||}{0} & \multicolumn{3}{|c|}{1} \\
\hline
power $p$ & $3/4$ & $2/3$ & $5/8$ & $3/4$ & $2/3$ & $5/8$ \\
\hline
\hline
${\rm Var_{int}}$ & 1.812 & 0.266 & 0.026 & 2.078 & 0.258 & 0.066 \\
\hline
${\rm Var_{int}^{(n)}}$ & 0.159 & 0.036 & 0.007 & 0.204 & 0.055 & 0.019 \\
\hline
\end{tabular}
\caption{\it A measure for the agreement between the cumulative
densities of the re-scaled eigenvalues $s_{i} = \lambda_{i} V^{p}$, at 
$m=0.01$ and $ |\nu | = 0, \ 1$. We show %the integrated variance
%\protect\newline
${\rm Var_{int}} = \sum_{i=1}^{4} \int ds_{i} \, 
[\sum_{k=1}^{\rm max} (\rho_{L_{k}}(s_{i}) - \rho_{\rm m}(s_{i}))^{2}] 
/ ({\rm max}-1)$,
the integrated variance,
where the sum over $k$ includes $L_{k}= 16, \, 20, \, 32$ at $\nu =0$
(${\rm max}=3$), and 
$L_{k}= 16, \, 20, \, 28, \, 32$ at $| \nu |=1$ (${\rm max}=4$).
$\rho_{\rm m}(s_{i})$ is the mean value over the volumes involved.
The quantity ${\rm Var_{int}^{(n)}}$ is normalised by dividing
through the relevant interval in $s$, where 
$0.01 < \rho_{\rm m}(s) < 0.99$. 
We confirm that the power $p=5/8$ yields by far the least
variance, {\it i.e.}\ the best agreement.}
\label{intvar}
\vspace*{-2mm}
\end{table}
Figure \ref{lam1Z} compares the Airy function density (\ref{airyfun})
to the histograms that we obtained in various volumes at $m=0.01$ and 
$\nu =0$. Our data exhibit a far more marked wiggle structure, hence 
the agreement is not really convincing.

The finite size scaling quality of $Z_{1}$ is also captured in 
Table \ref{intvar}. It is significantly better than the one of 
$\zeta_{1}$, but still not fully satisfactory.\\

\begin{figure}[h!]
\hspace*{-6mm}
\includegraphics[angle=270,width=.54\linewidth]{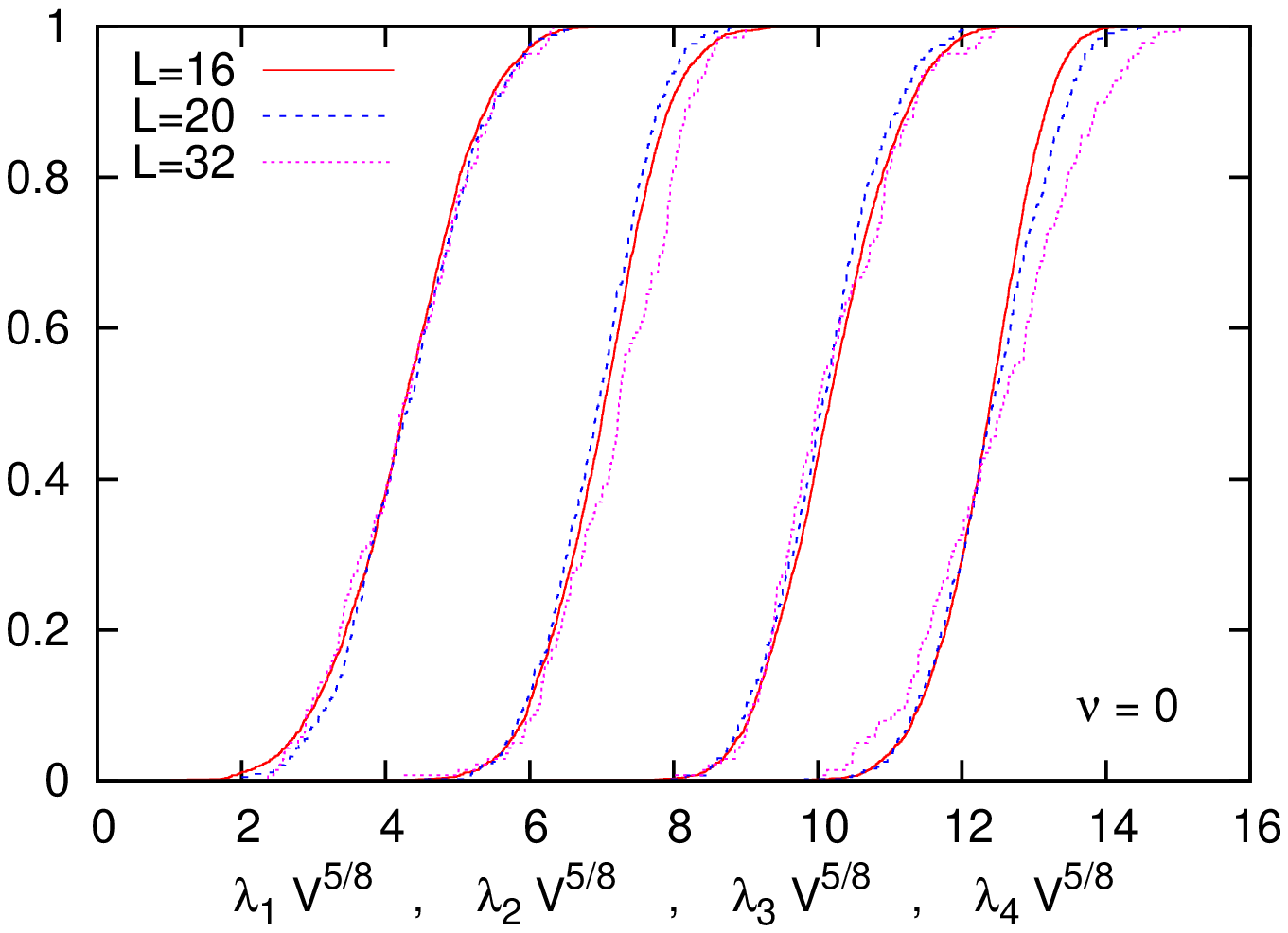}
\hspace*{-5mm}
\includegraphics[angle=270,width=.54\linewidth]{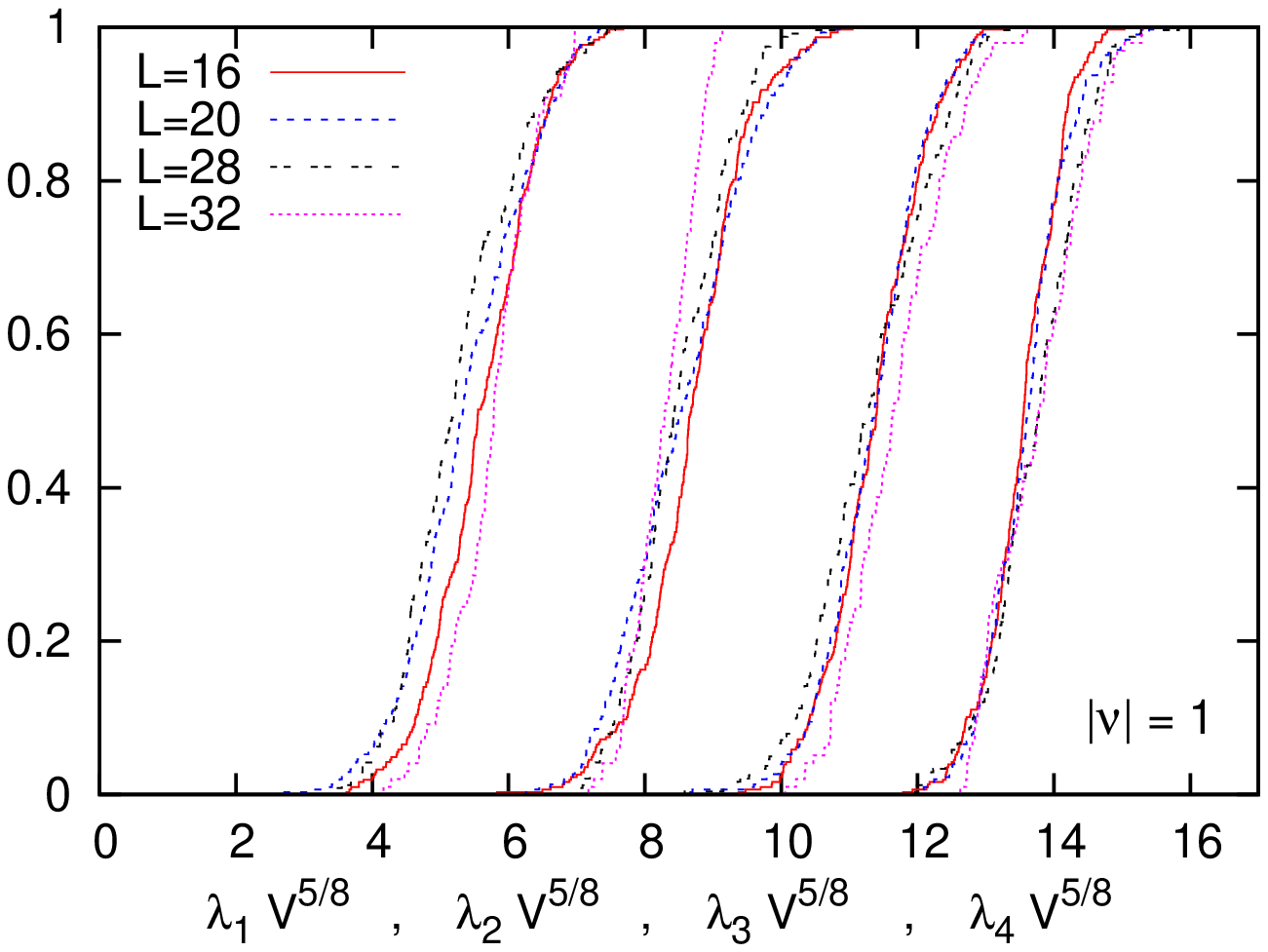}
\caption{\it{Cumulative densities of $\lambda_{i} V^{5/8} \propto z_{i}$, 
for $i=1 \dots 4$, at mass $m=0.01$ and topological charge 
$\nu =0$ (on the left) resp.\ $|\nu | = 1$ (on the right). 
In contrast to $\zeta_{i}$ and $Z_{i}$, the %properly re-scaled 
variable $z_{i}$ is scale-invariant to an impressive precision.}}
\label{lam1rs.625}
\end{figure}

Let us finally proceed to the most successful approach, 
which was identified empirically.
It turns out that our data are in excellent
agreement with a scale-invariant parameter
\be
z_{i} = \lambda_{i} V^{5/8} W_{z} \qquad
(W_{z} ~ : ~ {\rm constant~of~dimension~[mass]^{1/4}}) \ ,
\ee
which implies a microscopic spectral density 
$\rho (\lambda ) \propto \lambda^{3/5}$. 
The scale invariance of $z_{1} \dots z_{4}$ is
illustrated in Figure \ref{lam1rs.625}, again
for our lightest fermion mass, $m=0.01$, in the sectors of 
topological charge $\nu =0$ and $|\nu |= 1$, which can be compared
directly to Figures \ref{lam1rs.75}. A quantitative confrontation
with the previous two ans\"{a}tze is included in Table \ref{intvar}.

\begin{figure}[h!]
\hspace*{-6mm}
\includegraphics[angle=270,width=.54\linewidth]{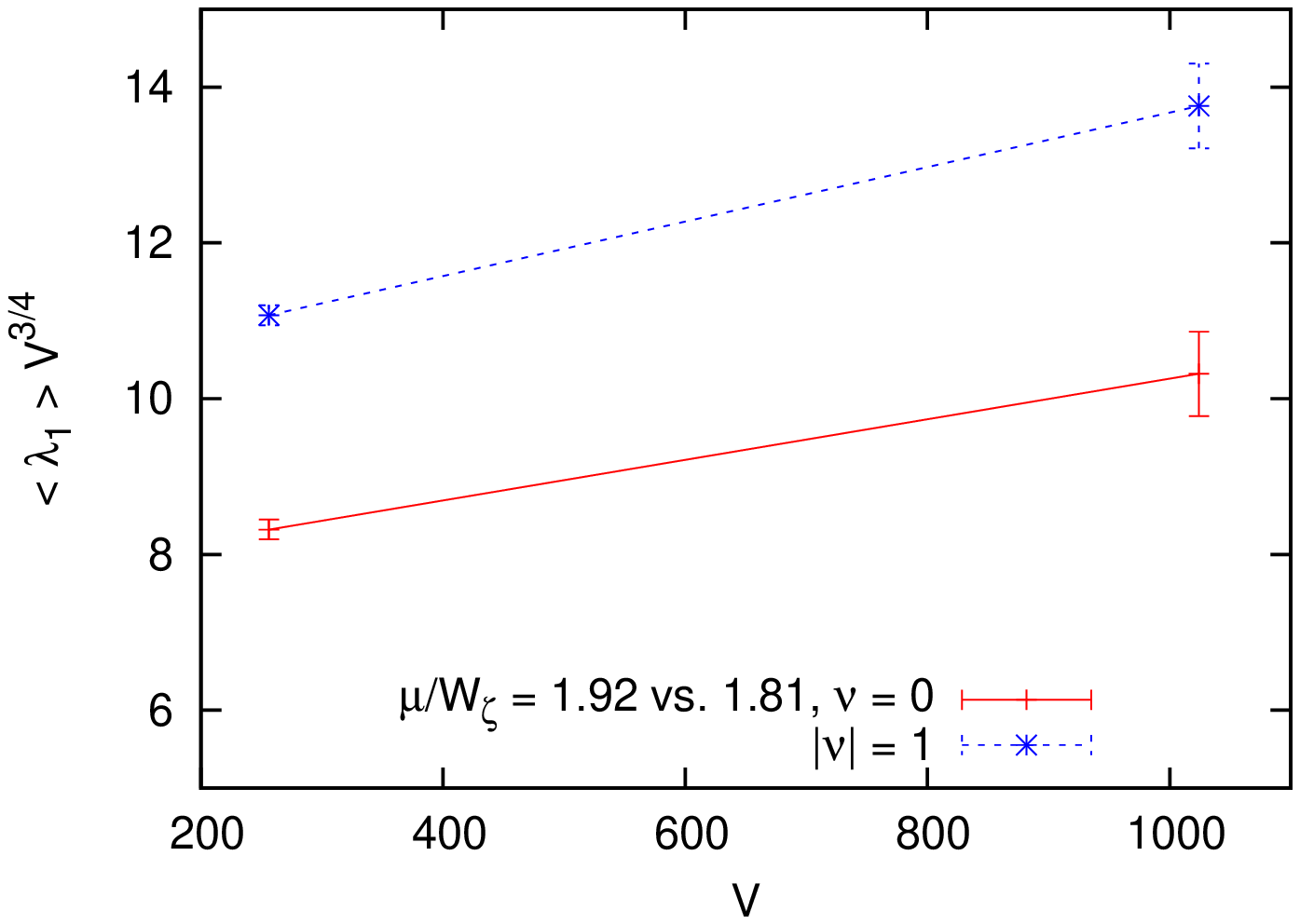}
\hspace*{-5mm}
\includegraphics[angle=270,width=.54\linewidth]{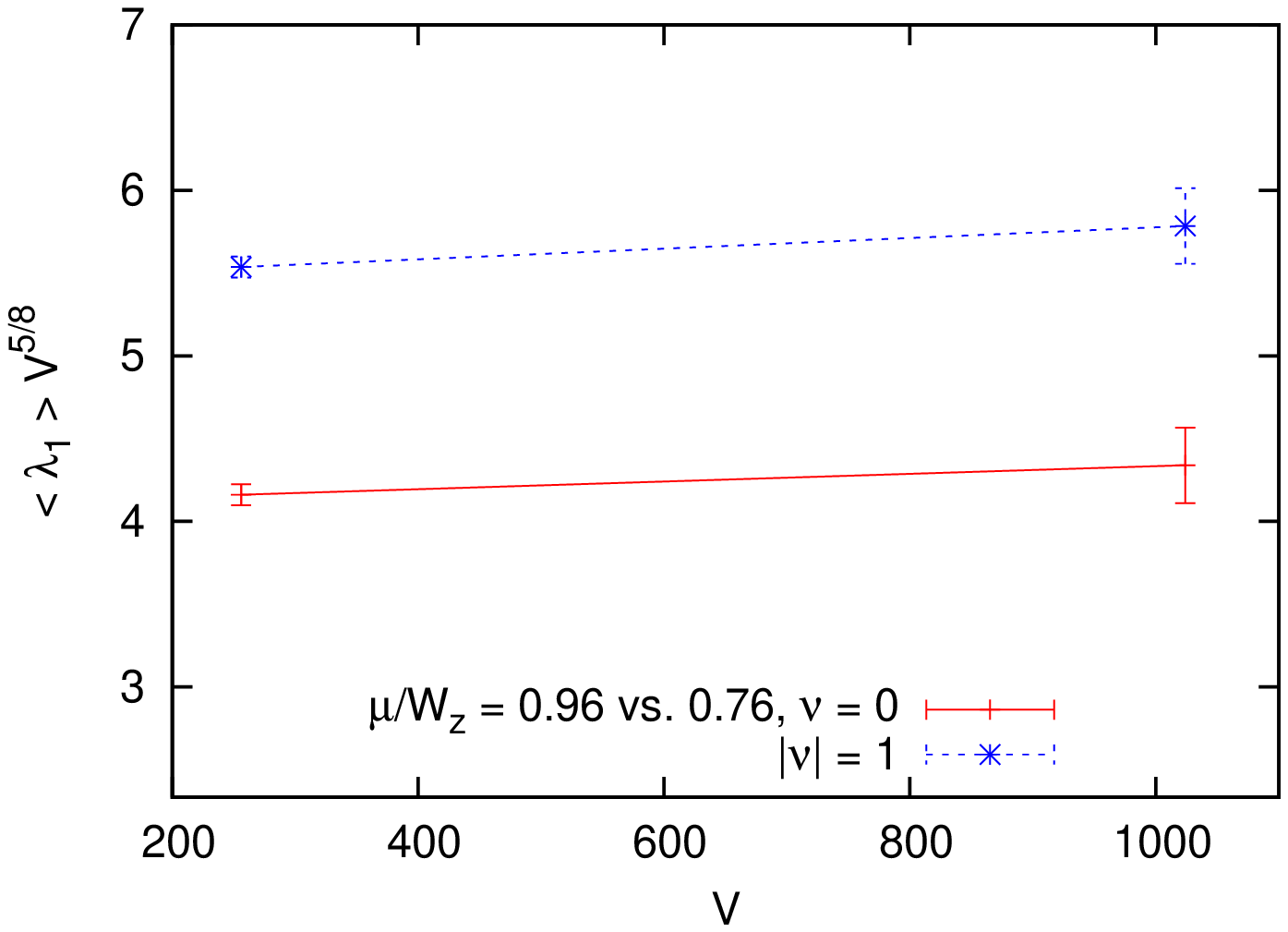}
\caption{\it{Finite size scaling for $\langle \zeta_{1} \rangle$ 
at $\mu_{\zeta} \propto mV^{3/4} \approx$~{\rm const.}\ (left) and 
$\langle z_{1} \rangle$ at $\mu_{z} \propto mV^{5/8} 
\approx$~{\rm const.}\ (right). These plots confirm again that $z_{1}$ 
performs much better as a scale-invariant variable.}}
\label{lam1m}
\vspace*{-3mm}
\end{figure}

We also tested the behaviour if the re-scaled mass is kept 
approximately constant, as an alternative to just keeping
$m$ small. In Figure \ref{lam1m} we compare 
$\langle \zeta_{i} \rangle$ for different lattice sizes, 
$L=16$ and $32$, again in the sectors $| \nu | =0$ and $ 1$, for
$\mu_{\zeta} \approx$~const. We add the corresponding test
with $\langle z_{i} \rangle$ and $\mu_{z} = m V^{5/8}  W_{z} 
\approx$~const., which reveals again a clearly superior finite size 
scaling. 

Our data favour this last scenario unambiguously.
A hint for a possible explanation can be found in
Ref.\ \cite{HHI}, which introduced the dimensionless parameter
\begin{equation}
l := \sqrt{2} m L^{3/2}/(\beta \pi)^{1/4}
\end{equation}
to distinguish different regimes. % It specifies that 
The aforementioned behaviour $\Sigma \propto m^{1/3}$ is 
expected for $l \gg 1$, whereas 
$l \ll 1 \ll 2L / \sqrt{\pi \beta}$ implies $\Sigma \propto mL$. 
For $m=0.01$ we are in an {\em intermediate regime,} $l = 0.5 \dots
1.3$ \ (and \ $2L / \sqrt{\pi \beta} = 8.1 \dots 16.2$), which 
renders our exponent in $\Sigma \propto m^{3/5}$ plausible.

However, a precise explanation for this behaviour 
remains to be worked out. In particular in the framework of RMT ---
extended to this extraordinary setting --- this might be feasible,
but it is far from trivial \cite{Poul}.

\subsection{Eigenvalues in the bulk of the Dirac spectrum}

\begin{figure}[h!]
\hspace*{-6mm}
\includegraphics[angle=270,width=.54\linewidth]{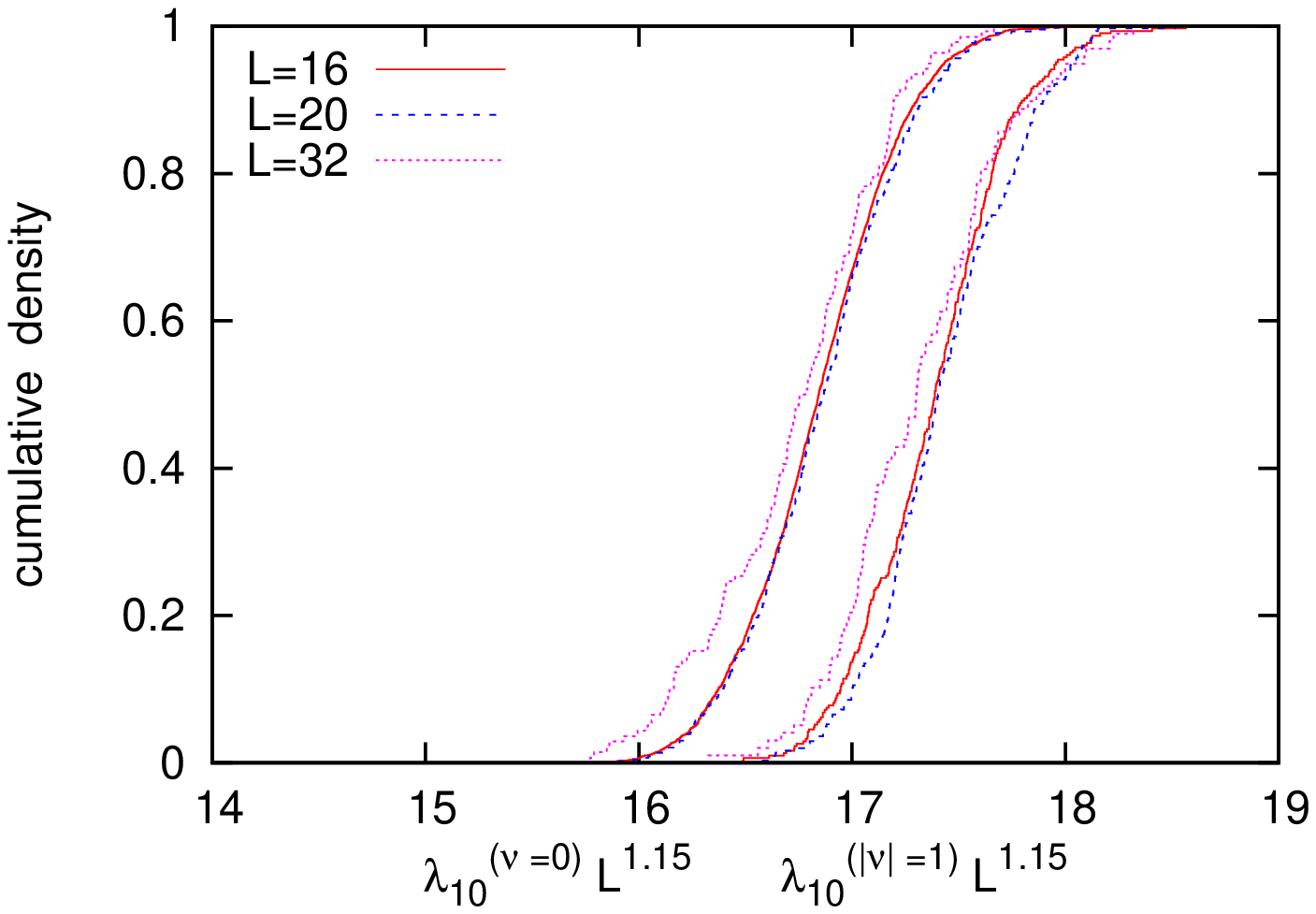}
\hspace*{-5mm}
\includegraphics[angle=270,width=.54\linewidth]{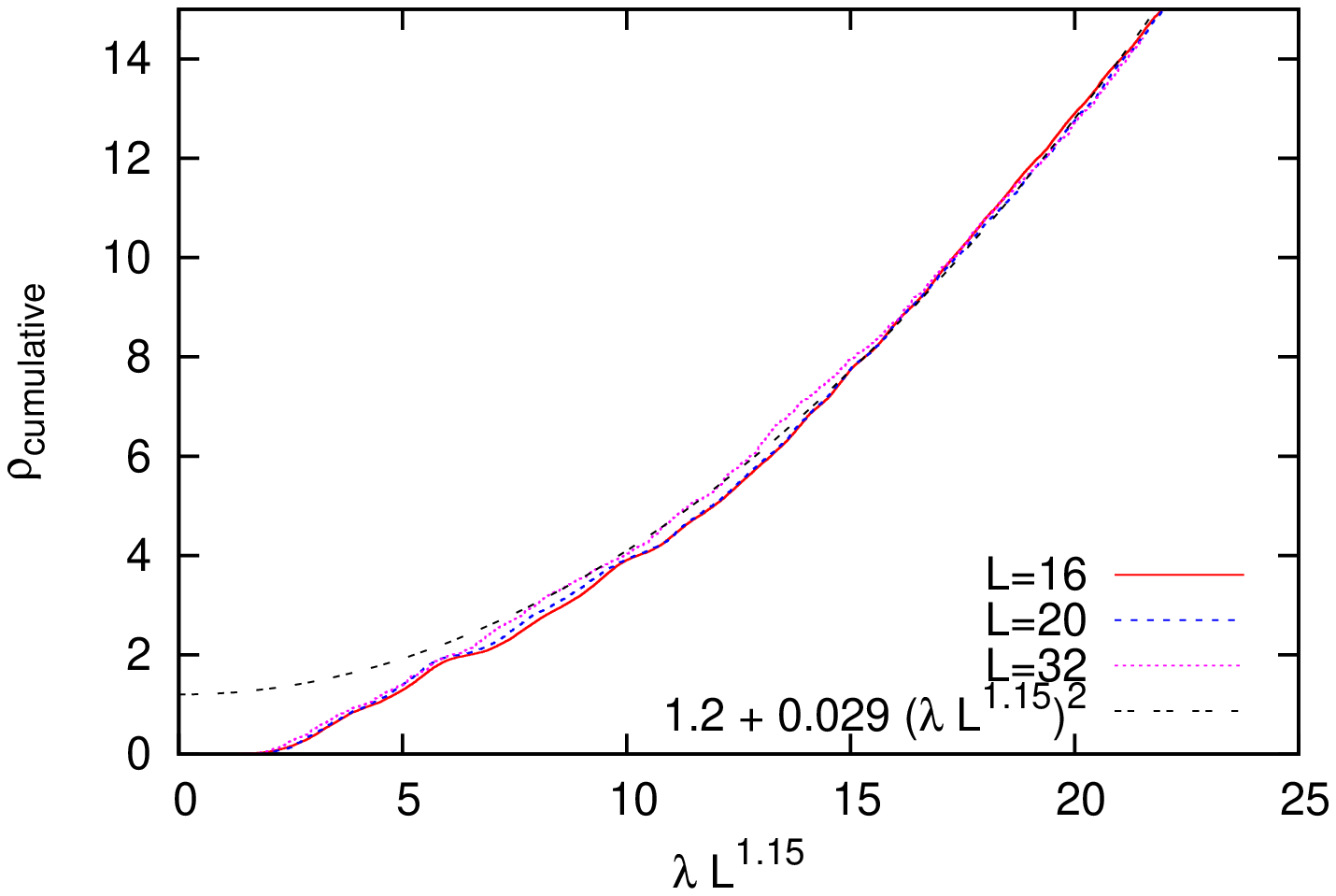}
\caption{\it{For $\lambda_{10}$, a low-lying bulk eigenvalue, the 
scaling factor is shifted to $L^{1.15}$. The re-scaled full spectral
cumulative densities in different volumes (for $m=0.01$, $\nu = 0$) 
agree well, and turn into the bulk behaviour $\rho (\lambda ) 
\propto \lambda$ (resp.\ $\rho_{\rm cumulative} (\lambda ) 
\propto \lambda^{2}$), which is expected in $d=2$.}}
\label{lam10}
\end{figure}

At last we take a look at $\lambda_{10}$ as one of the bulk
eigenvalues, and we find an optimal finite size scaling for
$\lambda_{10} L^{1.15}$, see Figure \ref{lam10} (left). 
The plot on the right shows that
this factor works well also for the re-scaled full cumulative 
density (including all eigenvalues up to the considered value).
Based on the fact that the spectral cutoff $\lambda_{\rm max}=2$ 
is fixed in any volume, it is now tempting to speculate that the
volume factor for a good finite size scaling gradually decreases
from $V^{3/5} \dots V^{0}$. However, considering eigenvalues
above the regime shown in Figure \ref{lam10}, but below the
cutoff regime, we could not find any consistent scaling 
factor. Indeed there is no compelling reason for such a factor 
to exist.

\section{Topological summation and susceptibility}

The chiral condensate $\Sigma$ can be measured using the complete 
Dirac spectrum,
\be  \label{sigmaspec}
\Sigma = \frac{1}{V} \sum_{i} \frac{1}{\lambda_{i} + m} \ ,
%\quad , \quad (V = L^{2}) \ ,
\ee
where we still refer to the eigenvalues $\lambda_{i}$ after
the projection (\ref{stereo}). Here we do not need any assumptions 
from RMT or the $\epsilon$-regime. We want to investigate the
link to the analytical formula for $\Sigma (m)$ in eq.\ (\ref{Sigmam}).

However, since there are only few topological transitions in the
HMC history (cf.\ Table \ref{stat-tabL16}), we can only measure 
\be
\Sigma_{|\nu |} = - \langle \bar \psi \, \psi \rangle\vert_{| \nu |} \ ,
\ee
{\it i.e.}\ the chiral condensate in separate topological 
sectors. Hence an appropriate summation has to be performed.
This challenge is generic for HMC simulations with
dynamical light fermions on fine lattices, so it is relevant
--- in particular in view of QCD % in the $\epsilon$-regime 
--- to explore such topological summations. Here we encounter
an interesting test bed to probe various methods for this purpose.

For very light fermions, the dominant contribution to 
$\Sigma_{\nu \neq 0}$ is due to the zero modes.
Hence a suitable notation is
\be  \label{epsnot}
\Sigma_{\nu} = \frac{| \nu |}{m V} + \varepsilon_{|\nu |} \ ,
\qquad %\frac{| \nu |}{m V} \gg 
\varepsilon_{0} > \varepsilon_{1}
> \varepsilon_{2} > \dots > 0 \ .
\ee
The inequalities at the end correspond to a general property of 
stochastic Hermitian matrices (such as $\gamma_{3} D_{\rm ovHF}$): the 
presence of zero modes pushes the small non-zero eigenvalues to higher
(absolute) values.

\begin{figure}[h!]
\vspace*{-5mm}
\hspace*{-11mm} 
\includegraphics[angle=270,width=.58\linewidth]{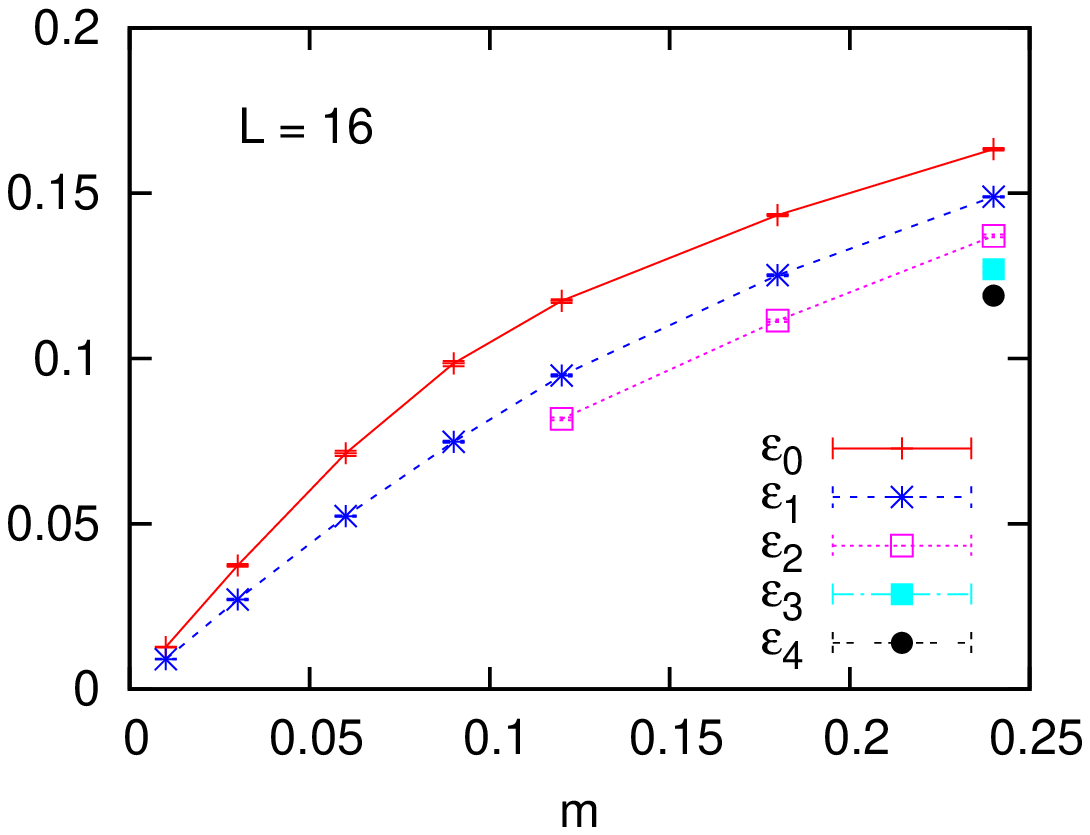}
\hspace*{-10mm}  
\includegraphics[angle=270,width=.58\linewidth]{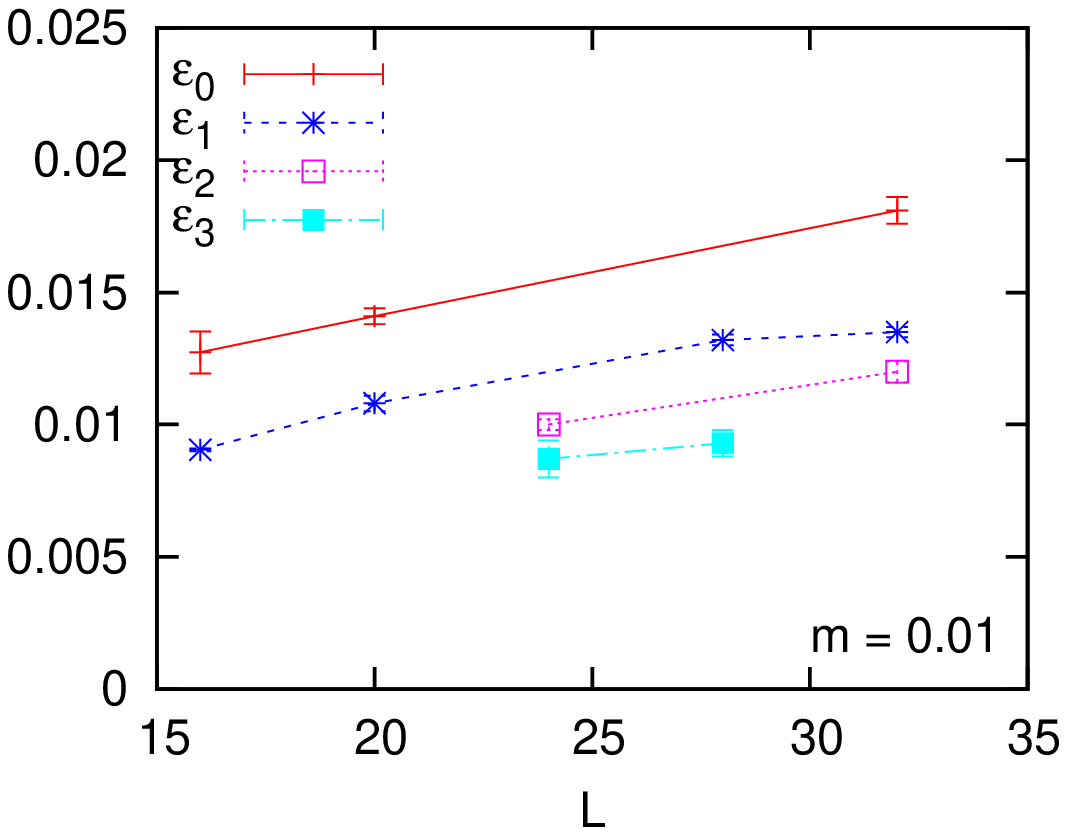}
\caption{\it{The terms $\varepsilon_{| \nu |}$ introduced in
eq.\ (\ref{epsnot}), which represent the contribution of
the non-zero modes to the chiral condensate in a fixed
topological sector. These plots reveal a smooth and monotonous
dependence on the fermion mass (on the left, at $L=16$) and
on the volume (on the right, at $m=0.01$).}}
\label{epsfig}
\end{figure}

Our results for the direct measurement of
$\Sigma_{\nu}$ are given in Table \ref{sigmatab}.
The hierarchy anticipated in relation (\ref{epsnot}) is consistently 
confirmed. In addition we observe the inequality
\be  \label{Vineq}
\varepsilon_{|\nu |}(V_{1}) >  \varepsilon_{|\nu |} (V_{2})
\quad {\rm if} \quad V_{1} > V_{2}
\ee
to hold. If the volume is enlarged, the non-zero
eigenvalues reach out to smaller values. We see that this
effect supersedes the pre-factor $1/V$ in eq.\ (\ref{sigmaspec}),
so that $\varepsilon_{|\nu |}$ increases.
The validity of inequalities (\ref{epsnot}), (\ref{Vineq})
is illustrated in Figure \ref{epsfig}. 
Moreover, we recognise a smooth mass and volume 
dependence of the terms $\varepsilon_{| \nu |}$.
\begin{table}[h!]
\centering
\begin{tabular}{|c|c||c|c|c|c|c|}
\hline
$L$ & $m$ & $\langle \Sigma_{\nu =0} \rangle$ &  
$\langle \Sigma_{|\nu | = 1} \rangle$ &
$\langle \Sigma_{|\nu | = 2} \rangle$ & 
$\langle \Sigma_{|\nu | = 3} \rangle$ & 
$\langle \Sigma_{|\nu | = 4} \rangle$ \\
    &     & $\langle \varepsilon_{\nu = 0} \rangle$ &
$\langle \varepsilon_{|\nu | = 1} \rangle$ &
$\langle \varepsilon_{|\nu | = 2} \rangle$ & 
$\langle \varepsilon_{|\nu | = 3} \rangle$ & 
$\langle \varepsilon_{|\nu | = 4} \rangle$ \\    
\hline
\hline
16 & 0.01 &  0.01273(8) &  0.39968(5)   &          &         & \\
   &      &  0.01273(8) &  0.00905(5)   &          &         & \\  
\hline
16 & 0.03 &  0.0374(4)  &  0.1573(2)   &          &         & \\
   &      &  0.0374(4)  &  0.0271(2)   &          &         & \\
\hline
16 & 0.06 &  0.0713(8)  &  0.1174(2)   &          &         & \\ 
   &      &  0.0713(8)  &  0.0523(2)   &          &         & \\
\hline
16 & 0.09 &  0.0985(8)  &  0.1182(3)   &          &         & \\
   &      &  0.0985(8)  &  0.0748(3)   &          &         & \\
\hline
16 & 0.12 &  0.1174(6)   &  0.1275(4)   &  0.1468(4)   &         & \\
   &      &  0.1174(6)   &  0.0949(4)   &  0.0817(4)   &         & \\
\hline
16 & 0.18 &  0.1434(3)   &  0.1469(3)  &  0.1548(4)  &         & \\
   &      &  0.1434(3)   &  0.1252(3)  &  0.1114(4)  &         & \\
\hline
16 & 0.24 & 0.1633(4) & 0.1652(2) &  0.1697(4) & 0.1758(9) & 0.1841(7) \\
   &      & 0.1633(4) & 0.1489(2) &  0.1371(4) & 0.1270(9) & 0.1190(7) \\
\hline
\hline
20 & 0.01 & 0.0141(3) & 0.2608(3) &          &           & \\
   &      & 0.0141(3) & 0.0108(3) &          &           & \\
\hline
24 & 0.01 &          &          & 0.3572(2) & 0.5296(2) & \\ 
   &      &          &          & 0.0100(2) & 0.0088(2) & \\
\hline
28 & 0.01 &          & 0.1408(2) &          & 0.3923(2) & \\ 
   &      &          & 0.0132(2) &          & 0.0096(2) & \\
\hline
32 & 0.01 & 0.0181(5) & 0.1112(2) & 0.2073(4) &           & \\ 
   &      & 0.0181(5) & 0.0135(2) & 0.0120(4) &           & \\
\hline
\hline
32 & 0.06 & 0.0883(7) & 0.093(1) &           &           & \\ 
   &      & 0.0883(7) & 0.077(1) &           &           & \\
\hline
\end{tabular}
\caption{\it Results for the directly measured chiral condensate
at different masses and lattice sizes, in distinct
topological sectors. We observe full agreement with inequalities
(\ref{epsnot}) and (\ref{Vineq}). 
As $m$ increases at fixed $L$, the dominant r\^{o}le of the
zero mode contributions to $\Sigma_{\nu \neq 0}$ is diminished.
As $L$ increases at fixed $m$, however, 
$\varepsilon_{|\nu |}$ is enhanced.}
\label{sigmatab}
\end{table}

\subsection{Gaussian evaluation of the topological susceptibility}

In this method we assume a Gaussian distribution 
of the topological charges, so that $\Sigma$ can be written as
\be
\Sigma = \sum_{\nu = -\infty}^{\infty} \, p (|\nu|) \
\Sigma_{|\nu|} \ , \quad
 p (|\nu|) = \frac{
\exp \Big( - \frac{\nu^{2}}{2 V \chi_{t}} \Big)}
{\sum_{\nu = -\infty}^{\infty} \,
\exp \Big( - \frac{\nu^{2}}{2 V \chi_{t}} \Big)} \ ,
\label{fixchit}
\ee
where $\chi_{t}$ is the topological susceptibility. At least
in QCD the charge distribution is indeed Gaussian to a good
approximation (see for instance the index histograms
in Refs.\ \cite{BS06}). A high statistics study
only found a tiny deviation, which tends to vanish in the
large volume limit \cite{kurto}.

If we have data in the sectors up to charge $Q$,
{\it i.e.}\ $Q$ is the maximum of the simulated sectors $|\nu |$,
we insert the measured values $\Sigma_{0} \dots \Sigma_{Q}$,
with the maximum and minimum of the statistical error bar.
For higher charges we make use of inequality (\ref{epsnot})
to fix the minimal and maximal values as
\be
\Sigma_{|\nu |, \min} = \frac{| \nu|}{m V} \quad ,
\quad \Sigma_{|\nu |, \max} = \Sigma_{|\nu |, \min} +
\varepsilon_{Q} \ .
\ee
%The lower bound is simply the zero mode contribution.
%The addition due to the higher modes decreases if $|\nu |$ 
%increases, because the presence of zero modes pushes the 
%low lying non-zero modes to higher values (this is generic
%for Hermitian random matrices, and it is also confirmed
%by all our results). That property provides the upper bound.
In the cases $L=24$ and $28$ we do not have data for all
the sectors with $| \nu | < Q$. Here we employ in addition 
inequality (\ref{Vineq}) and the results in the next smaller 
(larger) volume to fix $\Sigma_{|\nu |, \min}$ 
($\Sigma_{|\nu |, \max}$). \\

\begin{table}
\centering
\begin{tabular}{|c|c|c||l|l|}
\hline
$L$ & $m$ & $\Sigma$ of Ref.\ \cite{Smilga} & ~~~~~~~$\chi_{t}$ & 
$ ~~~~ \langle \nu^{2} \rangle$ \\
\hline
\hline
16 & 0.01 & 0.04888  & 0.0006586(5) &  0.1686(1) \\ 
\hline
16 & 0.03 & 0.07050  & 0.00117(1)  &  0.299(3) \\ 
\hline
16 & 0.06 & 0.08883  & 0.00159(8)  &  0.407(20) \\ 
%\hline
%16 & 0.09 & 0.10168  & 0.00084(10)  &  0.0215(26)\\ 
\hline
\hline
20 & 0.01 & 0.04888  & 0.000500(2) &  0.200(1) \\ 
\hline
24 & 0.01 & 0.04888  & 0.000408(16) &  0.235(9) \\ 
\hline
28 & 0.01 & 0.04888  & 0.000367(15) &  0.288(12) \\ 
\hline
32 & 0.01 & 0.04888  & 0.000341(4) &  0.349(4) \\ 
\hline
%32 & 0.06 & 0.08883  & ??? &  ??? \\ 
%\hline
\end{tabular}
\caption{\it{Results for the topological susceptibility $\chi_{t}$ 
based on the method described in Subsection 5.1, which assumes
Gaussian charge distribution and the chiral condensate according
to Ref.\ \cite{Smilga}. On the $L=16$ lattice this method does
not work for $m\geq 0.09$, due to the latter assumption.}}
\label{chitab}
\end{table}

\begin{figure}[h!]
\hspace*{-6mm} 
\includegraphics[angle=270,width=.53\linewidth]{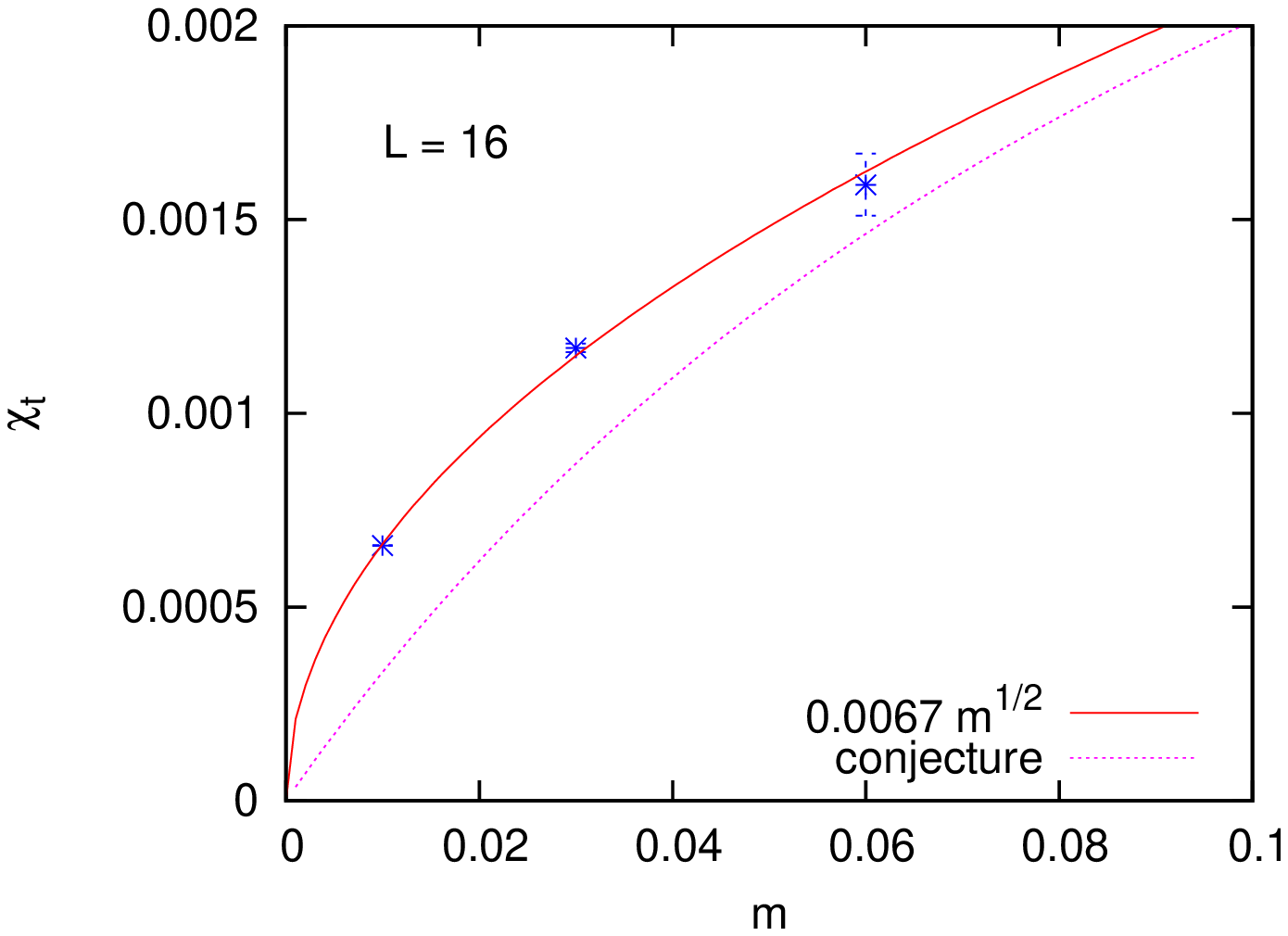}
\hspace*{-6mm}  
\includegraphics[angle=270,width=.53\linewidth]{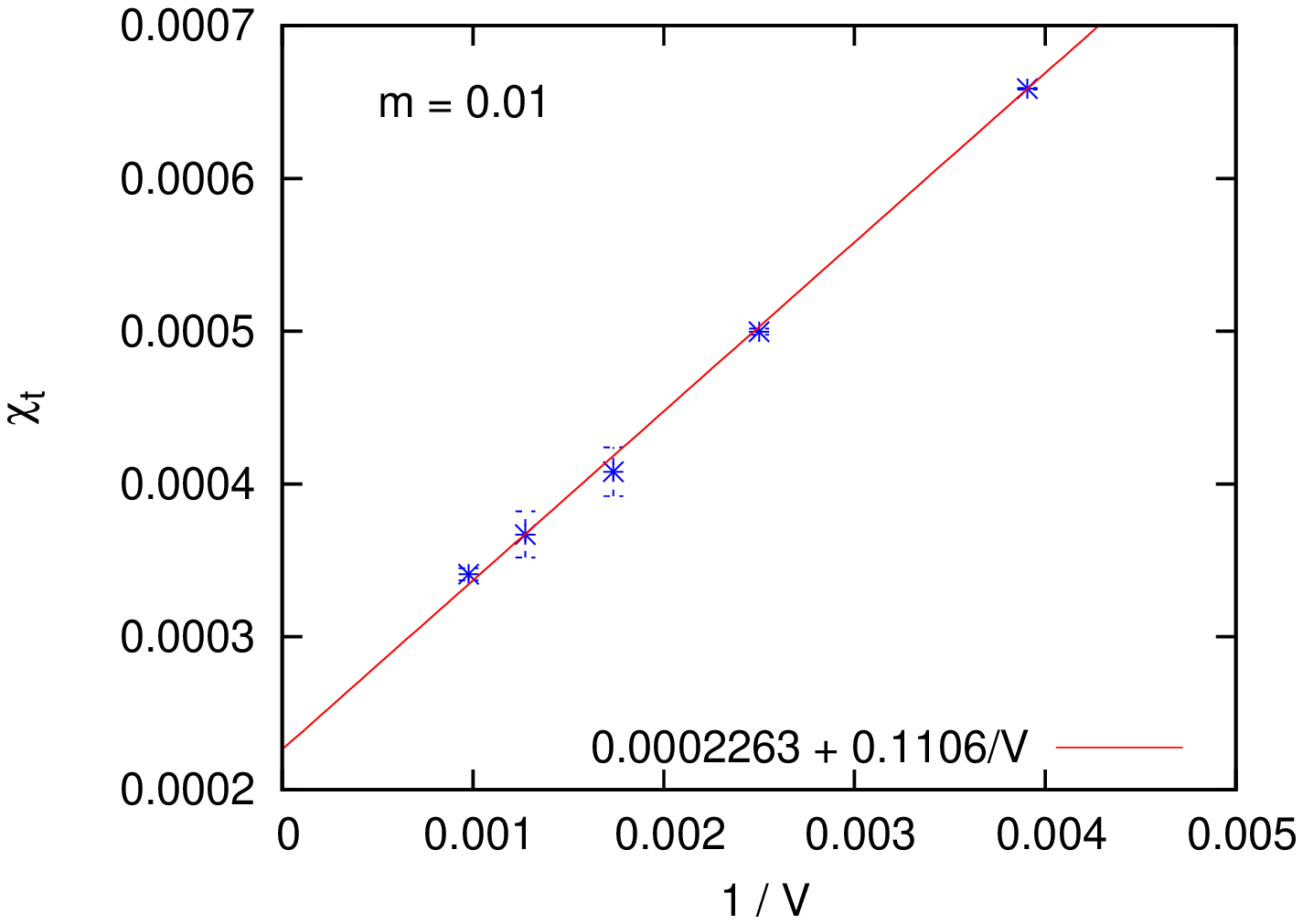}
\caption{\it{The topological susceptibility $\chi_{t}$ 
determined by the method described in Subsection 5.1.
On the $L=16$ lattice we see that Smilga's
formula cannot be valid at $m \geq 0.09$, hence
also this method is not applicable anymore.
For the smaller masses our results for the susceptibility suggest
a behaviour $\chi_{t} \propto \sqrt{m}$. It is compared to
the conjecture in eq.\ (\ref{chidu}), based on Refs.\ 
\cite{DuHo,Durr}. The plot on the right refers to $m=0.01$.
An infinite volume extrapolation, which assumes
$\chi_{t}$ to be consistent with a linear dependence on $1/V$,
leads to $\chi_{t} = 0.0002263(53)$.}}
\label{chifig}
\end{figure}

If we insert some susceptibility $\chi_{t}$ into eq.\ (\ref{fixchit}),
we obtain a value for $\Sigma$. Its uncertainty is modest, 
because most sectors, which have not been measured, have exponentially
suppressed probabilities $p(|\nu |)$. By requiring agreement with
the theoretical prediction of Ref.\ \cite{Smilga} (given in eq.\
(\ref{Sigmam})) we now determine $\chi_{t}$; 
the results are given in Table \ref{chitab}
and Figure \ref{chifig}.

Table \ref{sigmatab} shows that
only in the case of light fermions the prediction of Ref.\ 
\cite{Smilga} can be reproduced. For $L=16$ and $m \geq 0.12$
{\em all} the $\Sigma_{\nu}$ are larger than the prediction,
hence there is no way to reproduce it with a weighted sum.
In the case $L=32$, $m=0.06$ the value of 
$\langle \Sigma_{\nu =0} \rangle$
is too close to the prediction for $\Sigma$ 
to extract a sensible result for $\chi_{t}\,$.
%(not even if we allow for deviation from the Gaussian distribution).
At $L=16$, $m=0.09$ a value for $\chi_{t}$ which achieves this can still
be found, but the result does not make sense either, since
it is below the $\chi_{t}$ values determined at $m=0.03$ and
$0.06$ (see Figure \ref{chifig}).
We conclude that Smilga's formula is not applicable
at $m \geq 0.09$, which assigns an explicit meaning to his
assumption $m \ll 1 / \sqrt{\beta} \simeq 0.45$.

Figure \ref{chifig} (on the left) shows the
consistency of our results with the behaviour
\be  \label{chitvsm}
\chi_{t} (m) \propto \sqrt{m} \ .
\ee
Alternative results (with quenched configurations and re-weighting)
were given in Ref.\ \cite{DuHo}. On the theoretical side, Ref.\ \cite{Durr}
conjectured for $N_{f}$ degenerate flavours in the large volume limit
\be  \label{chidu}
\frac{1}{\chi_{t}} = \frac{N_{f}}{\Sigma^{(1)}m} +
\frac{1}{\chi_{t,q}} \ , \qquad
\Sigma^{(1)} %:= \Sigma_{N_{f}=1}(m=0) 
\simeq 0.160 \, g \ ,
\ee
where % $\Sigma^{(1)} := \Sigma_{N_{f}=1}(m=0) \simeq 0.16 g$ and
$\Sigma^{(1)} := \Sigma_{N_{f}=1}(m=0)$ (cf.\ Section 1), and
$\chi_{t,q} = \chi_{t} (m \to \infty)$ is the quenched value.
Actually this mass dependence of $\chi_{t}$ was conjectured in
the framework of QCD. If we apply the same form in the
Schwinger model, and insert the value 
$\chi_{t,q} \approx 0.023 \, g^{2} = 0.0046$ 
from Ref.\ \cite{DuHo}, we obtain $\chi_{t} \approx 
0.00033$. Figure \ref{chifig} (on the left) shows a trend towards
agreement with this conjecture as we increase the mass at
$L=16$, thus suppressing the finite size effects.
In particular at $m=0.06$ eq.\ (\ref{chidu}) implies
$\chi_{t} \simeq 0.0015$, close to our value in Table \ref{chitab}.

%\footnote{The result is hardly affected by the finiteness
%of $\chi_{t,q}$. If we set it to infinite, we obtain 
%$\chi_{t} \approx 0.00036$.} 
Also when we increase the volume at $m=0.01$, our results
attain the vicinity of this prediction: at $L=32$ we obtained
$\chi_{t} \approx 0.00034$, next to the predicted value
($\chi_{t} \approx 0.00033$).

If we try an infinite volume extrapolation, however,
the plot in Figure \ref{chifig} on the right
suggests a linear dependence of $\chi_{t}$ on $1/V$; that assumption 
leads to a smaller value of $\chi_{t}(V = \infty ) = 0.0002263(53).$
%\be  \label{chitvsV}
%\chi_{t}(V) \simeq \chi_{t}(V = \infty ) + 
%\frac{{\rm const.}}{V} \ , \quad
%\chi_{t}(V = \infty ) = 0.0002263(53) \ .
%\ee

\subsection{An approximate formula for the topological summation}

The procedure of the previous subsection is robust 
for light fermions (baring finite size effects on $\Sigma$).
However, since it uses
the analytic result for $\Sigma$ as an input to determine $\chi_{t}$,
it does not evaluate $\Sigma$ itself from the numerical results
in distinct topological sectors.
In Subsections 5.3 and 5.4 we will test a method to do so, following 
Ref.\ \cite{fixtopo1}. For convenience we re-derive here in a concise
form the formula for an approximate topological summation that was
given in Ref.\ \cite{fixtopo1} for the pion mass; we 
generalise it to arbitrary observables. This consideration follows 
the lines of Refs.\ \cite{fixtopo1,fixtopo2}, pointing out in 
particular which assumptions are involved, so the subsequent
subsections can refer to them.

First we assume the fermion field to be integrated out.
Thus we refer to an effective gauge action $S_{\rm eff}[U]$,
which keeps track of the fermion determinant. So the
partition function takes the form 
$Z = \int {\cal D} U \, \exp \{ -S_{\rm eff} [U] \} $. Next we 
introduce a Kronecker $\delta$ as a filter of gauge 
configurations $[U]$ with a specific topological charge $\nu$,
\be
\delta_{\nu , \nu [U]} = \frac{1}{2 \pi}
\int_{-\pi}^{\pi} d \theta \, 
\exp \{ \ri \theta (\nu - \nu [U])\} \ .
\ee
This formulation involves the vacuum angle $\theta$, and
it allows us to write a ``partition function'' restricted to
one topological sector as
%\bea
%Z_{\nu} &=& \int {\cal D}U \, e^{-S_{\rm eff}[U]} \, 
%\delta_{\nu , \nu [U]} %\nn \\
%= \frac{1}{2 \pi} \int_{-\pi}^{\pi} d \theta \,
%e^{\ri \theta \nu} \int {\cal D}U \, 
%e^{-S_{\rm eff}[U] - \ri \theta \nu [U] } \nn \\
%&=& \frac{1}{2 \pi} \int_{-\pi}^{\pi} d \theta \, e^{\ri \theta \nu}
%Z(\theta ) \ , % \qquad \quad
%\label{Znu}
%\eea
\be
Z_{\nu} = \int {\cal D}U \, e^{-S_{\rm eff}[U]} \, 
\delta_{\nu , \nu [U]} =
\frac{1}{2 \pi} \int_{-\pi}^{\pi} d \theta \, e^{\ri \theta \nu}
Z(\theta ) \ , 
\label{Znu}
\ee
where $Z(\theta ) = \int {\cal D} U \, 
\exp \{ -S_{\rm eff} [U] - \ri \theta \nu [U] \}$ 
is the (complete) partition function for a general 
vacuum angle (and $Z = Z(0)$).

If some observable ${\cal O}$ is measured only in one topological
sector, the corresponding expectation value is given by\footnote{We
use the notation $\bar {\cal O}$, rather than $\langle {\cal O} 
\rangle$, because in the following this will be more practical 
for indicating the dependence on $\theta$.}
\be  \label{Onu}
\bar {\cal O}_{\nu} = \frac{1}{Z_{\nu}}
\int {\cal D}U \, e^{-S_{\rm eff}[U]} \delta_{\nu , \nu [U]}
{\cal O} [U]
= \frac{1}{2 \pi Z_{\nu}} \int_{-\pi}^{\pi} d \theta \,
e^{\ri \theta \nu} Z(\theta ) \bar {\cal O} ( \theta ) \ .
\ee

The relation
\be
- Z''(\theta )\vert_{\theta = 0} = \langle \nu^2 \rangle =
V \chi_{t}
\ee
%where $\chi_{t}$ is the topological susceptibility, 
suggests 
% \cite{fixtopo1}
\be  \label{expansatz}
Z(\theta ) = Z \, \exp ( -V \chi_{t} \theta^{2} /2 ) \ .
\ee
Inserting this ansatz into eq.\ (\ref{Znu}) leads to
\be  \label{Znuint}
\frac{Z_{\nu}}{Z} = \frac{1}{2\pi } \int_{-\pi}^{\pi} d \theta \,
\exp \Big( - \frac{V \chi_{t}}{2} \theta^{2} + \ri \theta \nu \Big) \ ,
\ee
where we recognise the stationary phase
\be
\theta_{s,\nu} = \ri \frac{\nu}{V \chi_{t}} 
= \ri \frac{\nu}{\langle \nu^{2} \rangle} \ .
\ee
As an approximation, we extend the bounds in the integral
(\ref{Znuint}) to $\pm \infty$. For this step,
it is favourable if $V \chi_{t}$ is large.
%Substituting $\theta' = \theta - \theta_{s,\nu}$ yields an
%integral from $\theta ' = - \pi - \theta_{s,\nu}$ to
%$\pi - \theta_{s,\nu}$, which we approximate by a Gaussian
%integral $-\infty \dots + \infty$. For this approximation,
%it is favourable if $V \chi_{t}$ is large (such that
%$\exp ( - V \chi_{t} \pi^{2} /2)$ is small).
%To justify neglecting the imaginary part requires in
%addition a small $| \theta_{s,\nu}|$, so that small charges
%$| \nu |$ are favourable. 
Then we obtain
\be  \label{Znu2}
\frac{Z_{\nu}}{Z} \simeq \frac{1}{\sqrt{ 2 \pi V \chi_{t}}} 
\ \exp \Big( - \frac{\nu^{2}}{2 V \chi_{t}} \Big) \ .
\ee
This formula is obviously consistent with an integral
approximation for $Z = \sum_{\nu} Z_{\nu}$, which is again
best justified for large $V \chi_{t}$.

If we insert ansatz (\ref{expansatz}) into eq.\ (\ref{Onu}),
we obtain another integral with the stationary phase
$\theta_{s,\nu}$. By repeating its approximation as a Gaussian
integral, and employing relation (\ref{Znu2}), we arrive at
\bea
\bar {\cal O}_{\nu} &=& \frac{Z}{2 \pi Z_{\nu}}
\int_{-\pi}^{\pi} d \theta \, \bar {\cal O} ( \theta )
\exp \Big( - \frac{V \chi_{t}}{2} \theta^{2} + \ri \theta \nu \Big)
\nn \\
& \simeq & \sqrt{\frac{V \chi_{t}}{2 \pi}} \int_{-\infty}^{\infty}
d \theta \, \bar {\cal O} ( \theta ) 
\exp \Big( - \frac{V \chi_{t}}{2} (\theta - \theta_{s,\nu})^{2}
\Big) \ .
\label{Onu2}
\eea
Since we assume only %the vicinity of $|\theta_{s,\nu}|$ 
small $| \theta - \theta_{s,\nu}|$
to contribute significantly to this integral, we  may also
approximate
\be
\bar {\cal O} ( \theta ) \simeq
\bar {\cal O} (\theta_{s, \nu} ) + \frac{1}{2} 
\bar {\cal O}''(\theta )\vert_{ \theta_{s, \nu}}
( \theta - \theta_{s, \nu} )^{2} \ .
\ee
%where the primes are derivatives with respect to $\theta$.
Let us further assume $| \theta_{s}|$ to be small, so we
can replace the first term in this formula 
for $\bar {\cal O} ( \theta )$ by
\be  \label{approxO}
\bar {\cal O} (\theta_{s, \nu}) \simeq
\bar {\cal O}(0) + \frac{1}{2} 
\bar {\cal O}''(\theta )\vert_{0} \
\theta_{s, \nu}^{2} \ .
\ee
Hence a further property, which is favourable for our approximation,
is a small topological charge $|\nu |$, in addition to a large 
value of $V \chi_{t}$.

The approximation (\ref{approxO}) also implies
$\bar {\cal O}''(\theta )\vert_{\theta_{s}} \simeq
\bar {\cal O}''(\theta )\vert_{0}$. So we can express the
(numerically measurable!) restricted expectation value as
\be
\bar {\cal O}_{\nu} 
% & \simeq & \frac{1}{2 \pi Z_{\nu}} \int_{-\pi}^{\pi} 
% d \theta \, \Big[ \bar {\cal O} + \frac{1}{2} \bar {\cal O}''
% \Big\{ (\theta - \theta_{s})^{2}
% + \theta_{s}^{2} \Big\} \Big] \times \nn \\
% && \exp \Big( - \frac{V \chi_{t}}{2} 
% (\theta - \theta_{s})^{2} - \frac{\nu^{2}}{2 V \chi_{t}} \Big) \nn \\
 \simeq  \sqrt{\frac{ V \chi_{t}}{2 \pi}} \int_{-\pi}^{\pi} d \theta \, 
\Big[ \bar {\cal O} + \frac{1}{2} \bar {\cal O}'' 
[  (\theta - \theta_{s, \nu} )^{2} 
+ \theta_{s, \nu}^{2} ]
% - \Big( \frac{\nu}{V\chi_{t}} \Big)^{2} 
\Big] % \times \nn \\
\exp \Big( - \frac{V \chi_{t}}{2} 
(\theta - \theta_{s, \nu})^{2} \Big) \ ,
\ee
where $\bar {\cal O}$ and $\bar {\cal O}''$ are taken at $\theta =0$.
Extending once more the boundaries to $\pm \infty$ leads to the
final form
\be  \label{finapprox}
\bar {\cal O}_{\nu} \approx \bar {\cal O} + 
\frac{1}{2} \bar {\cal O}'' \, \frac{1}{V \chi_{t}} 
\, \Big( 1 - \frac{ \nu^{2}}{V \chi_{t}} \Big) \ .
\ee
This is the same structure as Ref.\ \cite{fixtopo1} obtained
for the pion mass. It is consistent that the limit 
$V \chi_{t} \to \infty$ renders all topological sectors equivalent, 
so that all $\bar {\cal O}_{\nu}$ coincide with $\bar {\cal O}$.

The numerical measurement with few topological transitions
provides results for the left-hand-side of eq.\ (\ref{finapprox}).
On the right-hand-side $\bar {\cal O}$, $\bar {\cal O}''$ and 
$\chi_{t}$ are unknown. We are interested in $\bar {\cal O}$ and 
$\chi_{t}$, and measurements of $\bar {\cal O}_{\nu}$ in
various topological sectors and volumes allow in principle
for their evaluation (as far as the above approximations make
sense). Up to now, this intriguing and possibly powerful
technique has not been tested with simulation data.
This will be pioneered in the next two subsections.

\subsection{Topological summation of the chiral condensate}

Now we insert the chiral condensate $\Sigma$ as our observable 
$\bar {\cal O}$. As our input we have data for some $\Sigma_{\nu}$,
{\it i.e.}\ for the left-hand-side of eq.\ (\ref{finapprox}), but
on the right-hand side $\Sigma$, $\Sigma''$ and $\chi_{t}$ are unknown.
%We are interested in $\Sigma$ and $\chi_{t}$.
In view of their evaluation, % of these unknown quantities, 
it is convenient to re-write eq.\ (\ref{finapprox}) as
\bea
\Sigma_{\nu} & \approx & \Sigma - \frac{A}{V}
+ \frac{B}{V^{2}} \, \nu^{2} \nn \\
&& A := - \frac{\Sigma''}{2 \, \chi_{t}} \ , \quad
B := - \frac{\Sigma''}{2 \, \chi_{t}^{2}} \ , \quad
\chi_{t} = \frac{A}{B}  \label{ABdef} \ .
\eea
%The new variables $A$ and $B$ have dimension [mass]$^{-1}$ resp.\
%[mass]$^{-3}$.
%
%From this structure, we clearly see the following property:
Based on data from different topological sectors in a fixed volume $V$
we can only evaluate $B$; it is not possible to obtain $\Sigma$
and $\chi$ without including different volumes.

By considering two sectors with charges $| \nu | =
k$ and $\ell$ (and $k \neq \ell$)
at fixed $V$ and $m$, a result for $B$ is obtained as
\be
\frac{1}{V} \, B_{k, \ell} = V \, \frac{\Sigma_{k} - \Sigma_{\ell}}
{k^{2} - \ell^{2}}
 = \frac{1}{m (k+ \ell )} + V \, \frac{\varepsilon_{k} -
\varepsilon_{\ell}}{k^{2} - \ell^{2}} \ . \label{Beval}
\ee
If more than two $\Sigma_{| \nu |}$ values have been measured,
the approximate agreement between the emerging $B_{k, \ell}$
represents a consistency condition.
In Table \ref{tabB} we give corresponding results for 
$B_{k, \ell}/V$, derived from the data in Table \ref{sigmatab}.

\begin{table}[h!]
\centering
%\begin{tabular}{|c|c||c|c|c|c|c|c|}
\begin{tabular}{|c|c||c|c|c|c|c|}
\hline
$m$ & $L$ & $B_{1,0}/V$ & $B_{2,0}/V$ & $B_{2,1}/V$ & 
$B_{3,0}/V$ & $B_{4,0}/V$ \\
%& $B_{3,1}$ & $B_{3,2}$ \\
\hline
\hline
0.01 & 32 & $95.33(55)$ & $48.44(16)$  & $32.80(15)$ & & \\ 
\hline
0.12 & 16 &  $2.59(18) $  & $1.882(50)$ & $1.647(48)$ & & \\
\hline
0.18 & 16 &  $0.90(11)$  & $0.730(32)$ & $0.674(43)$ & & \\
\hline
0.24 & 16 &  $0.49(11) $  & $0.410(36)$ & $0.384(38) $ &
$0.356(28) $ & $0.333(13) $ \\
\hline
\end{tabular}
\caption{\it Results for the term $B_{k,\ell }/V$ obtained
in each case from two values of $\Sigma_{|\nu |}$ in a fixed
volume $V$ and at a fixed mass $m$, according to eq.\ (\ref{Beval}).}
\label{tabB}
\end{table}

\begin{figure}
\begin{center}
\includegraphics[angle=270,width=0.55\linewidth]{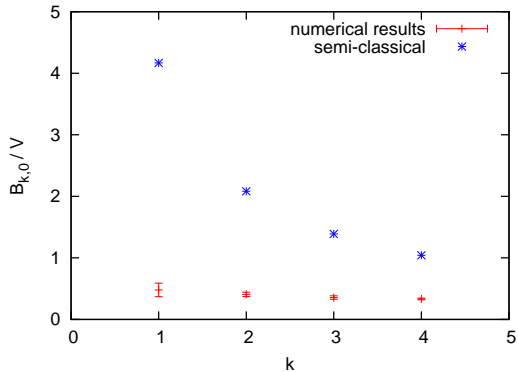}
\end{center}
\caption{\it{The terms $B_{k,0}/V$ for $k=1 \dots 4$,
at $m=0.24$, $L=16$. For the numerical result we observe a 
remarkable stability in $k$, in contrast to the semi-classical 
values.}}
\label{Bfig}
\end{figure}

For small fermion masses, the results for $B_{k,\ell}$
are dominated by the semi-classical term $1/ [m(k+\ell )]$,
and thus strongly dependent on the topological sectors involved.
This feature is suppressed, however, when $m$ increases;
%However, as $m$ increases this effect is suppressed; 
in particular for $L=16, \ m=0.24$ we do observe approximate agreement
for different choices of $k$ and $\ell$, in striking 
contrast to the semi-classical result. This similarity, which 
is illustrated in Figure \ref{Bfig}, is a remarkable quantum effect,
due to the fluctuation terms $\varepsilon_{|\nu |}$.
%The approximate consistency of $B_{k,\ell}$ evaluated at fixed
%$V$ and $m$ from different pairs of topological sectors seems
%satisfactory, although 
Still we observe systematically that 
the $B_{k,\ell }$ values decrease if higher
topological charges are involved. That also reduces the reliability 
of our assumption of a small $|\theta_{s, \nu}|$ --- which
favours the approximation (\ref{approxO}) --- so we
consider $B_{1,0}$ most reliable.
%This consistency test is non-trivial
%since it essentially depends on the small terms $\varepsilon_{|\nu |}$
%in notation (\ref{epsnot}) (see lower line of eq.\ (\ref{Beval})),
%hence it cannot be explained simply from a semi-classical picture
%(where all $\varepsilon_{|\nu |}$ vanish).

In order to proceed, {\it i.e.}\ to get access also to $A$ and thus 
%in principle 
to $\chi$ and $\Sigma$, we have to confront data from
different volumes. Considering two volumes $V_{1}$ and $V_{2}$,
but keeping $m$ and $|\nu |$ fixed, eq.\ (\ref{ABdef}) implies
\be  \label{Anu}
\Sigma_{|\nu |}(V_{1}) - \Sigma_{|\nu |}(V_{2}) =
A \, \Big( \frac{1}{V_{2}} - \frac{1}{V_{1}} \Big)
+ B \nu^{2} \, \Big( \frac{1}{V_{1}^{2}} 
- \frac{1}{V_{2}^{2}} \Big) \ .
\ee
%In practice we obtain values $A^{(|\nu |)}_{V_{1},V_{2}}$
%(similar to $B_{k,\ell}$). 
Most convenient is the %neutral
sector $\nu =0$, where we can read off $A$
%$A^{(0)}_{V_{1},V_{2}}$ 
without involving the variable $B$. At $|\nu| = 1$ it is most 
obvious to insert $B$ as
\be  \label{Anu1}
\Sigma_{1}(V_{1}) - \Sigma_{1}(V_{2}) =
A \, \Big( \frac{1}{V_{2}} - \frac{1}{V_{1}} \Big)
+ \frac{B_{1,0}(V_{1})}{V_{1}^{2}} - \frac{B_{1,0}(V_{2})}{V_{2}^{2}} 
\ ,
\ee
which is identical to the result obtained from $\Sigma_{0}(V_{1})$,
$\Sigma_{0}(V_{2})$. This is expected to be the most reliable 
value for $A$.
Our results obtained in this way are listed in Table \ref{tabA}.

\begin{table}[h!]
\centering
\begin{tabular}{|c||c|c|c||c|}
\hline 
$m$ & \multicolumn{3}{|c||}{$m =0.01$} & $m =0.06$ \\
\hline
$(L_{1}, \, L_{2})$ & $(20,\, 16)$ & $(32,\, 16)$ & $(32,\, 20)$ &
$(32,\, 16)$ \\
\hline
\hline
$A$ & 0.974(221) & 1.833(173) & 2.626(383) & 5.803(363) \\
\hline
%\begin{table}[h!]
%\centering
%\begin{tabular}{|c||c|c|c|c|}
%\hline 
%$(L_{1}, \, L_{2})$ & $(20,\, 16)$ & $(32,\, 16)$ & $(32,\, 20)$\\
%\hline
%\hline
%\multicolumn{4}{|c|}{$m =0.01$} \\
%\hline
%$A$ & 0.974(221) & 1.833 (173) & 2.626(383) \\
%%\hline
%%$A$ at $| \nu | = 1$ & 0.974(378) & 1.834(199) & 2.628(508) \\
%\hline
%\hline
%\multicolumn{4}{|c|}{$m =0.06$} \\
%\hline
%$A$ & & 5.803(363) & \\
%\hline
%%$A$ at $| \nu | = 1$ & & 5.805(600) & \\
%%\hline
\end{tabular}
\caption{\it Results for the term $A$ in eq.\ (\ref{ABdef}), 
obtained from $\Sigma_{0}$ or $\Sigma_{1}$ in two volumes,
at a fixed mass $m$, according to eqs.\ (\ref{Anu})
and (\ref{Anu1}).}
\label{tabA}
\end{table}

The idea is to use these results in both volumes 
involved. This is sensible if $A$ is approximately %more or less
constant in the volume, but that is not confirmed in
our data set for $m=0.01$. 
Since the assumptions tend to hold better for larger volumes,
we use the value of $A$ obtained in $(L_{1} , L_{2}) = (32,16)$
to determine $\Sigma$ in $L=16$, and  $A$ from
$(32,20)$ for $\Sigma$ in $L=20$ and $32$.
%The simplest way to extract $\Sigma$ is to apply
%$\Sigma_{0}$, without involving the term $B$,
%\be  \label{Sig0}
%\Sigma = \Sigma_{0} + \frac{A^{(0)}}{V} \ ,
%\ee
%which leads to
This leads to
\be  \label{Sig0r}
\Sigma \simeq 0.0199(7) \quad (L=16) \ , \quad
\Sigma \simeq 0.0207(12) \quad (L=20 ~ {\rm or} ~ 32) \ .
\ee
%Alternatively we can insert $\Sigma_{1}$, $A^{(1)}$ and $B_{1,0}$ into
%\be
%\Sigma \simeq \Sigma_{1} + \frac{1}{V} \Big( A - \frac{B}{V} \Big) \ ,
%\ee
%which yields
%\be \label{Sig1}
%\Sigma \simeq 0.0199(8) \quad (L=16) \quad , \qquad
%\Sigma \simeq 0.0207(13) \quad (L=20 ~ {\rm or} ~ 32) \ .
%\ee
%In this case the term $A$ plays an almost negligible r\^{o}le,
%so this determination is quite independent from (\ref{Sig0r}).
These results for $\Sigma$ are nicely consistent, but more than 
a factor of $2$ below the value expected in infinite volume. 
% $\Sigma =0.0489$ . 
Hence for $m=0.01$ this method works in an intrinsically consistent 
way, but the results for $\Sigma$ are reduced by strong finite size 
effects.
%\footnote{We recall that the correlation length for this fermion
%mass (in infinite volume) is $\xi \simeq 14$, cf.\ 
%Figure \ref{correfig}, hence this problem is not too surprising.}

Similarly, if we now evaluate $\chi_{t}$ by referring to eq.\ 
(\ref{ABdef}) we obtain values between $10^{-5}$ and 
$10^{-4}$, {\it i.e.}\ well below the results 
%obtained with the method of 
in Subsection 5.1. Note that the applicability
of the method used here --- based on the approximations 
in Subsection 5.2 ---  is indeed questionable for $m=0.01$,
since we are dealing with a small value of $V \chi_{t}$.
%$V \chi_{t} \ll 1$. % (according to Table \ref{chitab}).

So the mass $m=0.06$ is more promising.
We cannot test the volume independence of $A$ from
our data, but based on $L=16$ and $32$ we obtain
%\bea
%\nu = 0 &:& A = 5.80(36) \ , \qquad \Sigma = 0.0940(16) %\\
%| \nu | = 1 &:& A = 5.80(60) \ , \qquad \Sigma = 0.0940(40)
%\eea
\be
m=0.06 \quad : \quad A = 5.80(36) \ , \quad \Sigma = 0.0940(16) \ .
\ee
Let us focus on the %(more reliable) 
size $L=32$ and involve $B_{1,0}$, which leads to
\be
%m=0.06 \quad : \quad 
\chi_{t} = 0.00118(30) \ .
\ee
The method that we used in Subsection 5.1 to evaluate 
$\chi_{t}$ does not work in this case as we mentioned before
(due to the sizable uncertainty of $\Sigma_{1}$ it basically just 
constrains $\chi_{t} < 0.03$). However, if we
now insert the preferred value $\chi_{t} \simeq 0.00118$
and evaluate $\Sigma$ by topological summation, we
arrive at
\be
%m=0.06 \quad : \quad 
\Sigma = 0.0883(69) \ .
\ee
This result is in excellent agreement with the analytical predictions 
based on Ref.\ \cite{Smilga}, $\Sigma =0.08883$. \\

We conclude that some consistency tests are passed well.
However, for $m=0.01$ the ultimate results for $\chi_{t}$ and
the $\Sigma$ (summed over all topological sectors) 
seem to be affected by strong finite size effects.
%are probably not realistic; the strong finite size effects seem to
%prevent the applicability of this method. 
%Note that it is designed 
%for low temperature, {\it i.e.}\ a large extent in the Euclidean 
%time, which we are not dealing with.

The situation improves as we
proceed to $m=0.06$. In this case the correlation
length (in infinite volume) is only $4.2$ lattice spacings,
so that the finite size effects are suppressed quite
well, and we arrive at sensible results for $\chi_{t}$ and
$\Sigma$. 

\subsection{Topological summation of the ``pion'' mass}

For practical reasons, it is favourable to consider the decay of 
a current correlation function for measurements of the ``pion'' 
mass, rather than the pseudoscalar density \cite{Ivan}.
In this way we obtained the results in Table \ref{pimasstab}.
%the last column gives the prediction of eq.\ (\ref{mpieta}).

\begin{table}[h!]
\centering
\begin{tabular}{|c|c||c|c|c|c||c|}
\hline
$L$ & $m$ & $M_{\pi ,0}$ & $M_{\pi ,1}$ & $M_{\pi ,2}$ &$M_{\pi ,3}$ 
&  $M_{\pi}^{\rm theory}(L=\infty )$ \\
\hline
\hline
16 & 0.01 & 0.041(1) & 0.271(4) & & & 0.071 \\
\hline
16 & 0.03 & 0.123(5) & 0.275(4) & & & 0.148 \\
\hline
16 & 0.06 & 0.214(6) & 0.310(3) & & & 0.235 \\
\hline
16 & 0.09 & 0.302(5) & 0.358(2) & & & 0.308 \\
\hline
16 & 0.12 & 0.359(5) & 0.413(2) & 0.494(5) & & 0.374 \\
\hline
16 & 0.18 & 0.498(4) & 0.525(2) & 0.589(5) & & 0.490 \\
\hline
16 & 0.24 & 0.631(6) & 0.648(2) & 0.700(3) & & 0.593 \\
\hline
\hline
20 & 0.01 & 0.038(2) & 0.209(4) & & & 0.071 \\
\hline
24 & 0.01 & & & 0.257(8) & 0.32(1) & 0.071 \\
\hline
28 & 0.01 & & 0.146(4) & & 0.25(1) & 0.071 \\
\hline
\hline
32 & 0.01 & 0.05(1) & 0.160(8) & 0.192(6) & & 0.071 \\
\hline
32 & 0.06 & 0.23(1) & 0.232(7) & & & 0.235 \\
\hline
\end{tabular}
\caption{\it The ``pion'' masses measured in various volumes,
at different fermion masses, in the topological sectors
$|\nu| = 0 \dots 3$. The last column displays the 
(infinite volume) prediction of Ref.\ \cite{Smilga}
(cf.\ eq.\ (\ref{mpieta})).}
\label{pimasstab}
\end{table}

Again we can test the topological summation for the fermion
masses $m=0.01$ and $0.06$. Here we proceed in a manner
different from Subsection 5.3: if we insert the data
of Table \ref{pimasstab} into the formula
\be
M_{\pi , |\nu |} = M_{\pi } - \frac{A}{V} + \frac{B}{V^{2}} \nu^{2} \ ,
\ee
the unknown terms $M_{\pi}$, $A$ and $B$ are over-determined.
We choose the optimal values for them by a least square fit. 

Let us start again with $m=0.01$: if we include all 11
sector that we simulated, we  obtain $M_{\pi } = 0.115(3)$,
and if we skip the 4 sectors with $|\nu | > 1$ (which
are problematic for our approximations) the result is 
slightly reduced to $M_{\pi } = 0.113(4)$.
Thus we confirm the observation of Subsection 5.3 that the 
summation formula runs into trouble for this light mass.
Actually this is not surprising:
we recall that the correlation length for this fermion
mass (in infinite volume) is $\xi \simeq 14$, cf.\ 
Figure \ref{correfig}, hence strong finite size effects
are expected, and they generically enhance the mass gap.

Therefore, we now skip the smallest volumes. 
We need at least two volumes to determine the term 
$A$, so we restrict the consideration to $L=28$ and $32$,
and we include $|\nu | \leq 2$; thus we are left with four
sectors. This small number of input data causes a large error, 
but the optimal value of the least square fit moves very close
to the theoretical prediction (see Table \ref{pimasstab}),
\bea
m=0.01 &:&  \ M_{\pi } = 0.073(25) \ . \nn
\eea

For $m=0.06$ we only have four sectors to deal with,
but the finite size effects are much less severe,
and we arrive again at a sensible result,
\bea
m=0.06 &:&  \ M_{\pi } = 0.232(8) \ , \quad
\chi_{t} = 0.0007(4) \ .
\eea
As we already saw in Subsection 5.3, this method is
not very useful for the determination of $\chi_{t}$;
it is always plagued by a large uncertainty, as in
the example given here (for $m=0.01$ this is even worse).
The three results that we obtained for $\chi_{t}$ at
$m=0.06$ (in Subsection 5.1, 5.3 and 5.4) differ within the
same magnitude.\footnote{The result in Table \ref{chitab}
is larger, but only based on $L=16$.}

Regarding the ``pion'' mass, we confirm that this method
can work, if it is applied in an appropriate regime.
Hence this approach may in fact be promising for further 
applications, including QCD, if finite size effects 
are under control --- in particular the term 
$\langle \nu^{2} \rangle$ should not be too small.

\subsection{Correlation of the topological charge density}

Members of
the JLQCD Collaboration proposed an alternative method to evaluate
$\chi_{t}$ even in one single topological sector based on
the topological charge {\em density}, $\rho_{t}(x)$. 
%In fact, their QCD configurations, which were
%generated with two flavours of dynamical overlap
%quarks, are restricted to the sector $\nu =0$.
Ref.\ \cite{fixtopo2} derived the ``model independent formula''
\be \label{Jap}
^{~~\lim}_{|x| \to \infty} \ \langle \rho_{t}(x) \rho_{t}(0) 
\rangle\vert_{| \nu |}
\simeq - \frac{\chi_{t}}{V} +
\frac{1}{V^{2}} \Big( \nu^{2} + \frac{c_{4}}{2 \chi_{t}} \Big) 
+ O(V^{-3}) \ ,
\ee
which captures even a possible deviation from the Gaussian
charge distribution by a non-vanishing kurtosis
$c_{4} = ( \langle \nu^{4} \rangle - 
3 \langle \nu^{2} \rangle ) / V \,$.\footnote{A variant of this
method, which uses the $\eta'$-correlation of the
pseudo-scalar density, has been applied to two-flavour QCD 
in Ref.\ \cite{chitQCDNf2}.} However, this
term is known to be tiny, so we neglect it in the following.

The issue is to search for a plateau value of the charge
density correlation at large distances, which differs from the 
constant $\nu^2 /V^{2}$ (in the sector with topological charge 
$|\nu|$). This shift $\Delta = - \chi_{t}/V$ is 
negative because fluctuations in $\rho_{t}(x)$ have to compensate.
However, $\Delta$ tends to be small and hard to
resolve numerically. For the absolute value $| \Delta |$ it
is favourable if $m$ increases (although we are actually
interested in quasi-chiral fermions), but it is unfavourable
if $V$ increases (although we need access to the asymptotic 
value in eq.\ (\ref{Jap}), and $O(V^{-3})$ should be negligible).
Moreover, in analogy to the approximations that lead to
eq.\ (\ref{finapprox}), the derivation of eq.\ 
(\ref{Jap}) involves the assumptions
\be  \label{asunu}
\langle \nu^{2} \rangle = V \chi_{t} {\rm ~~ is ~ large}
\ , \qquad 
\frac{|\nu |}{\langle \nu^{2} \rangle}  {\rm ~~ is ~ small} \ .
\ee

%According to our results in the preceding subsections,
Our estimates for $\chi_{t}$ suggest 
that we do not have any setting which is really
adequate for the first of these two conditions.
%as Table \ref{chitab} shows. 
Sometimes, however, such approximations
apply reasonably well even if the assumptions do not strictly
hold (this is the experience in QCD with simulation data matched 
to formulae of the $\epsilon$- or $\delta$-regime, and in the 
preceding two subsections to some extent). 

\begin{figure}[h!]
%\vspace*{-3mm}
\center
\includegraphics[angle=270,width=.62\linewidth]{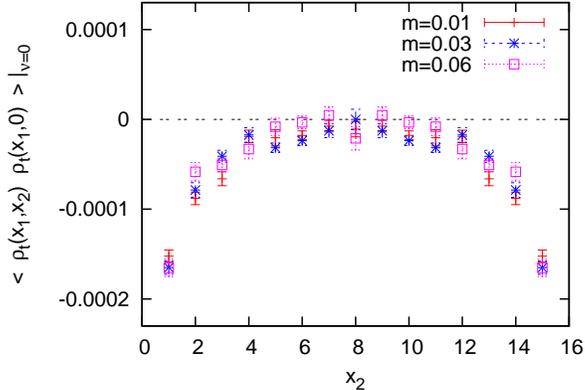}
\vspace*{-3mm}
\caption{\it{The topological charge correlation function for 
$L=16$, $\nu =0$ and $m = 0.01$, $0.03$ and $0.06$.}}
\label{topcharcor}
\vspace*{-3mm}
\end{figure}

Since our configurations are smooth, it is not problematic to 
use the naive lattice formulation of the topological charge density,
$\rho_{t} = \epsilon_{12} F_{12}$. (There is a cleaner formulation for
Ginsparg-Wilson fermions \cite{ML}, but it is tedious in practice.)
%As in Ref.\ \cite{chitQCDNf2} we sum over the spatial component and
%measure the correlation of the averaged time layers as a function of 
%their separation in Euclidean time, $t_{\rm E}$. 
As examples, the results for $L=16$ in the topologically neutral
sector for three masses are shown
in Figure \ref{topcharcor}. In particular we see that the correlation
over short (but non-vanishing) distances is negative, 
since a given link variable
contributes with opposite signs to $\rho_{t}$ in its adjacent 
plaquettes. The jackknife errors for the quantity
$\langle \rho_{t}(x) \rho_{t}(0) \rangle\vert_{0}$ are typically
in the range $(1 \dots 2) \cdot 10^{-5}$.

In the most promising case, $m=0.06$, we evaluated 741 topologically
neutral configurations and we have
identified the magnitude for $\chi_{t} \approx 0.0015$ %= O(10^{-3})$, 
which implies $\Delta \approx -6 \cdot 10^{-6}$. 
In order to safely resolve this shift
we would therefore need about $50\, 000$ to $100 \, 000$ 
%$O(10^{5})$ 
configurations. We conclude that
this method requires a very large statistics, 
so we cannot apply it.

\section{Conclusions}

We presented a study of the 2-flavour Schwinger
model with dynamical chiral fermions. We applied the
overlap Hypercube Fermion (HF), and confirmed its features
as a promising formulation of a chiral lattice fermion. This is
manifest by its excellent locality and scaling 
behaviour. Moreover, our study allowed us to explore
a number of conceptually interesting and relevant issues,
and to test new methods to handle them. 

This is certainly of interest, even in a toy model,
since simulations with dynamical overlap fermions
have been explored only poorly so far.\\

For our simulation we designed  
a specifically suitable variant of the Hybrid Monte Carlo (HMC)
algorithm, demonstrated its correctness and tested its efficiency. 
It is sufficient to insert a low polynomial approximation 
for the overlap operator to compute the force term,
while the high precision overlap operator is employed in
the accept/reject step. It is an open question if this strategy 
--- perhaps with further refinements --- can be carried over
to the simulation of QCD with dynamical chiral quarks.\\

Next we discussed the spectrum of the Dirac operator.
Random Matrix Theory is not yet worked out for this type of model, 
with a vanishing chiral condensate at zero fermion mass. 
Nevertheless the unfolded level spacing density follows the 
standard RMT formula for the unitary ensemble.

The prediction $\Sigma (m) \propto m^{1/3}$ \cite{Smilga92,HHI,Smilga}
suggests a microscopic spectral density $\rho (\lambda \gsim 0) \propto 
\lambda^{1/3}$, and the scale-invariant variable 
$\zeta \propto \lambda V^{3/4}$ (if we assume %that there is 
no explicit mass dependence of $\rho$). %, $\rho (\lambda, m)$).
This conjecture does, however, not agree with our data.
An alternative scenario with %a microscopic density 
$\rho (\lambda \gsim 0) \propto \lambda^{1/2}$
has a theoretical background as well (Gaussian distribution), 
but the data do not really support it either.
Instead they favour $z \propto \lambda V^{5/8}$
as the scale-invariant variable, and therefore 
$\rho (\lambda \gsim 0) \propto \lambda^{3/5}$.

A hint for an interpretation of this surprising result 
can be found in Ref.\ \cite{HHI}, 
which derived the behaviours $\Sigma \propto m^{1/3}$ and
$\Sigma \propto mL$ in two limiting cases. Regarding the
parameter that characterises these extreme cases, our settings
are in an intermediate regime, which appears compatible with the 
relation $\Sigma \propto m^{3/5}$ that we observed. For a precise 
theoretical clarification, we hope for the corresponding RMT 
formulae to be elaborated. \\

Direct measurements of the chiral condensate 
%--- based on a summation over the entire Dirac spectrum ---
and of the mass of the iso-triplet (``pion'')
%and the iso-singlet (``pions'' and 
%``$\eta$-particle'') 
could only be performed in fixed topological sectors, since
the HMC histories contain only few topological transitions.
This limitation is generic for the simulation of
light fermions close to the continuum limit. 
It is therefore a major challenge to
explore techniques for the topological summation of such
measurements. Assuming a Gaussian distribution of the topological
charges, the data can be used to evaluate the topological
susceptibility. In order to determine the actual observable,
we explored approximate summation techniques, which
led to sensible results in some parameter window, 
where the term $\langle \nu^2 \rangle = V \chi_{t}$
is not too small.

Topological summations could become relevant for future 
QCD simulations.
Nowadays they are carried out with very light dynamical quarks 
--- such that the pion mass is close to its physical value ---
on finer and finer lattices. In particular in applications of
dynamical overlap fermions \cite{RMTdQCD2} the HMC history
tends to be trapped in the topologically trivial sector
for the entire simulation, which endangers ergodicity
and does not provide the physical result (unless the volume
is very large). For Wilson-type lattice fermions
(which break the chiral symmetry explicitly), the problem is less 
obvious on the currently used lattice spacings, but on still
finer lattices (with $a \lsim 0.05 ~ {\rm fm}$)
it is expected to show up as well \cite{MLtopo}.
Hence topological summation methods are of interest for
non-perturbative studies of low energy nuclear physics
based on first principles of QCD. Our results tell us to be 
cautious with such summations, but at the same time they 
provide hope for their feasibility.

\vspace*{7mm}

\noindent
{\small
{\bf Acknowledgements} \ We are indebted to Martin Hasenbusch for
helpful advice on algorithmic aspects, to Poul Damgaard
and Jacques Verbaarschot for sharing their deep 
insight into Random Matrix Theory, and to Stephan D\"{u}rr,
Hidenori Fukaya and Jim Hetrick for very useful comments. 
Most production runs were performed on the clusters of the 
``Norddeutscher Verbund f\"{u}r Hoch- und H\"{o}chstleistungsrechnen'' 
(HLRN). We thank Hinnerk St\"{u}ben for technical assistance,
as well as Edwin Laermann and Michael M\"{u}ller-Preu\ss ker for 
their support at Bielefeld and Humboldt University, respectively.
This work was supported in part by the Croatian Ministry of
Science, Education and Sports (project No.\ 0160013), and by 
the Deutsche Forschungsgemeinschaft (DFG) through 
Sonderforschungsbereich Transregio 55 (SFB/TR55) 
``Hadron Physics from Lattice QCD'', which is coordinated 
by the University of Regensburg.}

\end{document}